\documentclass{achemso}
\usepackage[version=3]{mhchem, xcolor} 
\usepackage{graphicx}
\usepackage{xcolor}
\usepackage[normalem]{ulem}
\usepackage{float}
\usepackage[outdir=./]{epstopdf}

\author{Miriam Marqués}
\affiliation{Centre for Science at Extreme Conditions and School of Physics and Astronomy, University of Edinburgh, Edinburgh, U.K.}

\author{Miriam Pe\~{n}a Alvarez}
\affiliation{Centre for Science at Extreme Conditions and School of Physics and Astronomy, University of Edinburgh, Edinburgh, U.K.}

\author{Miguel Martinez-Canales}
\affiliation{Centre for Science at Extreme Conditions and School of Physics and Astronomy, University of Edinburgh, Edinburgh, U.K.}

\author{Graeme J. Ackland}
\affiliation{Centre for Science at Extreme Conditions and School of Physics and Astronomy, University of Edinburgh, Edinburgh, U.K.}

\title{Breaking the H$_2$ chemical bond in a crystalline environment}


\begin{document}

\begin{tocentry}
\includegraphics[width=1\columnwidth]{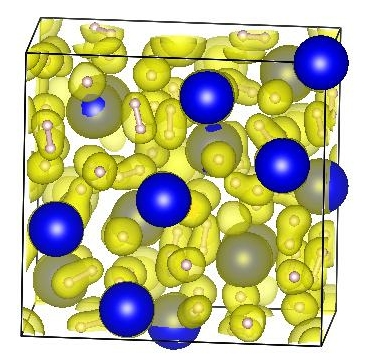}
\end{tocentry}

\begin{abstract}

Through density functional theory and molecular dynamics calculations, we have analysed various metal polyhydrides to understand whether hydrogen is present in its molecular or atomic form - tetrahydrides of Ba,Sr,Ra, Cs and La; Ba$_8$H$_{46}$ and BaH$_{12}$.
  We show that, in experimentally reported binary barium hydrides (BaH$_x$), molecular H$_2$ and atomic H$^-$ can coexist with the metallic cations.
  In this thorough study of differences between BaH$_4$, higher barium hydrides, and other binary tetrahydrides we find the number of atomic hydrogens is equal to the formal charge of the cations.  The remaining hydrogen forms molecules in proportions yielding, e.g. BaH$_2$(H$_2)_x$, at pressures as high as 200 GPa.  At room temperature these are highly dynamic structures with the hydrogens switching between H$^-$ and H$_2$ while retaining the 2:x ratio.

We find some qualitative differences between our static DFT calculations and previously reported structural and spectroscopic experimental results. Two factors allow us to resolve such discrepancies: Firstly, in static relaxation H$_2$ must be regarded as a non-spherical object, which breaks symmetry in a way invisible to X-rays; 
Secondly the required number of molecules $x$ may be incompatible with the experimental space group (e.g. $BaH_2(H_2)_5$). In molecular dynamics, bond-breaking transitions between various structural symmetry configurations happen on a picosecond timescale via an H$_3^-$ intermediate.  Rebonding is slow enough to allow a spectroscopic signal but frequent enough to average out over the lengthscale involved in diffraction.  

\end{abstract}

\maketitle

Recent 
reports of superconductivity approaching room temperature \cite{Drozdov2015,drozdov2018superconductivity,drozdov2019,nakao2019superconductivity,osmond2022clean}
have galvanised attempts to synthesise hydrogen-rich materials. Much effort is being invested to understand their crystal structures and properties\cite{roman2022silico,errea2022superconducting,errea2020,pickard2020superconducting,zhang2022superconducting,duan2017structure,flores2020perspective,sun2020second}.   
However, high-temperature hydride superconductivity is associated not with a particular crystal structure, but with the stabilisation of dense atomic hydrogen over molecular dimers \cite{ashcroft1968, Ashcroft2004,zurek2009}.
In this paper we address the fundamental issue of whether hydrogen in metal hydrides is atomic or molecular. \cite{pena2020,pena2022chemically,chen2021synthesis}

Molecular crystals provide a unique challenge for structural studies, especially under high pressure.  
From the experimental side, a major challenge is the 
very weak X-ray scattering by hydrogen atoms.  Spectroscopy can only give indirect information about the atomic structure, e.g. Raman signal related to H$_2$ molecules whose shift can be related to their intra-molecular bond length.

Density Functional Theory (DFT) is the dominant theoretical method applied for structural determination in these materials. An implicit assumption here is that the lowest enthalpy static arrangement of the ions will be good candidates for the the lowest Gibbs free energy state.
In static relaxations, the point symmetry of molecules may constrain the  molecular orientation in a given crystal symmetry.  However, in practice, due to quantum or classical rotation, H$_2$ molecules may behave as spherical rather than linear objects, which in turn allows high symmetry crystals\cite{van1983solid}.

Molecular hydrogen is understood as two hydrogen atoms joined by a covalent bond.  In compression of pure hydrogen, the H--H bond lengthens from 0.718 at 6 GPa in phase I to 0.84 {\AA} when entering phase IV at 225 GPa\cite{magduau2013identification,ackland2020structures,howie2012}.  Although there is no unique definition of a covalent bond in DFT, many measures can be indicative.  For example, a well-defined interatomic distance which persists long enough for the bond to oscillate;  a high value of electron localization function (ELF) between atoms; a build up of charge in the region between the atoms; a narrow Kohn-Sham band of energies which when projected onto localised basis functions is localised on the bond. 
We will utilise all these methods to build a coherent picture of the presence or absence of covalent bond in a binary hydride solid.

\begin{figure}[t!]
\includegraphics[width=0.15\columnwidth]{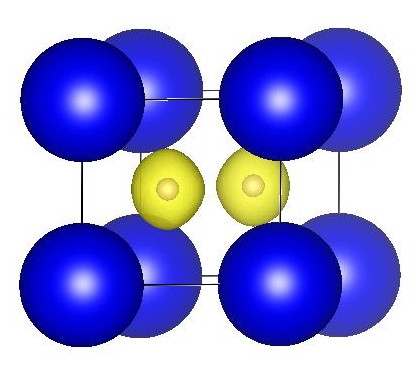}
\includegraphics[width=0.2\columnwidth]{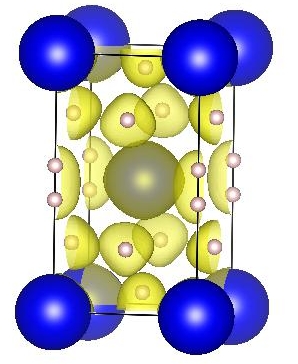}
\includegraphics[width=0.2\columnwidth]{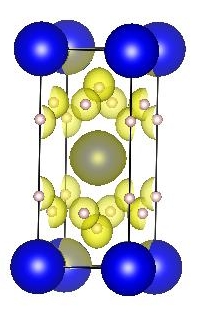}
\includegraphics[width=0.35\columnwidth]{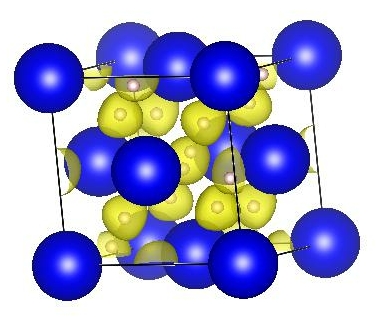}
\includegraphics[width=0.25\columnwidth]{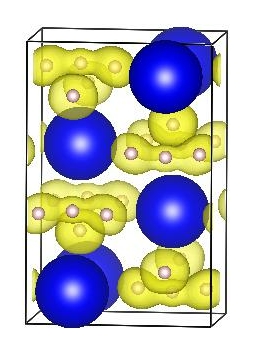}
\includegraphics[width=0.4\columnwidth]{50pctotisocrop.jpg}
\includegraphics[width=0.32\columnwidth]{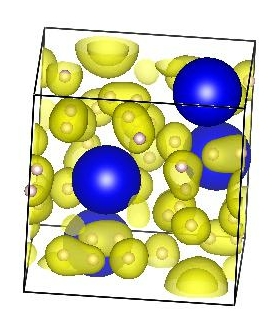}
\caption{Crystal structures, with Ba and H atoms represented as blue and pink spheres, respectively. ELF isosurfaces (ELF=0.85) shown in yellow are associated with anionic hydrogens and molecules.  From top left $P6/mmm$-BaH$_2$, $I4/mmm$-BaH$_4$ (low c/a), $I4/mmm$-BaH$_4$ (high c/a), $R\bar{3}m$-BaH$_4$, $Cmcm$-H3-BaH$_4$, $Pc$-BaH$_{5.75}$, $P2_1$-BaH$_{12}$ at 50 GPa.
\label{fig:structures}}
\end{figure}

Stable alkaline-earth metals (marked as M hereafter) are known to form binary hydrides (Fig.\ref{fig:structures}). The lowest hydrogen content is experimentally seen in the dihydrides, MH$_2$. These satisfy the octet rule, and no molecular hydrogen is expected\cite{tse2007structural, smith2009sr, Kinoshita2007, Smith2007, Tse2009structure}. 
Tetrahydrides have been  experimentally reported for Ca, Sr and Ba, containing their binary hydride or salt (divalent metal atom, M$^{2+}$, anionic hydrogens, H$^-$), and molecular hydrogen which was interpreted as slightly charged, H$^{{\delta}-}_2$\cite{Bi2021}.
Ca\cite{Mishra2018, Wu2019, pena2022chemically} and Sr\cite{hooper2014, pena2022chemically}  tetrahydrides adopt the $I4/mmm$ structure, both with $c/a$ ratios around 1.6, while BaH$_4$ forms an $I4/mmm$ structure with $c/a$ of 2.2, as well as an $fcc$ structure \cite{pena2022chemically}. 
Raman spectroscopy supports H--H bonds in these tetrahydrides\cite{Mishra2018, Bi2021, pena2022chemically}. Moreover, synthesis of $Pm\bar{3}n$ Ba$_8$H$_{46}$ (BaH$_{5.75}$) at 50~GPa\cite{pena2021synthesis}, and $Cmc2_1$ BaH$_{12}$ at 90~GPa\cite{chen2021synthesis} has also been reported. Neither compound has an identifiable Raman contribution from the stretching H$_2$ mode. Therefore, barium hydrides represent an ideal system to study how the change in stoichiometry alters the H$_2$ molecule within the solid \cite{pena2020, pena2022chemically, chen2021synthesis}.
Thus, a systematic investigation of the mechanisms involved in the hydrogen intra-molecular bond weakening and the influence of the metallic atomic number  is here undertaken as the route to understand the concepts  of bond-weakening and chemical pre-compression. 

In this work we conduct a thorough theoretical analysis of BaH$_4$ and compare  with other binary metal tetrahydrides with $I4/mmm$ symmetry, along with Ba$_8$H$_{46}$ and BaH$_{12}$. We have run an extensive series of DFT calculations. Electronic structure is explored using the ELF and Bader (Quantum Theory of Atoms in Molecules, QTAIM\cite{Bader94}) topologies. Details of the calculations, as well as further data, may be found in the supplemental material (SM). In this paper we will start by discussing the DFT approach on molecular hydrogen. We then discuss quantitative results from static relaxations for all materials, followed by qualitative results from molecular dynamics.

Breaking the H$_2$ bond may occur due to mechanical pressure or electronic effects. To illustrate the electronic effects in isolation, we calculated the hydrogen molecule in a homogeneous electron gas 
\cite{bonev2001hydrogen}, 
(SM4.3).
For low electron density the band structure comprises a clearly-defined H$_2$ bonding state and a free electron-like density of states which lies much higher in energy. At electronic densities above 0.06~$e/\textrm{\AA}^{3}$ ($r_s\approx 3.0~a_0$), the bond breaks spontaneously. This is evidenced by a discontinuous jump in the
H-H separation, the vanishing of the flat Kohn-Sham band defining the molecule and the
appearance of atomic-hydrogen (H$^-$) states at the bottom of the free-electron band. This suggests an upper limit on the electron density where molecular hydrogen can be found. Interestinglym the maximum bond length before the transition is less than 0.9 \AA.

The easiest approach to identify a bond is to define some interatomic cutoff distance $R_{\textrm{bond}}$, and declare that any pair of hydrogen atoms closer than this are \emph{``bonded''}.  
Molecular dynamics  requires $R_{bond}$ to be large enough that an oscillating molecule is not marked as ``breaking'' whenever the bond-length briefly extends beyond $R_{\textrm{bond}}$ at its largest extent.  Consequently $R_{\textrm{bond}}$ must be set larger than one might expect a H-H bond to be. Based on the observation of a minimum in radial distribution functions, $R_{\textrm{bond}} = 1~$\AA\ is a pragmatic choice. We do not draw conclusions which are highly sensitive to this choice.
We further refine our definition by counting the number of atoms with precisely one neighbour within 1 \AA\,, and dividing by 2 to get the number of molecules. In practice we find that bond-breaking usually occurs via the approach of a third atom and formation of an intermediate H$_3^-$ complex, a process which our definition counts as one molecule throughout.
Nevertheless, we record every instance where an H-H distance passes through the 1~\AA\, threshold --- this is used for an estimate of bond lifetimes.  Such a coarse lifetime estimate is not quantitatively meaningful - we use it only for comparisons between structures and not, e.g. for calculating Raman linewidths \cite{ackland2014efficacious,cooke2020raman,cooke2022calculating}.
A detailed discussion of the technical definition of a bond is given in the SM.

In static relaxation, including from MD snapshots, other measures from electronic structure can be applied to detect molecules. In $k$-space, a discrete set of electron bands may represent the bond, and this can be tested using a projection of the Kohn-Sham bands onto maximally-localised Wannier functions. On the other hand, in real space, and within Bader topology, H$_2$ molecules are identified by the presence of a covalent bond between the hydrogen atoms. It is characterized by the high value of the electron density and the negative sign of the laplacian at the bond critical point (first-order saddle point of the electron density connecting the two maxima located on the atoms) (see SM). The charge of the atoms forming a molecule can be unambiguosly calculated by integration of the electron density within the topological atoms forming the molecule. They are defined by the union of the electron density maxima (located on the atoms) with their attraction basins and delimited by perfectly defined surfaces obeying the zero-flux condition for the electron density. ELF is a relative measure of the electron localization with respect to the homogenous electron gas (HEG). In general, the ELF value approaches 1 in regions of the space where electron pairing occurs (e.g., atomic shells, bonds and lone pairs). In analogy with QTAIM, a partition of the space based on the ELF can be performed. It consists of non-overlapping basins with well-defined chemical interpretation (cores, bonds, lone pairs). Moreover, the basin populations come from integration of the electron density within these regions. The signature of a H$_2$ molecule within the ELF topology is the existence of a high ELF isosurface enclosing both atoms. Typically, a H$_2$ molecule is considered to exist if the minimum ELF value between the two hydrogen atoms is above 0.85. We find these methods to be consistent with one another, and consistent with the much faster bondlength criterion.

We performed extensive DFT geometry optimisation calculations, fully relaxing the atomic positions.  Important structures are shown in Figure~\ref{fig:structures}, and we will focus our detailed discussion on 50~GPa. Other structures, as well as details at other pressures, are described in the SM.   We emphasise none of the lowest-enthalpy structures have the symmetry reported from X-ray experiments.  

\begin{table}[hbt]
    \centering
    \begin{tabular}{|c|c|c|c|c|c|}
    \hline
    System & c/a  & nearest H--H, \AA & volume & o,t ELF basin population  & metal \\
    \hline
    BaH$_4$ $Cmcm$ H$_2$ & 1.716 & 0.806 &  68.00 & & no  \\
    BaH$_4$ $Cmcm$ H$_3$ & 1.469 & 0.944 (H$_3^-$) &  68.12 && no  \\
    BaH$_4$ $C2/c$ & 1.514 & 0.829 & 70.87 & & no   \\
    BaH$_4$ $C2/c$ I & 1.467 & 0.942 (H$_3^-$) & 68.17 & & no   \\
    BaH$_4$ $C2/c$ II & 2.000 & 0.822 & 68.756 && no   \\
    BaH$_4$ $R\bar{3}m$ & - & 0.831 & 72.44 & & no   \\
\hline
    BaH$_4$ (low) &  1.763  & 0.803 &  72.28 &   1.047 1.731 & no   \\
    BaH$_4$ (high) & 2.323 & 1.495 &  65.96 &1.367 1.337  & yes \\
    \hline
    SrH$_4$ (low) & 1.712 & 0.798 &  61.42 & 1.050 1.772 &   no         \\
    SrH$_4$ (high)& 2.083 & 1.629 &  57.14 & 1.362 1.416 & yes       \\
    RaH$_4$ (low) & 1.721 & 0.796 &    78.58 & 1.052 1.773 & no  \\
    RaH$_4$ (high) & 2.338 & 1.655 &    72.22 & 1.390 1.362 & yes  \\
    CsH$_4$ (low) & 1.642 & 0.768 &    69.38    &  1.029 1.451 &  yes  \\
    CsH$_4$ (high) & 2.821 & 1.489 &    67.98    &  1.258 1.098 &  yes  \\
    LaH$_4$ & 1.863 & 1.502 &     64.56   &    1.450 1.604 & yes   \\
    \hline
    \end{tabular}
    \caption{Calculated PBE properties of $I4/mmm$-XH$_4$ structures at 50GPa, and five other structures\cite{pena2022chemically}. c/a and volumes refer to the conventional  cells. ``high" and ``low" refer to the initial c/a ratio before relaxation. Atomic ELF population centered on the octahedral site (forming two independent atoms or a molecule) and located on the tetrahedral site respectively.  LaH$_4$ exhibits only one  minimum \label{tab:I4mmm} }
    \end{table}

$I4/mmm$ is the symmetry reported from X-ray diffraction measurements for the cation positions in a number of tetrahydrides\cite{pena2022chemically,bi2021electronic}.
For Sr, Ba, Ra and Cs $I4/mmm$ tetrahydrides,  we find two energy minima in $I4/mmm$ as a function of c/a ratio (Table \ref{tab:I4mmm}).  Despite having the same symmetry, there is a clear distinction between the two structures which have very different c/a ratios and H--H separations.  The high c/a version ($c/a \sim$ 2--2.8) is metallic and has atomic H$^-$. The low c/a version ($c/a\sim $1.7) has molecular H$_2$ and is semiconducting in alkaline earths, but metallic in Cs and La.  Typically, the atomic version has higher density and becomes stable as higher pressure. 

\begin{figure}[t!]
\includegraphics[width=0.47\columnwidth]{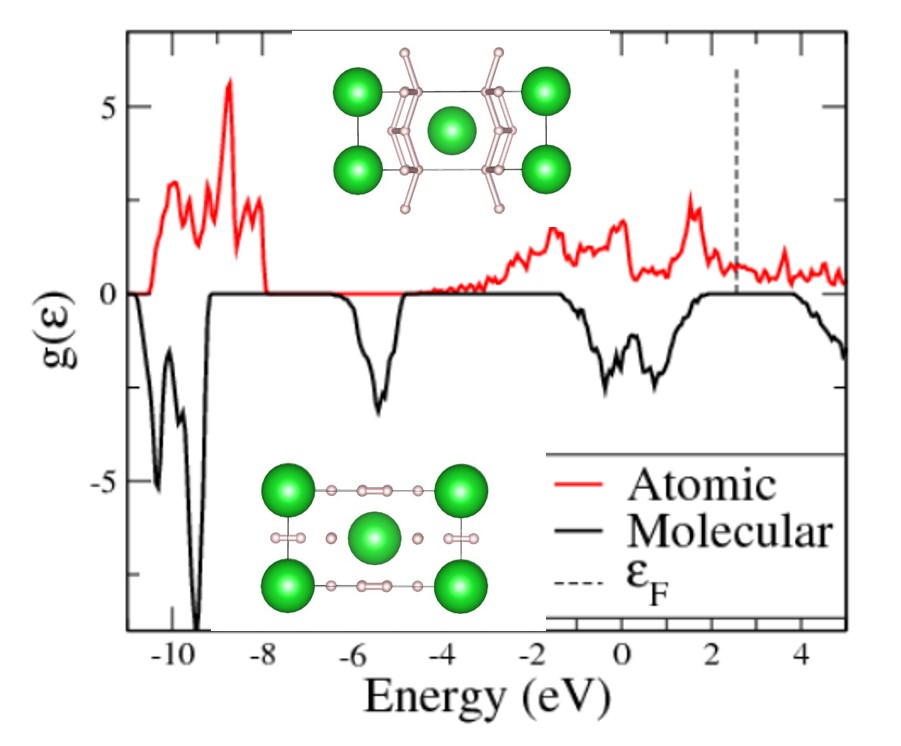}
\includegraphics[width=0.52\columnwidth]{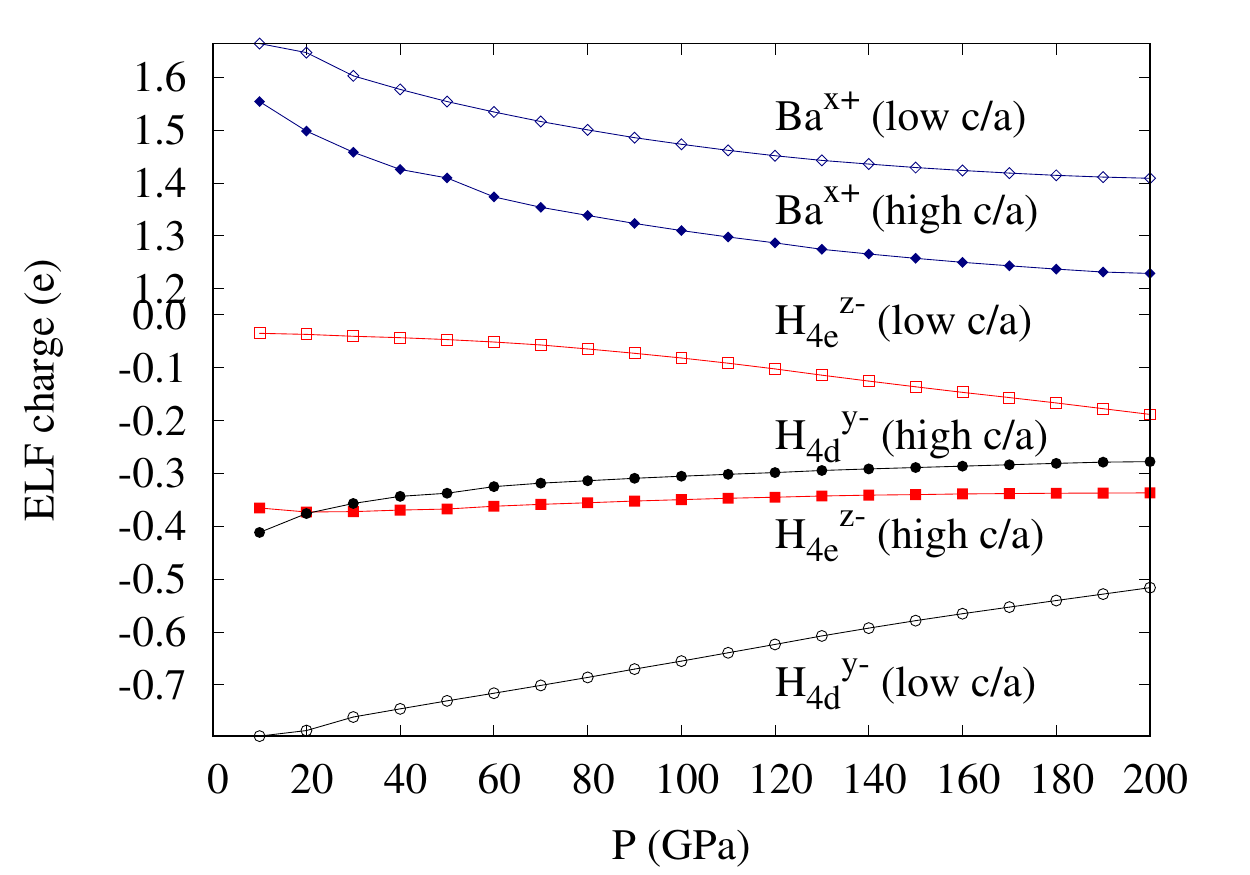}
\includegraphics[width=0.49\columnwidth]{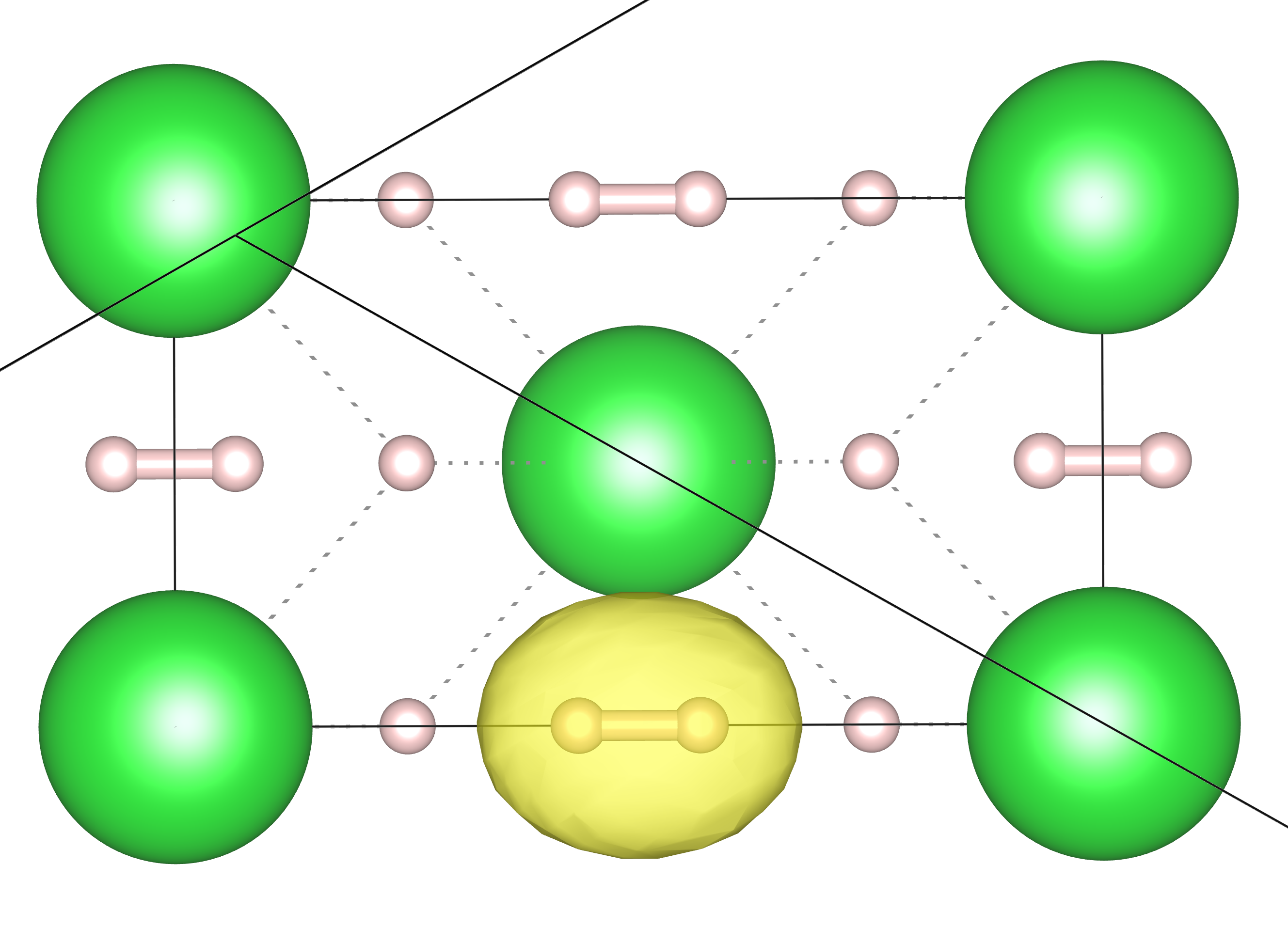}
\includegraphics[width=0.49\columnwidth]{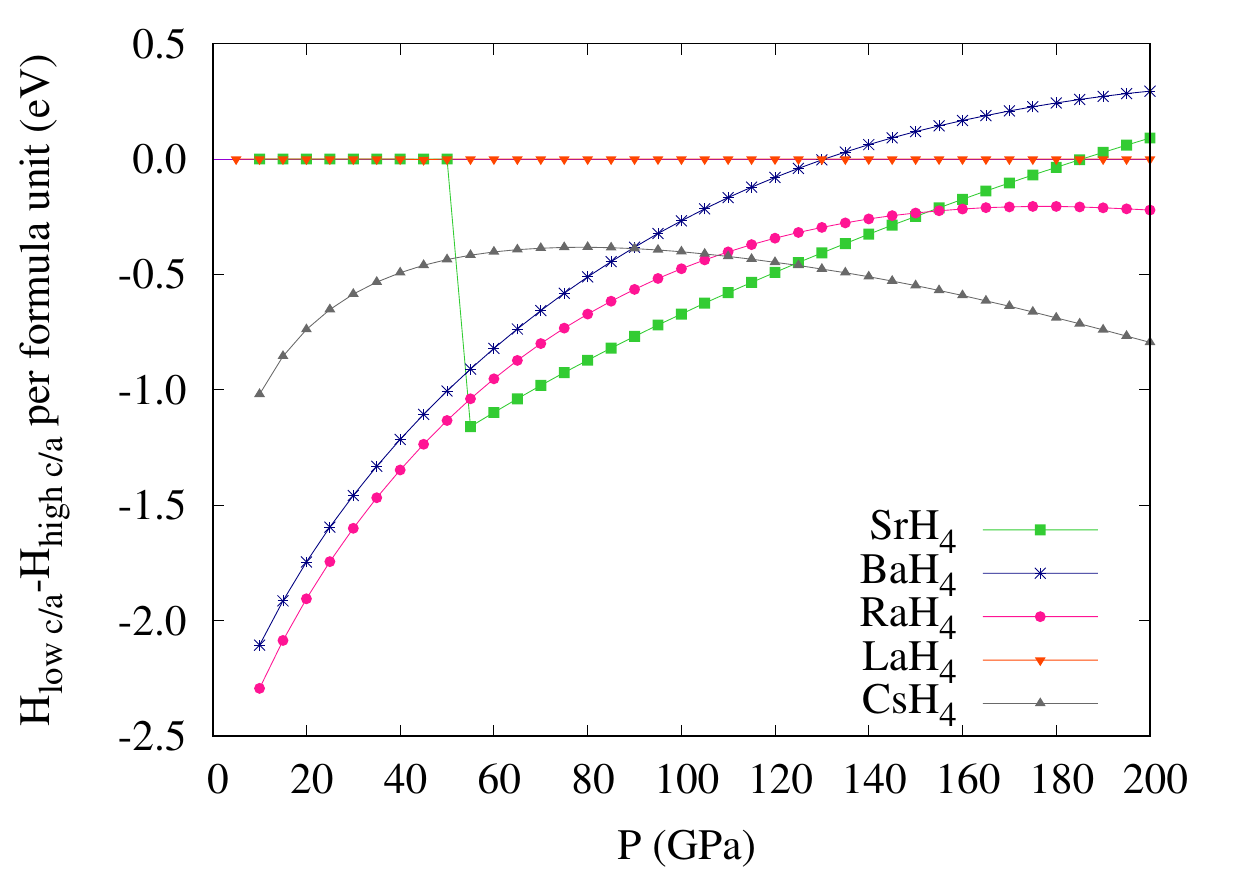}

\caption{Results of DFT calculations for  the isostructural forms of  $I4/mmm$-BaH$_4$.  Atomic $I4/mmm$-BaH$_4$ has high c/a $\approx$ 2.40 , closest hydrogen pairs are $\approx$1.5\AA\, apart in the xz plane.  Molecular $I4/mmm$-BaH$_4$ has  low c/a $\approx$ 1.8 at 100GPa:  bonds are along z-direction $<0.85$\AA\,  (a) Density of electronic states at 10 GPa (b) Charges based on ELF topology. (c) Maximally Localised Wannier Function whose projection corresponds to the covalent band  (d) Enthalpy differences between high c/a and low c/a for various $I4/mmm$ compounds, n.b. LaH$_4$ has only one minimum at low pressure
\label{fig:DFT}}

\end{figure}

Figure~\ref{fig:DFT} and Tables \ref{tab:I4mmm} and S1 show the two $I4/mmm$ structures, ELF populations and Bader charges.  All hydrogens in the high c/a phases have broadly similar charges, whereas, in the low c/a phases, the molecular hydrogens are essentially neutral (the integration of the electron density within the H$_2$ units according to both,  Bader and ELF topologies gives a value close to 2 electrons)  whereas atomic hydrogens have a sizeable negative charge. The evolution of the ELF charges with pressure is shown in Fig.\ref{fig:DFT}b and S4. In low c/a structures, in all compounds, a narrow band of states around 6~eV below the Fermi energy is clearly associated with the H$_2$ molecules (Fig.\ref{fig:DFT}). Projecting that band onto maximally localised Wannier functions (Fig \ref{fig:DFT}c) shows this band to be primarily associated with a H--H $\sigma$ bond.
Finally, Figure \ref{fig:DFT}d shows the enthalpy crossover from the low-c/a to the high c/a with pressure, (zero in LaH$_4$  below 50~GPa where only a single minimum exists).

\begin{figure}[htb]
\includegraphics[width=0.50\columnwidth]{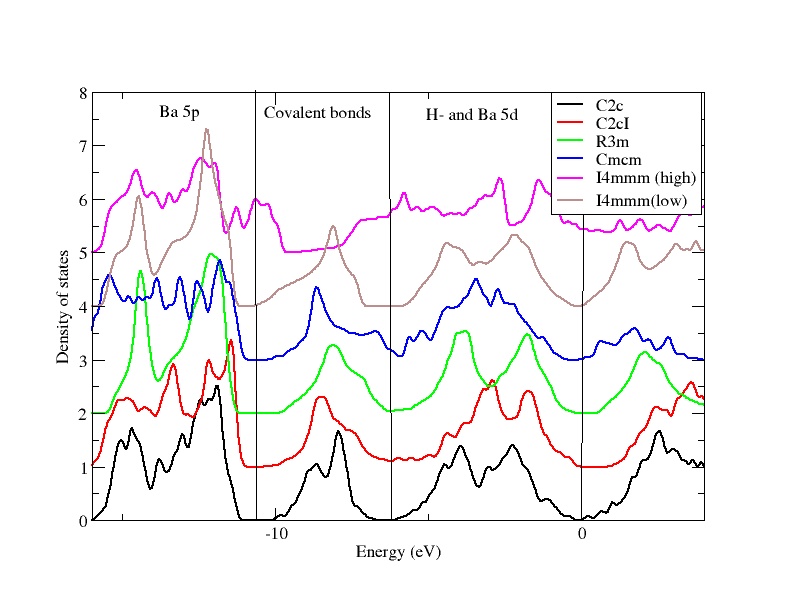}
\includegraphics[width=0.46\columnwidth]{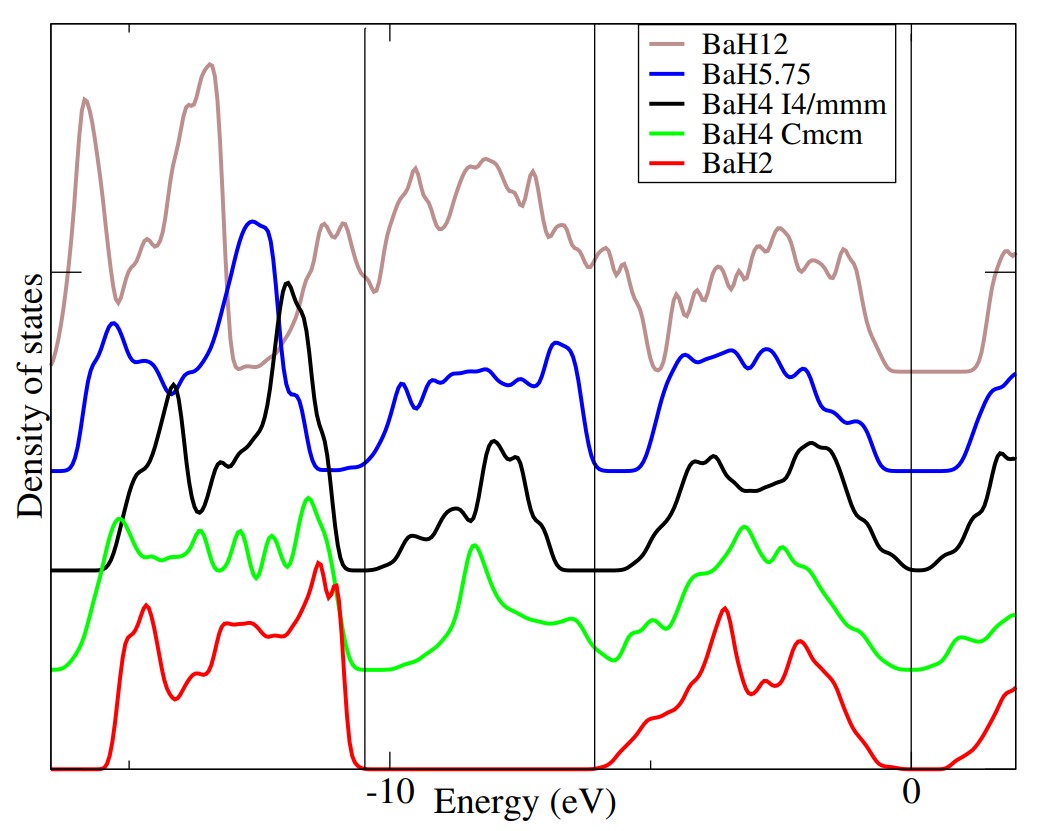}
\caption{Density of electron states from statically relaxed Barium hydrides at 50GPa normalised to BaH$_2$(H$_2)_x$  calculated using PBE. (a) BaH$_4$ candidates, with four distinct bands associated with their largest projection onto localised states 
 Ba $5p$,  H$_2$, Ba $5d$+H, Ba $6s+5d$. 
(b) Other stoichiometries.  Cmcm refers to Cmcm-H2, Note that the BaH$_2$ $P6_3/mmc$ (red line) has no H$_2$. \label{fig:dos}}
\end{figure}

In BaH$_4$ the static $I4/mmm$ calculation favours the low c/a $\approx 1.76$, but other BaH$_4$ structures are calculated to be more stable than any $I4/mmm$\cite{pena2022chemically} (see SM). Some of these are similar to $I4/mmm$, e.g. $R\bar{3}m$ is simply a rotation of the hydrogen molecules out of $I4/mmm$ symmetry leaving the Ba in very similar positions. However, two structures with $Cmcm$ symmetry have especially low enthalpy.  These differ in enthalpy by only a few meV at 50~GPa, but are chemically very different ($Cmcm$-H2 lattice angle $\gamma=141^o$ with H$_2$ molecules and $Cmcm$-H3 $\gamma=114^o$ with H$_3^-$ units, see Fig.\ref{fig:structures}). The $Cmcm$-H3 version is more stable with the PBE and BLYP functionals, whereas LDA favours  $Cmcm$-H2.  Both $Cmcm$ structures contribute two e$^-$/f.u.\ to the ``H2'' band (Fig.\ref{fig:dos}). In any case, these  structures are very different from either experimentally reported structure for BaH$_4$.\cite{pena2022chemically}


We now investigate how molecular dynamics (MD) simulations can illustrate the dynamics of these systems.
It would be prohibitively expensive to run accurate simulations of phase transformations or melting with conventional MD, because the Ba atoms are 137 times heavier than hydrogen and move too slowly. Therefore, an efficient way to sample the phase space is to run {\it ab initio} molecular dynamics in CASTEP with fictitious masses, NPT ensemble and ringing-mode quenching\cite{ackland2016rapid} (Details SM2.2, Table S3). The masses of H$_2$ and Ba are set to be equal, so that all atoms are able to move significantly. 
The MD runs are analysed using radial distribution functions, lattice parameters, mean squared displacements, and {\it ab oculo} using VESTA and VMD \cite{momma2011vesta,humphrey1996vmd}.  After each MD run, snapshots from the trajectory were relaxed to 0~K. These provided low-symmetry structure candidates, and allowed us to validate the bond assignment using ELF or charge density methods. 
Full details from the molecular dynamics runs and their analysis is given in the supplementary materials table S2. Here we describe features common to all barium hydrides.

The mean squared displacement (MSD) shows that the hydrogens move much further than Barium, but the MSD quickly flattens off, showing all the materials are crystalline up to 600~K; not superionic nor liquid. The one exception was  BaH$_{5.75}$ at 1000~K which went through a melting transformation after 1.2~ps.

\begin{figure}[t!]
\includegraphics[width=0.48\columnwidth]{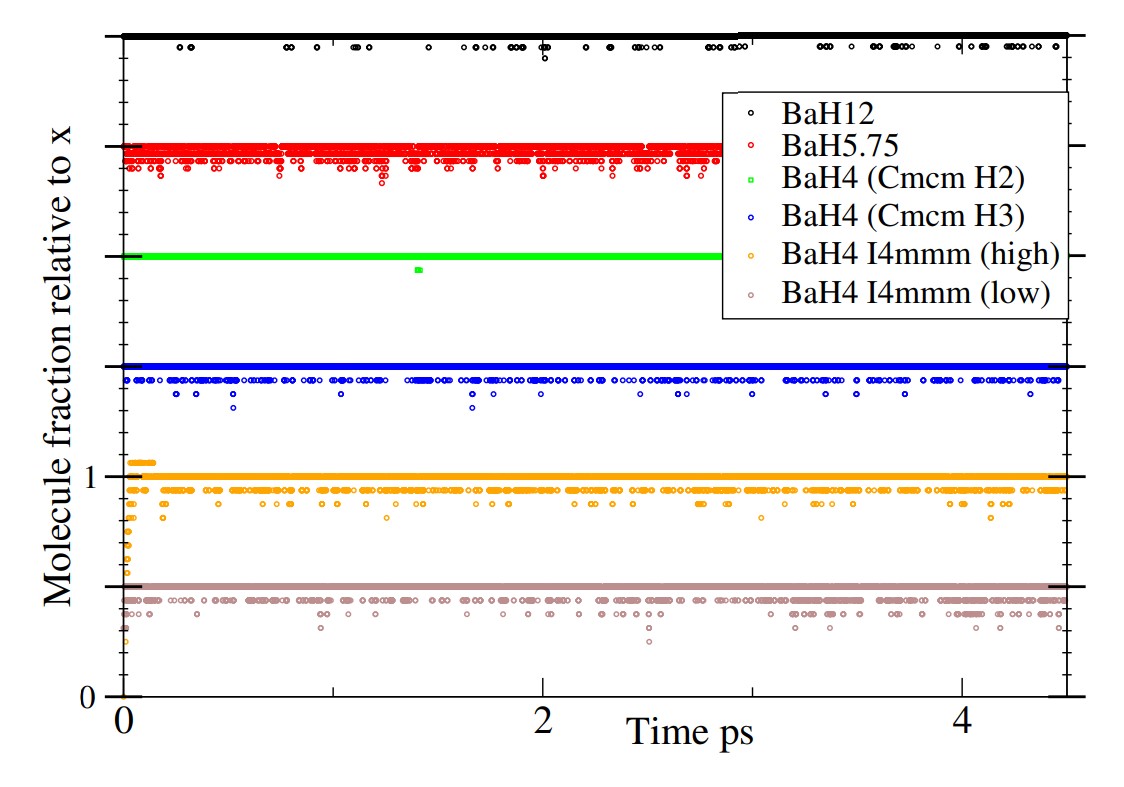}
\includegraphics[width=0.48\columnwidth]{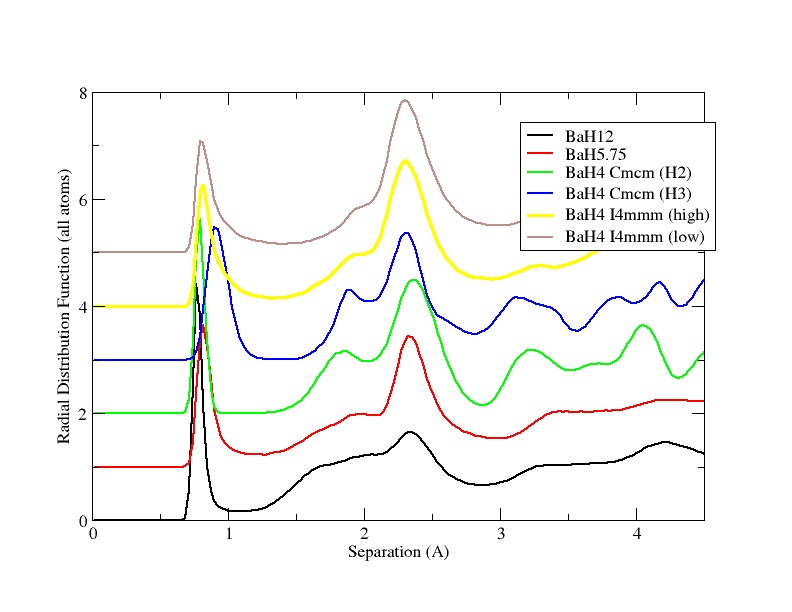}
\caption{ Analysis of MD at 50~GPa/300~K.  I4mmm/fcc indicates simulations in $I4/mmm$ for which the c/a ratio is close to $\sqrt{2}$ (a) Fraction of bonds relative to $x$ BaH$_2$(H$_2)_x$ in as a function of time for various structures. Each line is displaced by 0.5 for clarity (b) Radial distribution functions, normalized to number of atoms pfu.  The large peak at 2.4~\AA\, corresponds to BaH pairs.
}\label{fig:md}
\end{figure}

 Typical radial distribution functions (RDF) are shown in Fig.\ref{fig:md}.  They all have a well-defined peak between 0.7\AA\, and 1.0\AA\, which defines the molecule.  Integrating the first peak, or counting the number of bonded pairs,  gives a number of bonds in all cases consistent with BaH$_2$(H$_2)_x$, where $x$ is the number of bonds required to make up the stoichiometry. This observation holds for  BaH$_2$, all BaH$_4$ structures, BaH$_{5.75}$ and BaH$_{12}$. All peaks in the H-H RDF are in much the same place, regardless of composition. The shortest Ba-H distance represents atomic size, and the first H-H is as expected for a molecule, but the second H-H peak being similar is surprising. Under pressure, the molecular peak moves to larger bondlength  and subsequent peaks move to shorter separation as density increases. 
 
We estimate relative molecular lifetimes between structures by counting how many times an H-H distance is exactly 1\AA (See Table S2). In all MD structures, molecules have lifetimes of hundreds of femtoseconds, an order of magnitude larger than their vibrational frequencies. The H$2$ molecular version of $Cmcm$ has the longest lifetime, while $I4/mmm$ (high) has the shortest.  The bond breaking rate increases with temperature and pressure.
Bond breaking is normally mediated by a third hydrogen atom, which forms an H$^{-}_3$ complex, which then breaks back into a molecule and an isolated atom, which may originally have been in the molecule.  These  H$^{-}_3$ units are usually transient, but are also observed with two equal H-H distances in MD and static relaxation of e.g. Cmcm-H3, implying that they represent a stable structural motif. 
The number of singly-bonded atoms remains remarkably stable at the $2x$ in all BaH$_2($H$_2)_x$ compounds.  It is robust to the choice of bond cut-off, whereas the number of unbonded and double-bonded hydrogens is very system- and time- dependent.  Fourfold coordinated hydrogen is never observed.

Turning to specifics, for $I4/mmm$, the two distinct $c/a$ ratios behave very differently. 
  In the low $c/a$ the molecules persist at 1 pfu, average bondlength 0.81\AA.  They rotate continuously and the $c/a$ drops well below the static value.  To within sampling and finite size error, $c/a=\sqrt{2}$, which means that in X-ray diffraction from the Ba atoms will be indistinguishable from the reported fcc structure.
  At 1000K and 600K the unit cell remains orthorhombic, but at 300K there is a fluctuation between $\gamma=90^o$ and $100^o$.
  In the high c/a ratio version molecularization occurs between atoms in the same xy plane, not those sharing the octahedral interstice.  Our algorithm detects 0.84 molecules/fu, but this varies during the run and is sensitive to cutoff because these molecules are short-lived and long (0.95\AA\, on average).  

The two $Cmcm$ structures are intermediate in density between atomic and molecular $I4/mmm$.  In molecular dynamics, their hydrogen atoms show no rotation.  This offers an immediate explanation  for why the $I4/mmm$/$fcc$ structures are more stable at room temperature: there is a significant rotational contribution to the entropy.

BaH$_{5.75}$ has been synthesized experimentally and the Ba positions were identified via X-ray diffraction to correspond to the Weaire-Phelan (A15 or $\beta$-W) structure.  This is a topologically close packed structure ($Pm\bar{3}n$) in which all interstitial sites are tetrahedra.  Theory suggests that the unusual 1:5.75 stoichiometry comes from placing one hydrogen in each tetrahedron. If the hydrogens are placed in the tetrahedra, and relaxed preserving $Pm\bar{3}n$ symmetry, the octet rule cannot be satisfied and precisely one H$_2$ molecule per Ba forms.
To satisfy the octet rule (1.875 molecules per Ba) much larger supercells are needed, with at least 8 symmetry-inequivalent Ba atoms.

 The most stable structure we observe  has $P_c$ symmetry with 30 H$_2$ units (identified from both RDF and ELF, Fig.\ref{fig:structures}): $\textrm{BaH}_2(\textrm{H}_2)_{1.875}$. which has 3.75 electron bands per Ba in
”H$_2$” region of the bandstructure (Fig.{\ref{fig:dos}}).
Throughout the MD, bonds form between atoms in adjacent tetrahedra: as a consequence the H$_2$ molecules do not rotate, but the structure is dynamics because bond breaking takes place via  H$_3^-$ units.

We performed similar MD calculations on pseudocubic BaH$_{12}$\cite{chen2021synthesis}, at pressures up to 150~GPa.  We find that any structure consistent with cubic symmetry is, in DFT, highly unstable compared to H$_2$ bond formation.  Metastable structures with broken symmetry involve at least four formula units (e.g. Ba$_4$H$_{48}$,  P$2_1$ symmetry, or several variants with P1). The common theme of these structures is molecularization, confirmed by ELF analysis (Fig.\ref{fig:structures}): the molecular distances range from 0.77 to 0.9~\AA, with a clear gap to the next H-H distance (1.02~\AA) and the number of molecules  follows the octet rule: BaH$_2$(H$_2)_5$.

By contrast,  BaH$_2$ is an unremarkable ionic insulator, with no close H--H distances and no electrons in the energy region associated with H$_2$ molecules(Figs.\ref{fig:structures},\ref{fig:dos}). The ambient cotunnite structure transforms to P6$_2$/mmc at high temperature and above 1.6GPa, and around 50~GPa it transforms to a metallic P6/mmm 
\cite{kinoshita2007pressure,smith2007high,zhang2018hydride,novak2020pressure,novak2021temperature}

Figure \ref{fig:dos} shows a remarkable similarity in the electronic density of states   between the BaH$_4$ candidates and other barium hydrides of various compositions. The extra electrons from hydrogen-rich stoichiometry are all located in the H$_2$ band.   Projections of the wavefunctions onto atomic orbitals enables us to define distinct character to well-defined groups of bands, independent of crystal structure.
For BaH$_4$ distinct bands can be associated with Ba $5p$, molecular hydrogen and atomic hydrogen.  All these solids are small-gap semiconductors.   This suggests that even at 50GPa, all these compounds can be regarded as BaH$_2$(H$_2)_x$.  This does not hold for $I4/mmm$ with high $c/a$, which has no molecules and no associated peak in Fig.\ref{fig:dos}, and is highly unstable in MD.    

 High ELF values are found at the bond points connecting all pairs of atoms separated by $<1\textrm{\AA}$ (Fig.\ref{fig:structures}).
 If the bands localized in energy space at around $-6$--10~eV below the Fermi energy are projected onto maximally-localized Wannier functions they are observed to also be localised in real space. 

 We also studied the electron density and ELF by calculating ``empty" $I4/mmm$ structures comprising just Ba without any hydrogen, and $I4/mmm$ BaH$_2$ without the hydrogen molecules. We found that for the low $c/a$ ratio where interstitial hydrogen adopts a molecule form in the octahedral site, there is a single ELF maximum centered on the octahedral site and elongated along the direction of the hydrogens forming the molecule. At low pressures, there is no difference in the ELF value for the associated ELF basin, but at bigger pressures the ELF value at the ends of the basin is sligtly higher than that at the octahedral site signalling the stabilisation of the atomic phase at high pressure. Indeed, for the high $c/a$ phase stable at high pressures and where hydrogen forms two atoms the original ELF basin is broken in two well diferentiated ELF maxima with low ELF value in between. 
 Interestingly, these ELF maxima are not non-nuclear maxima of the electron density. Therefore, the ability of high-pressure metals to host molecular hydrogen and their corresponding position should be understood resorting to the ELF topology.
 
 Calculations of substoichiometric $I4/mmm$ showed that the most stable defect is to remove the hydrogen molecule altogether - replacing H$_2$ with H in the octahedral site is energetically unfavourable.


Localised H$_2$ molecular units are well defined in five different metrics: Bondlength, bandstructure,  Wannier wavefunction projection, and in both Bader and ELF topologies. 
The  number of H$_2$ molecules is determined by the octet rule  BaH$_2$(H$_2)_x$: with  x=0,1,1.875,5.  This value for $x$ persists in molecular dynamics of all these materials even after melting.
In related compounds, the octet rule holds and we observed
CaH$_2$(H$_2)_x$, SrH$_2$(H$_2)_x$ and RaH$_2$(H$_2)_x$ but CsH(H$_2)_x$ and LaH$_3$(H$_2)_x$ are  formed in molecular dynamics. 

Calculated hydrogen positions for stable structures of BaH$_2$(H$_2)_x$ are inconsistent with the reported room temperature experimental symmetry.  In the case of BaH$_2($H$_2)$ this can be explained by the MD calculations: molecular rotation in the $I4/mmm$ structure leads to a $c/a$ ratio of $\sqrt{2}$, for which the Ba positions are at the fcc sites as observed by diffraction. Rotation means that there is no symmetry-breaking due to molecule orientation.  The rotation also means $I4/mmm$ has high entropy, which stabilises it against the lower enthalpy $Cmcm$ phases, which are predicted to be stable at low temperature.
For the topologically close packed BaH$_{5.75}$ compound, molecular rotation cannot occur because each hydrogen atom is trapped in its own tetrahedra cage.   The bonds are continuously being made and broken - on a picosecond timescale across the face of the tetrahedra while  the Ba atoms remain in positions observed in experiment.   Remarkably, at any given time, the number of such bonds meets the BaH$_2$(H$_2)_{1.875}$ octet criterion.
For BaH$_{12}$ the decomposition into five molecules per unit cell is inconsistent with the available Wyckoff sites in cubic symmetry. Once again, molecular dynamics shows that continuous bond breaking and making occurs, such that a hydrogen is sometimes bonded to each of its neighbours and on average the high symmetry structure is recovered. 

The instability of high symmetry structures to forming electron pairs as bonds is likely to compete with the superconductivity instability towards Cooper pair formation.  Since our calculations are carried out using Kohn-Sham wavefunctions, Cooper-pair correlations cannot be observed (see SM7).

Overall, this work reveals the persistence of the H$_2$ bond in high pressure compounds. Molecular rotation and dynamic bond breaking allow this to occur without XRD-detectable long-range symmetry breaking.     Furthermore, the requirement from the octet rule to form  BaH$_2$(H$_2)_x$, and the incompatibility of $x$ with simple Wyckoff sites, leads to frustration in the bonding network and lowest-energy structures having large unit cells.   This study focused mainly on barium hydrides, but evidence from other hydrides suggest these principles will be general.

\begin{acknowledgement}
For the purpose of open access, the author has applied a Creative Commons Attribution (CC BY) licence to any Author Accepted Manuscript version arising from this submission.
 MM MPA and GJA would like to acknowledge the support of the European Research Council (ERC) Grant "Hecate" reference No. 695527. MPA acknowledges the support of the UKRI Future Leaders Fellowship Mrc-Mr/T043733/1.

\end{acknowledgement}

\newpage

{\bf\Huge Supplementary Material }

\section*{S1. How to define an H-H covalent bond?}

There are many measures which can be used to define a covalent bond in an {\it ab initio} molecular dynamics simulation. A good measure should be a parameter which can distinguish between bonded and unbonded pairs.

\subsection{S1.1 Bondlength}

The easiest approach is to define some cutoff distance $R_{bond}$, and declare that any pair of hydrogen atoms closer than this are "bonded".   This can be immediately evaluated from integrating the radial distribution function $g(r)$.  

For perfect crystals $g(r)$ is zero between shells of neighbours.  In many cases of md, including solid hydrogen, there of a large range where $g(r)=0$ and the hydrogen bonded pairs can be unambiguously identified, even though the choice of $R_{bond}$ is ambiguous, because  any  cutoff with $g(R_{bond})=0$ will give the same results.

In the present work, $g(r)$ does not go to zero.  This however there is often a distinct minimum at around 1~\AA. In the static structures we invariably finds that $g(r=1~$\AA\,$) = 0$. $R_{bond}=1~$\AA\,, therefore a pragmatic choice which we adopt as part of the definition. However, one must be careful that any measure of bonding is not highly sensitive to this choice.

The second ingredient of defining bonds from MD comes from investigating "Bond-breaking" events. In practice we see two types of event. 
\begin{itemize}
\item A sensible definition requires $R_{bond}$ to be large enough that an oscillating molecule is not marked as "breaking" if the bondlength briefly extends beyond $R_{bond}$ at its largest extent.  This is why our $R_{bond}$ must be set larger than one might expect a H-H bond to be.

\item Real bond breaking events in the current system come from reactions between atomic and molecular hydrogen. These involve an H$^-$ approaching an H$_2$ molecule, forming an intermediate state of an H$_3^+$ unit, which then breaks so that the ion is now part of a molecule, and one atom of the molecule is an ion.  We write this process as

H$^-$ + H$_2 \rightarrow $ H$_3^-  \rightarrow$
H$_2$ + H$^-$

where the ordering of the hydrogen atoms in the formula is assumed unchanged.   In isolation, this could be regarded as a double-well potential, but in the condensed phase each molecule as many option for rebonding.

Our $R_{bond}=1$\AA\,\ criterion is long enough to flag the 
H$_3^-$ unit as two bonds, and so we observe that the "number of bonds" increases during these reactions in a way that is very sensitive to the choice of  $R_{bond}$.  
Thus we refine the measure by counting the number of atoms with precisely one neighbour within 1\AA\,, and dividing by 2 to get the number of bonds.   One can readily verify that during the processes described above this measure give one covalent bond throughout.  It turns out that this measure is insensitive to choice of  $R_{bond}$.

\end{itemize}

\subsection{S1.2 Topological analysis of the electron density and ELF} 

There have been different approaches to analyze the chemical bonding in crystals. Probably, one of the most straightforward and mathematically rigorous is the Quantum Theory of Atoms in Molecules developed by Bader.\cite{Bader94} It is based on the topology of the electron density.
The topological analysis first identifies the critical points of the scalar fields, i.e, points where their gradient vanishes. These are then classified as maxima, first-order saddle points, second-order saddle points, or minima. The maxima of the electron density are usually located on the atoms, whereas the first-order saddle points are associated with chemical bonds (denoted bond critical points (b.c.p.)). The electron density and the Laplacian at these points are used to characterize the bond strength and its character. For instance, in a molecule such as H$_2$ the b.c.p. is characterized by high electron density and negative laplacian (electrons are locally concentrated).
The crystalline space can be also partitioned in topological atoms defined by the union of the electron density maxima with their attraction basins and delimited by surfaces obeying the zero-flux condition for the electron density. Their charge can be calculated from the integration of the electron density within these regions. These regions are non-overlapping and additive, recovering the total volume of the crystal. 

The electron localization function was introduced by Becke and Edgecombe in 1990 for the analysis in real space of the electron localization \cite{BE90}, and later reinterpreted by Savin in terms of the Pauli kinetic energy density ($t_p$) \cite{savin1997}. By definition, ELF is a relative measure of the electron localization with respect to the homogeneous electron gas (HEG). In general, the ELF value approaches 1 in regions of the space where electron pairing occurs (e.g., atomic shells, bonds and lone pairs). In analogy with QTAIM, a partition of the space based on the ELF can be performed. It consists of non-overlapping basins with well-defined chemical interpretation (cores, bonds, lone pairs). Moreover, the basin charges come from integration of the electron density within these regions. For example, a typical molecule will feature an ELF maxima in the middle of the interatomic distance associated with the bond. What is more, a 
region of the space (ELF maxima basin) can be attributed to it, with an electron population obtained by integration of the electron density in that region.
Hydrogen constitutes an exception. A maximum does not appear along the H-H bond since the hydrogen molecule only contains two electrons. In this case, a high ELF isosurface encapsulates both atoms. Typically, a H$_2$ molecule is predicted to exist if the hydrogen atoms are connected at a value of ELF higher than 0.85, i.e, if the minimum ELF value in between the hydrogen atoms is above 0.85.

\subsection{S1.3 Wavefunctions}

A quantum wavefunction can be expanded in any complete set of orthogonal basis functions. In a DFT calculation we use a plane wave basis set so there are no molecular orbitals.  Plane waves offer the big advantage that one can systematically approach a complete set by increasing the cutoff, and they are automatically orthogonal. A basis set of localised orbitals is typically non-orthogonal and therefore simultaneously overcomplete and undercomplete.

\begin{figure}[t!]
\includegraphics[width=0.49\columnwidth]{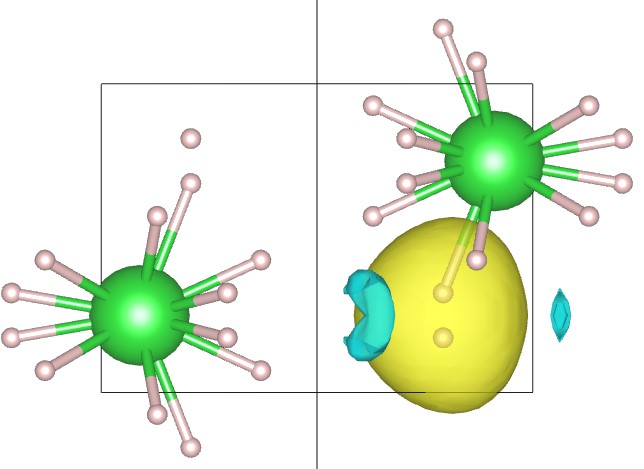}
\includegraphics[width=0.49\columnwidth]{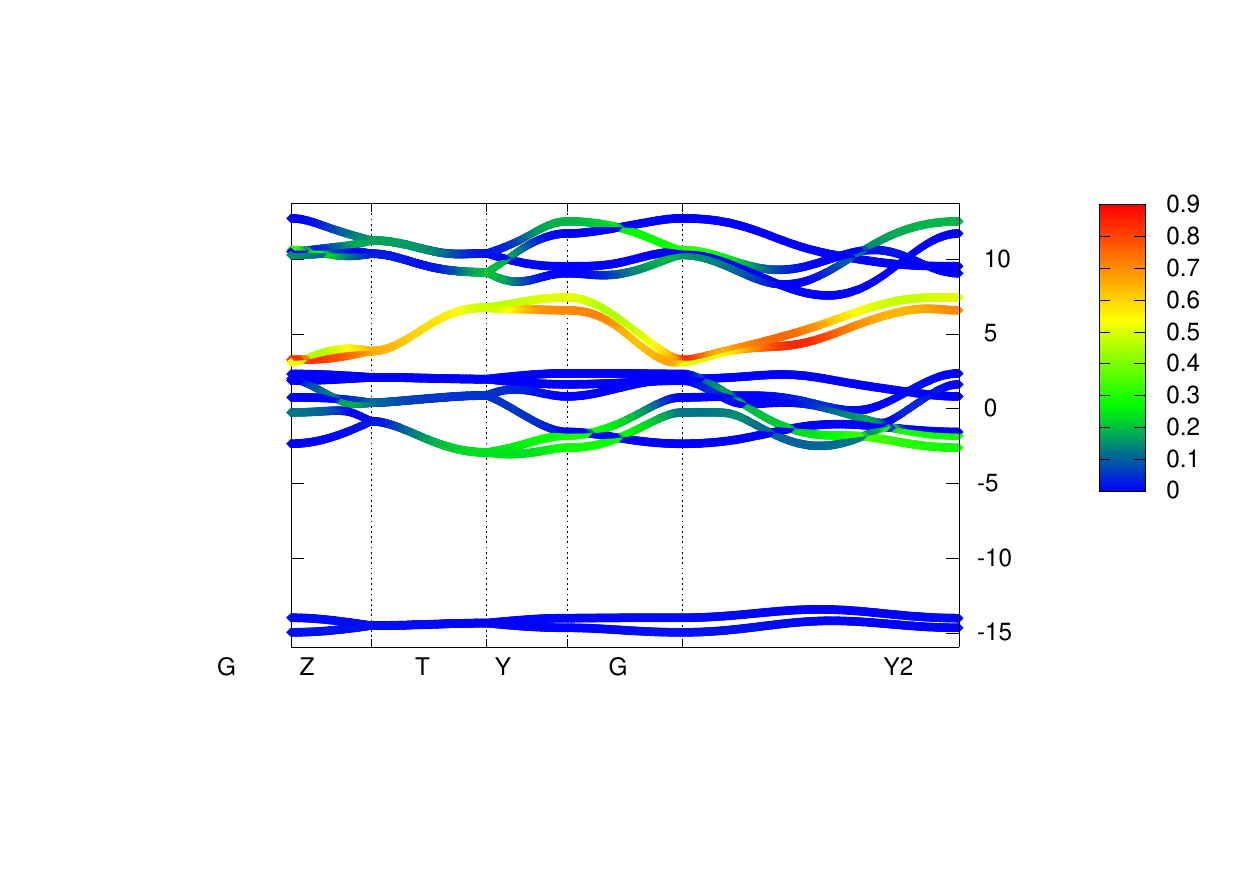}
\caption{ \label{fig:projbands}(right) Projection of Kohn-Sham band onto Maximally Localised Wannier function for BaH$_4$ $I4/mmm$ structure showing localisation on a single H$_2$ unit. (right) Bandstructure of BaH$_4$ with colours showing projection onto H$_2$ covalent bond orbitals.}
\end{figure}

\begin{figure}[t!]
\includegraphics[width=0.49\columnwidth]{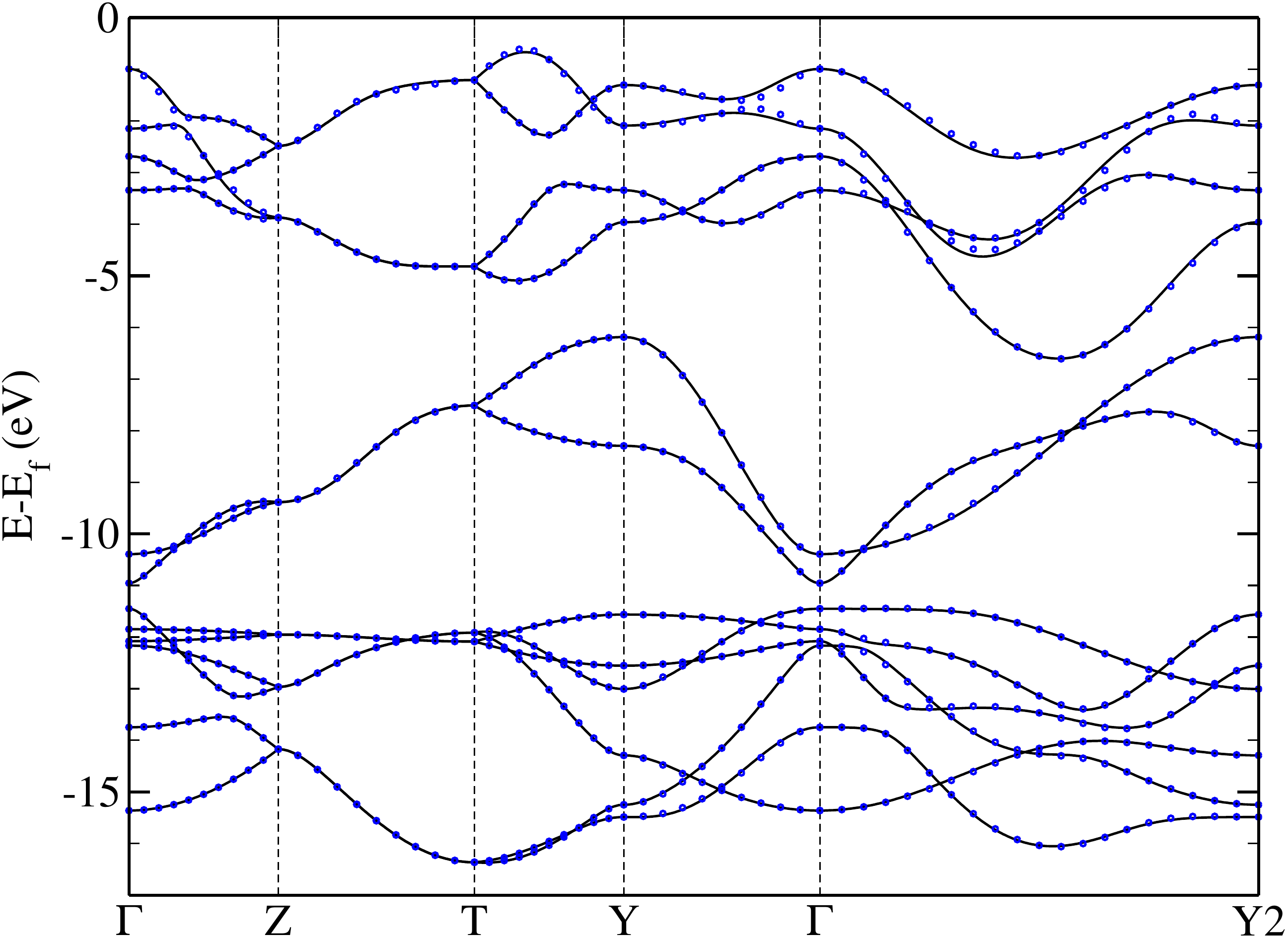}
\includegraphics[width=0.49\columnwidth]{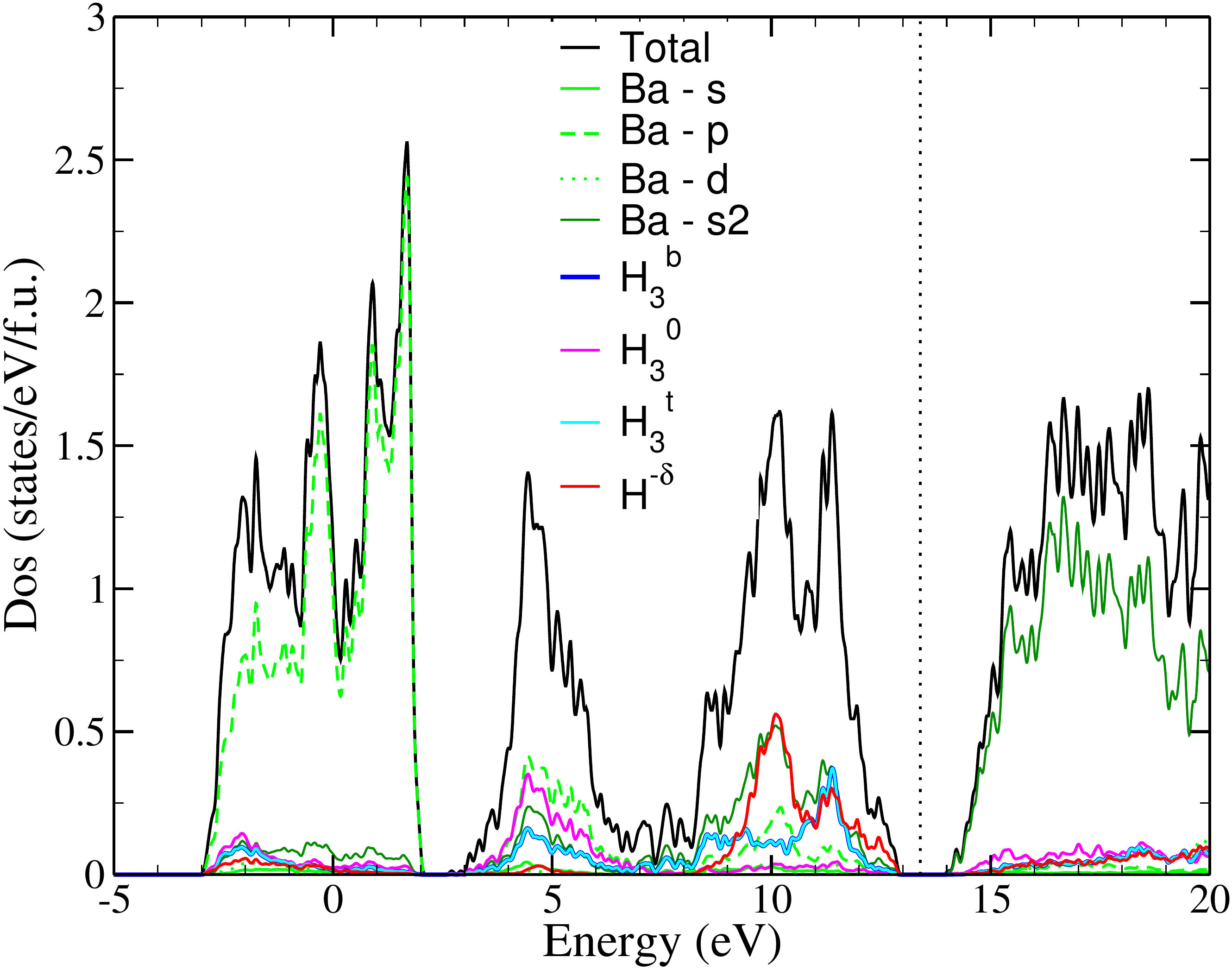}
\includegraphics[width=0.49\columnwidth]{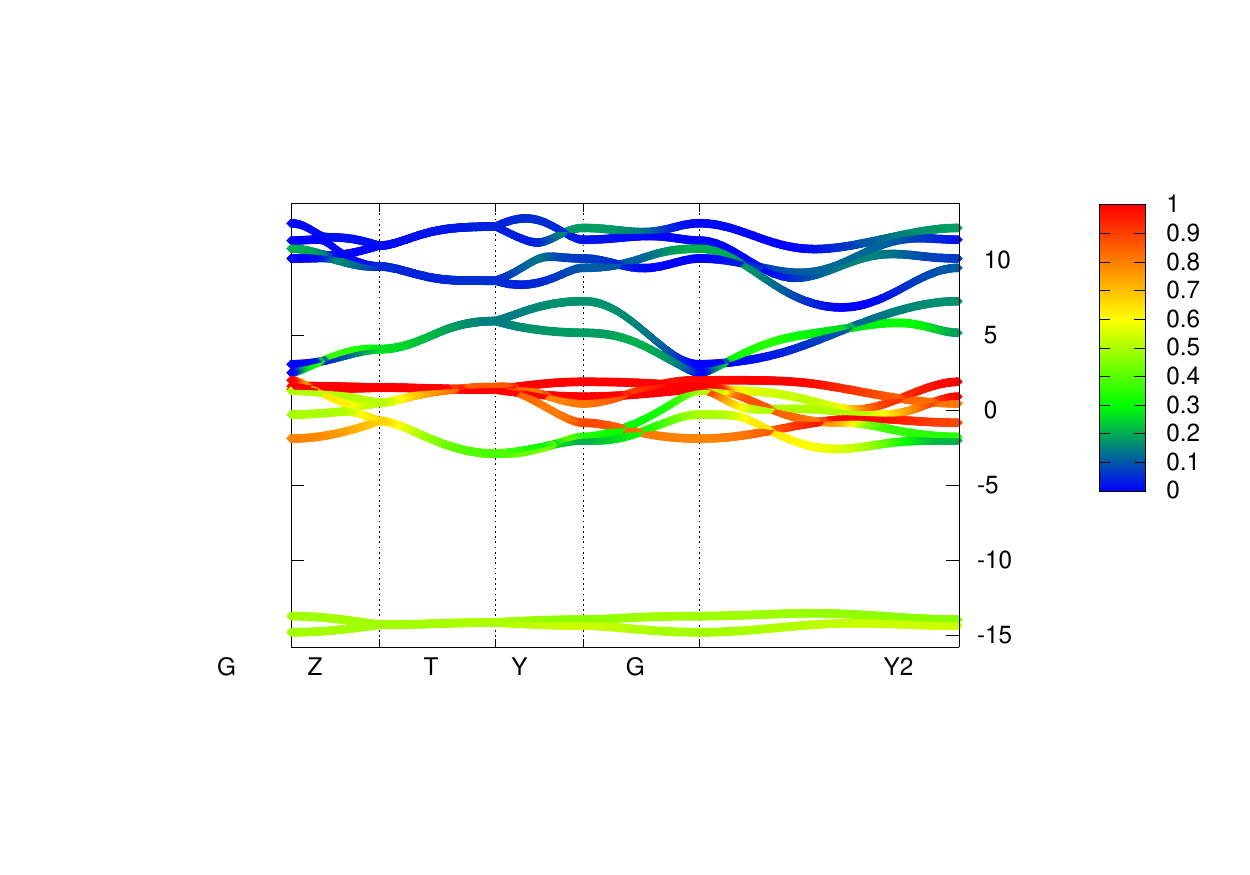}
\includegraphics[width=0.49\columnwidth]{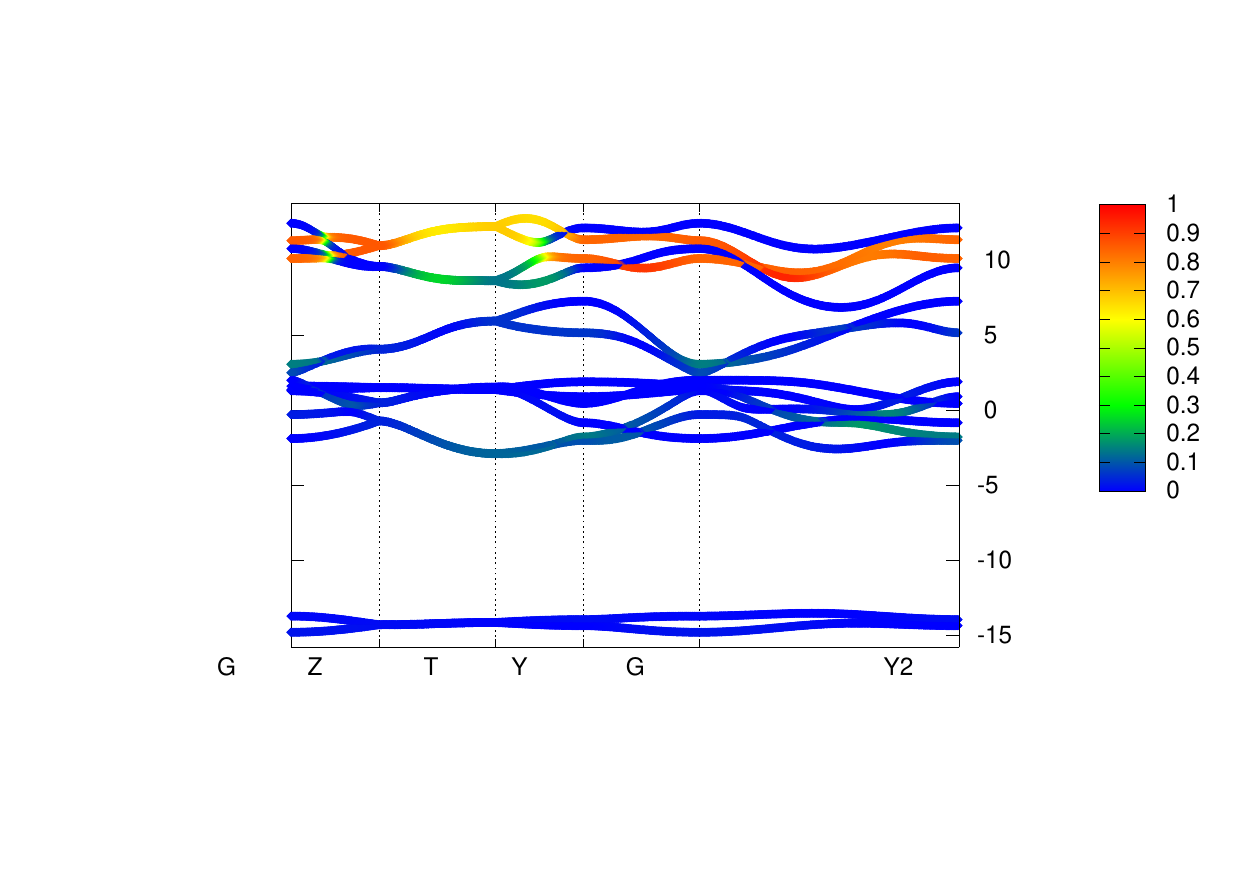}
\includegraphics[width=0.49\columnwidth]{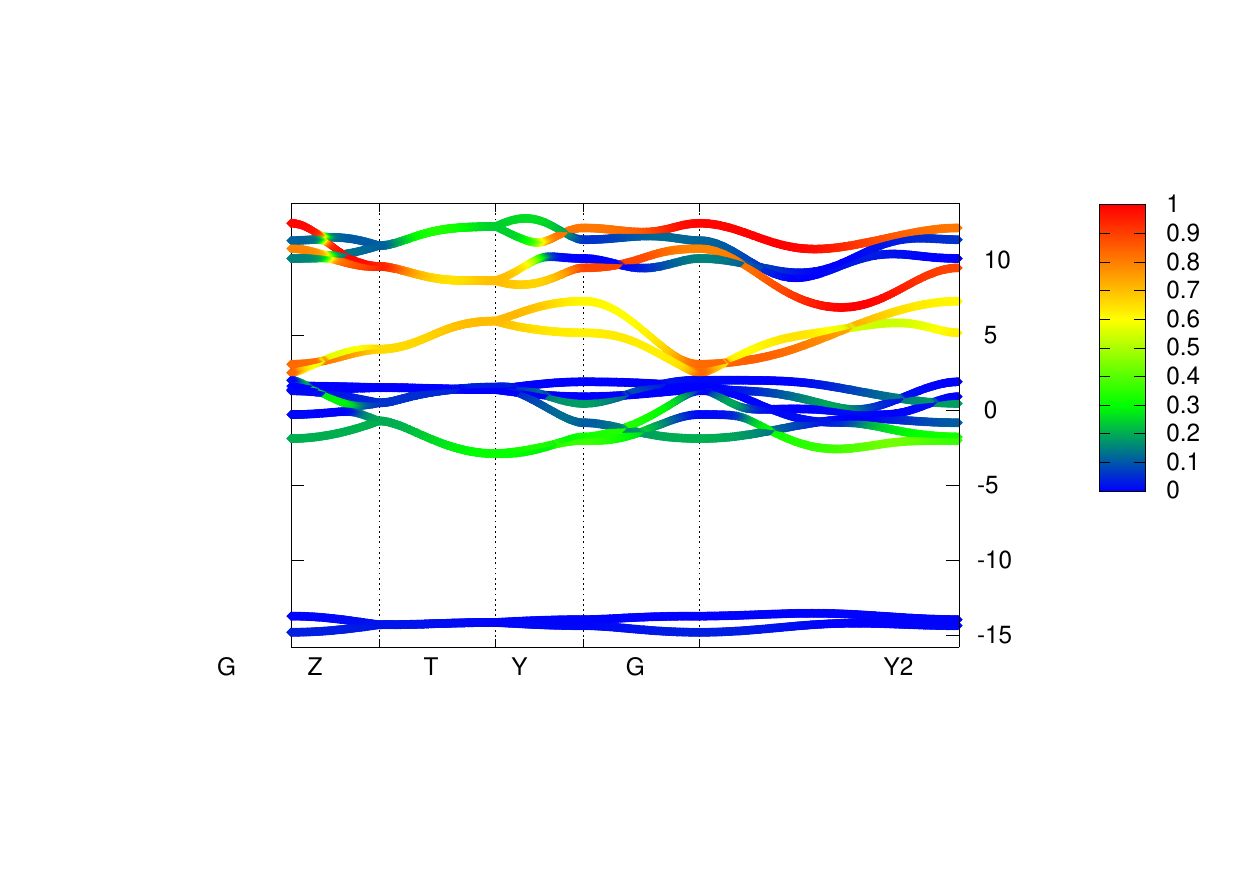}
\includegraphics[width=0.49\columnwidth]{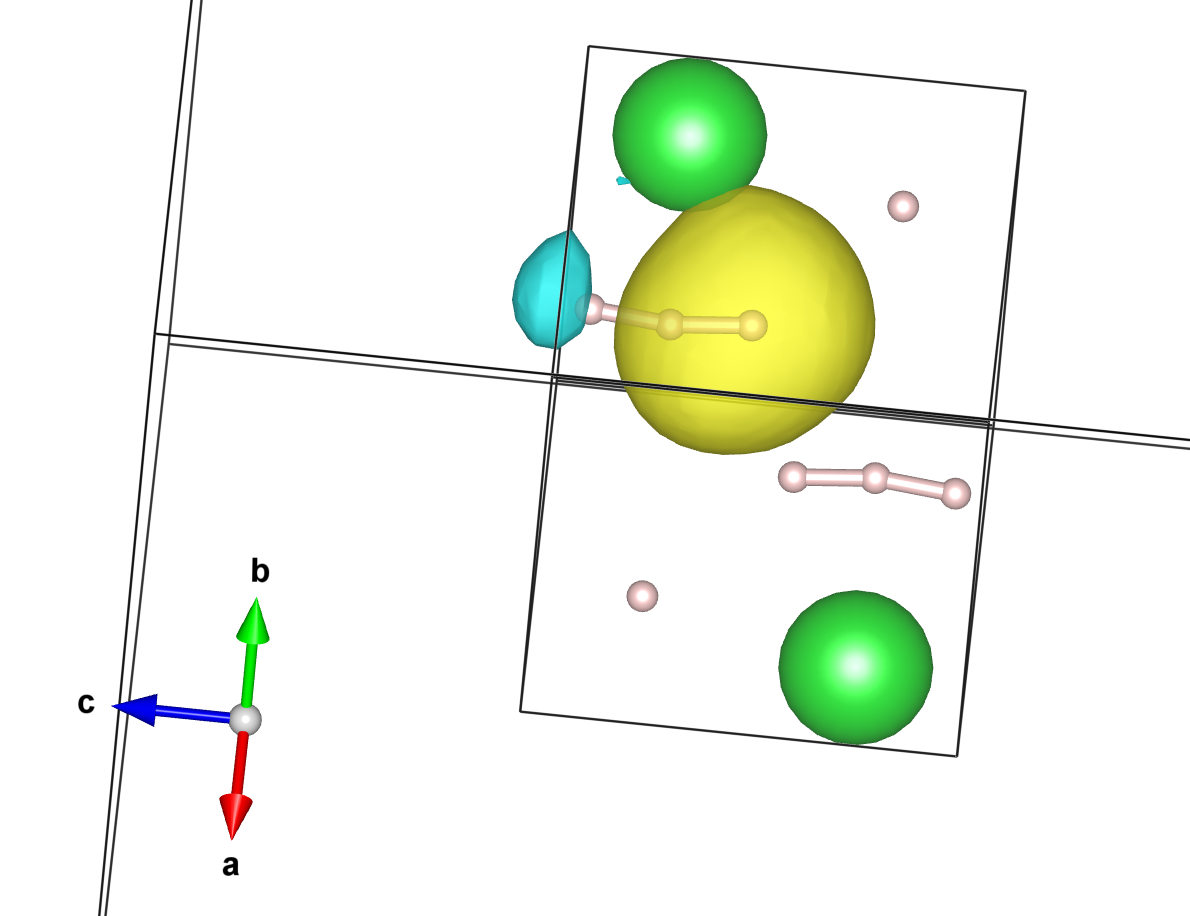}
\caption{(right) Projection of Kohn-Sham band onto Maximally Localised Wannier function for the $Cmcm$-H3 structure. (a) Band structure: bands are enumerated from lowest energy upward, including the Ba 6s (b) Projection of MLWF  onto Bloch states.
(c) Projection onto Ba 5p MLWF states.
(d) Projection onto atomic H 1s MLWF states (e) Projection onto MLWF states for H$_3^-$  (f) Isosurface of an H$_3$ MLWF} \label{fig:MLWF}
\end{figure}

To use the Kohn-Sham wavefunctions to identify molecules one must project them from plane waves onto localised basis functions.  This can be done using maximally localised Wannier functions. Sup. Figs \ref{fig:projbands}, and \ref{fig:MLWF} show the projection of the Kohn-Sham bands identified as bonds onto maximally localised Wannier functions, the localisation in real space and the identification in reciprocal space are clear.

   Some 20\% spillage of the low-lying Ba 6s states projected onto the Ba-p orbitals give some indication of the ambiguity in using the projection:  it is good for qualitative measure.  This identified localised states  on the Ba and the H$^+$ as one expects for ionic materials (Sup. Fig.\ref{fig:MLWF}c-d). 

The H$_2$ unit can be considered in terms of its two molecular orbitals, the bonding $1\sigma_g$
corresponds to the band some 5eV below the Fermi energy, while the antibonding $1\sigma_u$ state is above the Fermi energy.
The H$_3$ unit can also be considered in terms of molecular orbitals.
Since it has three atoms, H$_3$  has precisely three molecular $s$-orbitals, most simply written\footnote{Sometimes the three names "bonding, nonbonding and antibonding" are applied to these states.} as 1$\sigma_g$, 1$\sigma_u$ and 2$\sigma_g$, whose wavefunctions have zero, one and three nodes. H$_3^-$ implies that 1$\sigma_g$ and 1$\sigma_u$ are occupied.

The Wannier approach was also applied to the Cmcm-H3 structure, which has H$_3^-$ units (Fig.\ref{fig:MLWF}). 
The  H$_3$ 1$\sigma_g$, 1$\sigma_u$ electronic states are located in the same region of space.  This enables the MLWF to generate  combination of bonding and nonbonding states localised on one end of the H$_3$ unit.   Such a state is preferred by the MLWF procedure because it is more localised than either 1$\sigma_g$ or 1$\sigma_u$ states which are spread across three atoms. Projection of this MLWF state onto bandstructure shows it occupies both the 5eV region where the H$_2$ bands lie in other compounds, and the DoS peak just below the Fermi Energy degenerate with he H$^-$ states   (Sup. Fig.\ref{fig:MLWF}e-f).  We propose that the in the molecule orbital picture 1$\sigma_g$ and 1$\sigma_u$  can be assigned to these energy regions, while 2$\sigma_u$ lies above the Fermi energy.

\section{S2 Density functional and ab initio molecular dynamics calculations}

\subsection{S2.1 Codes and settings}

For static relaxations we carried out density functional theory calculations using the CASTEP code, with ultrasoft pseudopotentials and the PBE functional\cite{clark2005first,PBE}.  We used a plane-wave basis with a 700 eV cutoff and a k-point spacing better than 0.04 \AA$^{-1}$. 
Key results were repeated using LDA and BLYP functionals\cite{LDA,perdew1981self,blyp-B,blyp-LYP}, which give similar qualitative results but have some minor effect on transition pressures.
Molecular dynamics were run with a a cutoff energy of 463 eV and k-point grids up to 26$^3$ in CASTEP, with a timestep of 0.5fs in the NPT ensemble, and were followed by geometry optimisations of snapshots to zero temperature.

ELF and Bader analyses were carried out using the CRITIC2 program\cite{CRITIC2}, and applied to all single unit cell static structures

Wannier analysis was carried out using the wannier90 code\cite{mostofi2014updated} 

\subsection{S2.2  Fictitious Mass Molecular Dynamics}
 
 The partition function in the NPT ensemble:
$$ Z=\sum_i \exp(-H_i/kT)= \sum_i \exp(-U_i/kT)\exp(-PV_i/kT)\exp(-KE/kT)$$
Where  $U_i$ is the potential energy of the microstate, $V_i$ is the volume and KE is the Kinetic energy.   In the classical limit, the kinetic energy is proportional to temperature, so the final term factors out.  We observe that the first two terms are independent of the atomic masses, so we are free to choose them as we wish. 

In the case of barium hydride, the two orders of magnitude difference in mass means that in conventional MD the bariums would scarcely move. 
However, the fictitious mass approach enables us to correctly sample the partition function, so phase transformations involving the Ba atoms can be observed. Likewise a simulation run above the melting temperature will move into the phase-space region of the melt.  The partition function is correctly sampled, so equilibrium thermodynamic properties are correct.  However the dynamics of the melting process, or the vibrational spectra, will not be correct.

The most challenging thing to determine in this partition function is the definition of $i$ - this indicates "Microstates corresponding to the I4mmm structure". Of course, finite temperature microstates have no symmetry, and NPT simulation in designed to allow for a phase change.  In practice, we monitor the box dimensions and the fraction of molecules to determine distinct phases.

 We derived this method independently, but it seems such an obvious idea that we assume it is not new.

\subsection{S2.3 Hydrogen in jellium}

Breaking the H$_2$ bond can be due to physical confinement in an interstial site: "chemical pressure", or by electronic effects "antibonding".

\begin{figure}[t!]\includegraphics[width=0.48\columnwidth]{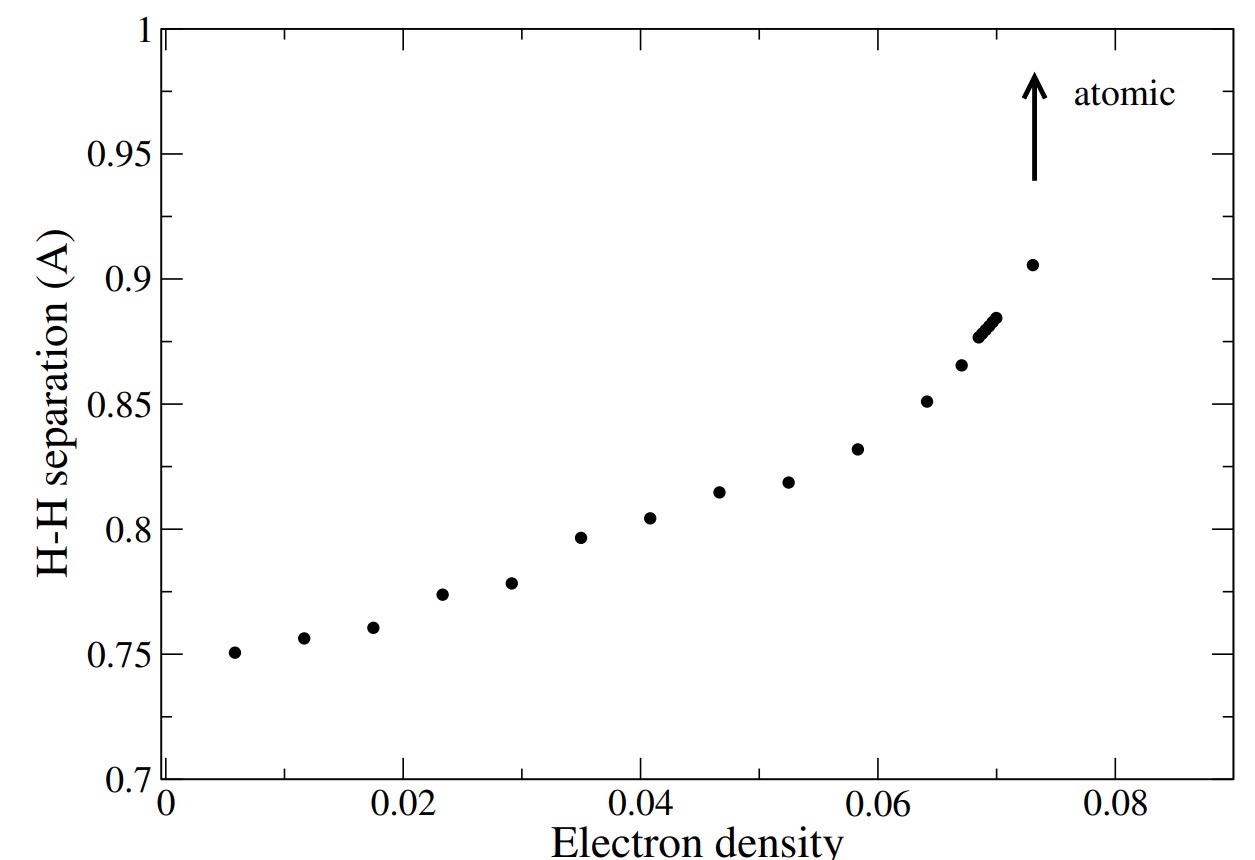}\includegraphics[width=0.52\columnwidth]{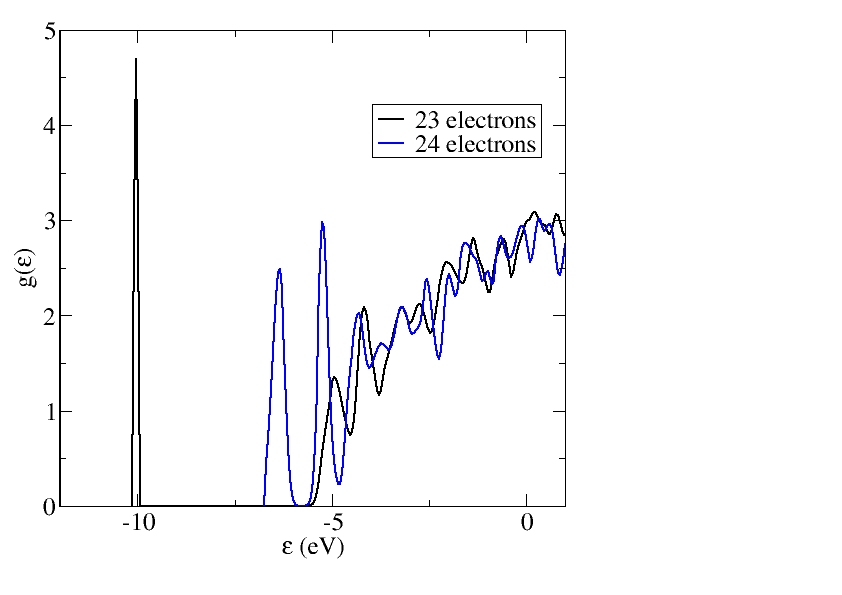}
\caption{ \label{fig:jellium} \textbf{The maximum electron density into which an H$_2$ molecule can be embedded is 0.07~\AA\,$^{-3}$.}
a) Plot of bondlength vs DFT Calculation using PBE hydrogen embedded in a jellium background at the shown electron density.  Calculations were done with one hydrogen molecule in a 343~\AA\,$^3$ box with a 16$^3$ k--point grid.  Similar calculations using BLYP give the bond breaking at the same density
b) Bandstructure calculations (density of states) for the system described above with a total of 23 or 24 electrons in the system.  The covalent bond in the 23 electron system lies some 10~eV below the Fermi energy, or 4.5~eV below the bottom of the free electron band. After bond-breaking, the atomic states lie at the bottom of the free electron band. Other oscillations in the free-electron band structure come from the cubic symmetry of the supercell and are not relevant.
 }
\end{figure}

The embedded atom method is one of the most enduring and successful techniques in materials modelling.   It was introduced with the hypothesis that the binding energy of hydrogen in a metal would depend on the electron density of the other atoms, and there was an optimal value for this.    Hence the electronic contribution to bond breaking can be calculated in isolation by embedding the H$_2$ molecule in a homogeneous electron gas.
This system has been studied in detail\cite{bonev2001hydrogen}, and here we repeat that work with the methods employed in this paper.  For low electron density the band structure comprises a free electron-like density of states which lies much higher in energy than a clearly-defined H$_2$ bond.  Above 0.06~s $e$~\AA$^{-3}$: the bond breaks spontaneously, evidenced by a discontinuous jump in the H-H separation, the vanishing of the flat Kohn-Sham band defining the molecule and the appearance of an atomic-hydrogen state at the bottom of the free-electron band (Sup.\ Sup. Fig.\ \ref{fig:jellium}). This sets an upper limit on where molecular hydrogen can be found.

\section{S3 $I4/mmm$ structures}

\begin{figure}[t!]
\includegraphics[width=0.8\columnwidth]{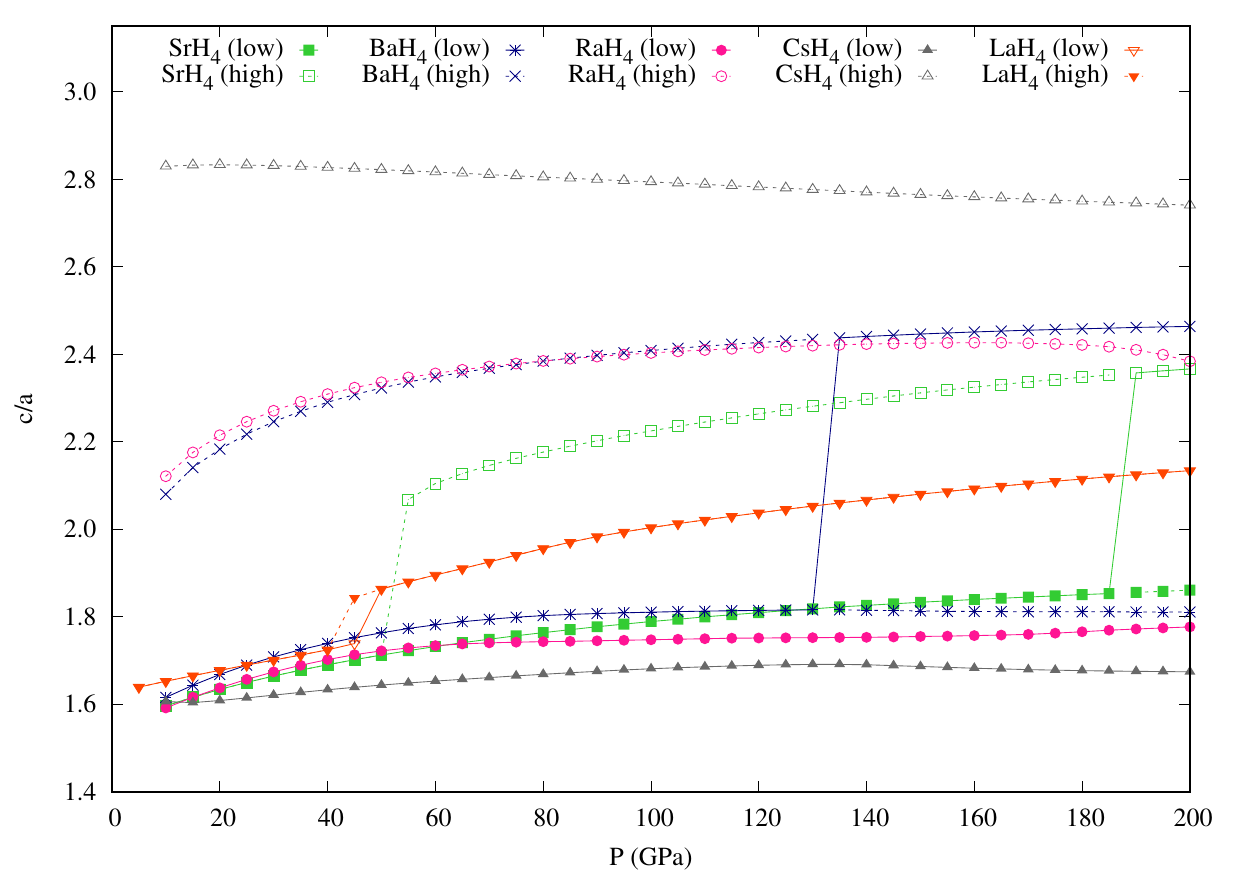}
\caption{ Evolution of $c/a$ with pressure for the $I4/mmm$ structures. Solid and open symbols correspond to the low and high $c/a$ structures. The $I4/mmm$ structures with the lowest enthalpies at each pressure are indicated by solid lines. \label{fig:ca}
}
\end{figure}

\begin{figure}[t!]
\includegraphics[width=0.8\columnwidth]{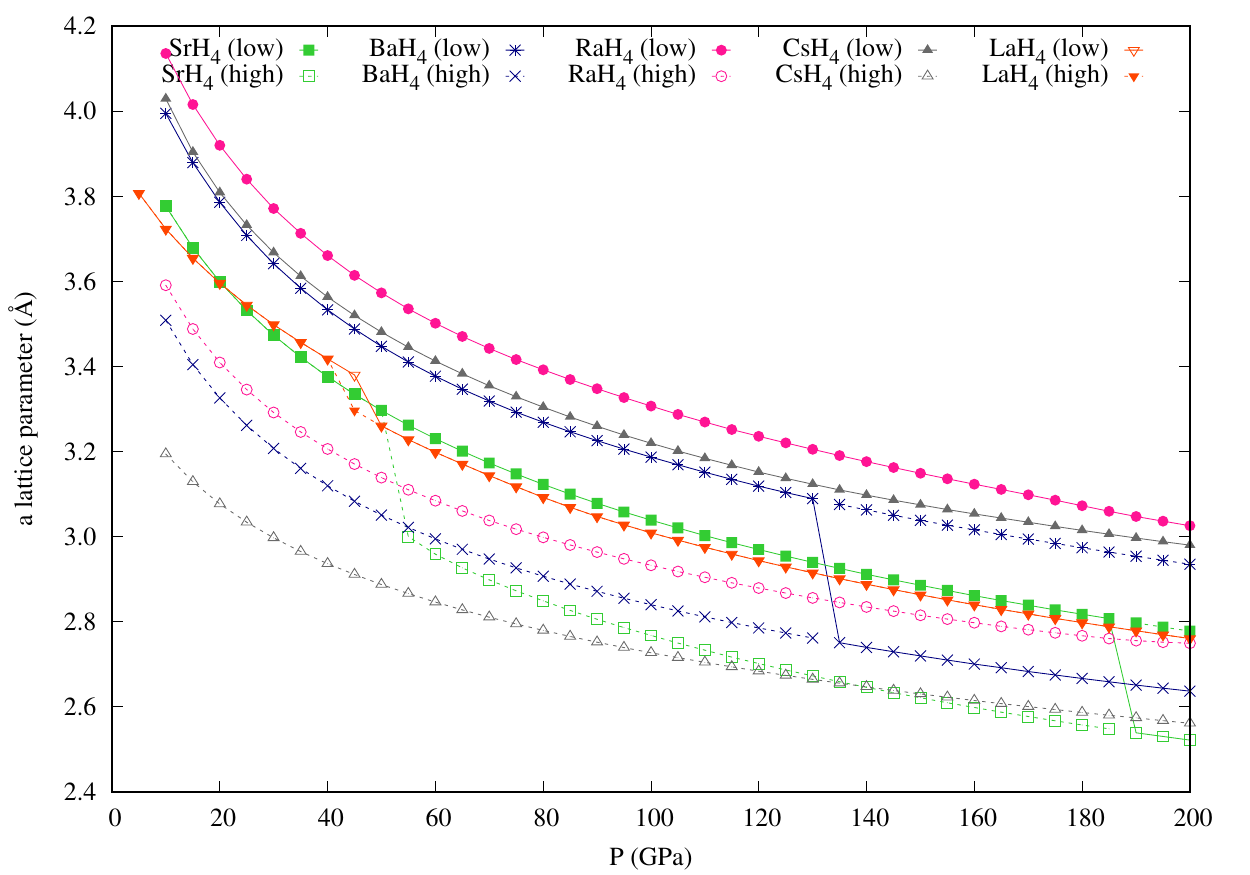}
\caption{ Evolution of $a$ lattice parameter with pressure for the $I4/mmm$ structures. Solid and open symbols correspond to the low and high $c/a$ structures. The $I4/mmm$ structures with the lowest enthalpies at each pressure are indicated by solid lines. \label{fig:canda}
}
\end{figure}

\begin{figure}[t!]
\includegraphics[width=0.49\columnwidth]{chargesbah4.pdf}
\includegraphics[width=0.49\columnwidth]{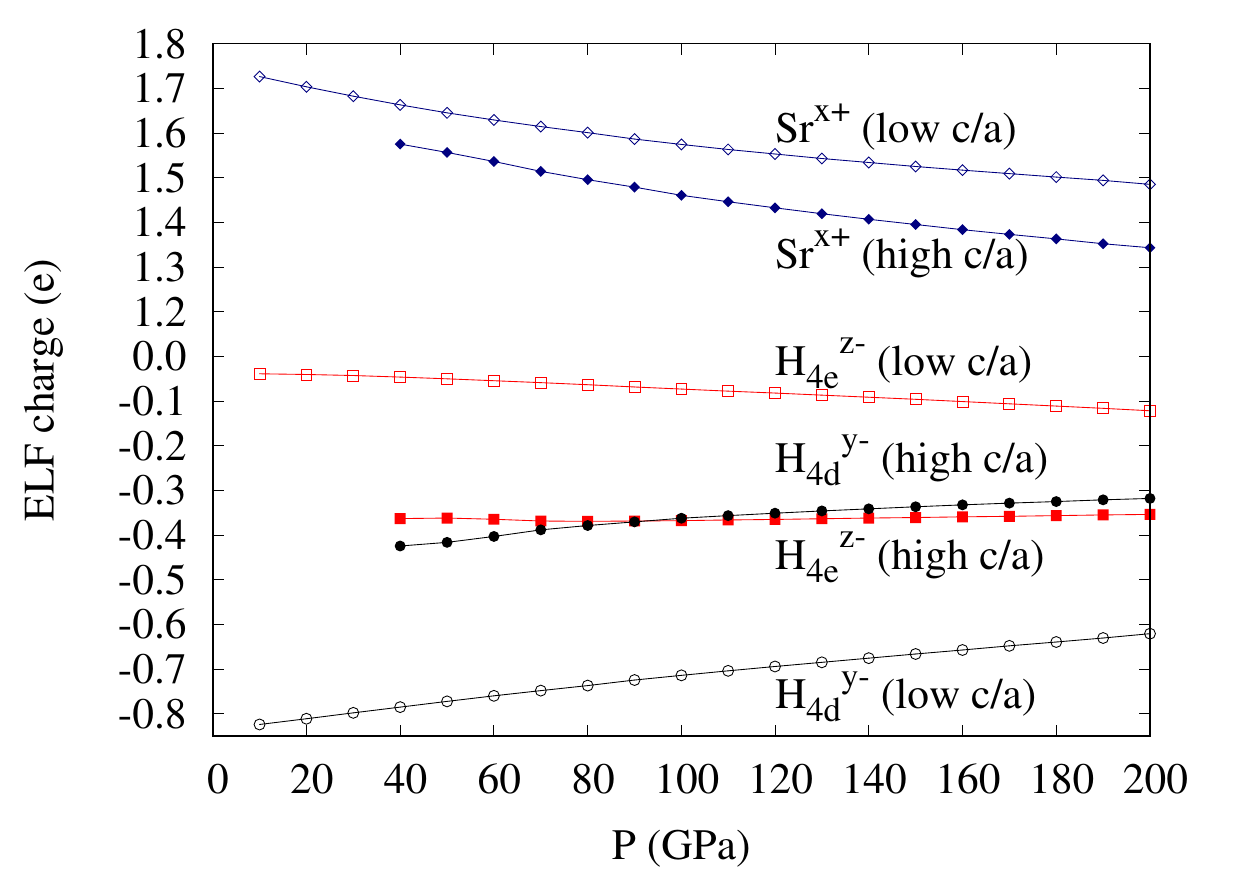}
\includegraphics[width=0.49\columnwidth]{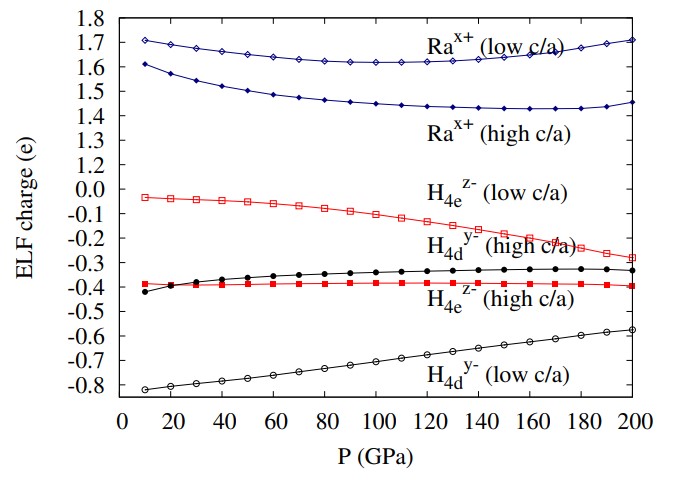}
\includegraphics[width=0.49\columnwidth]{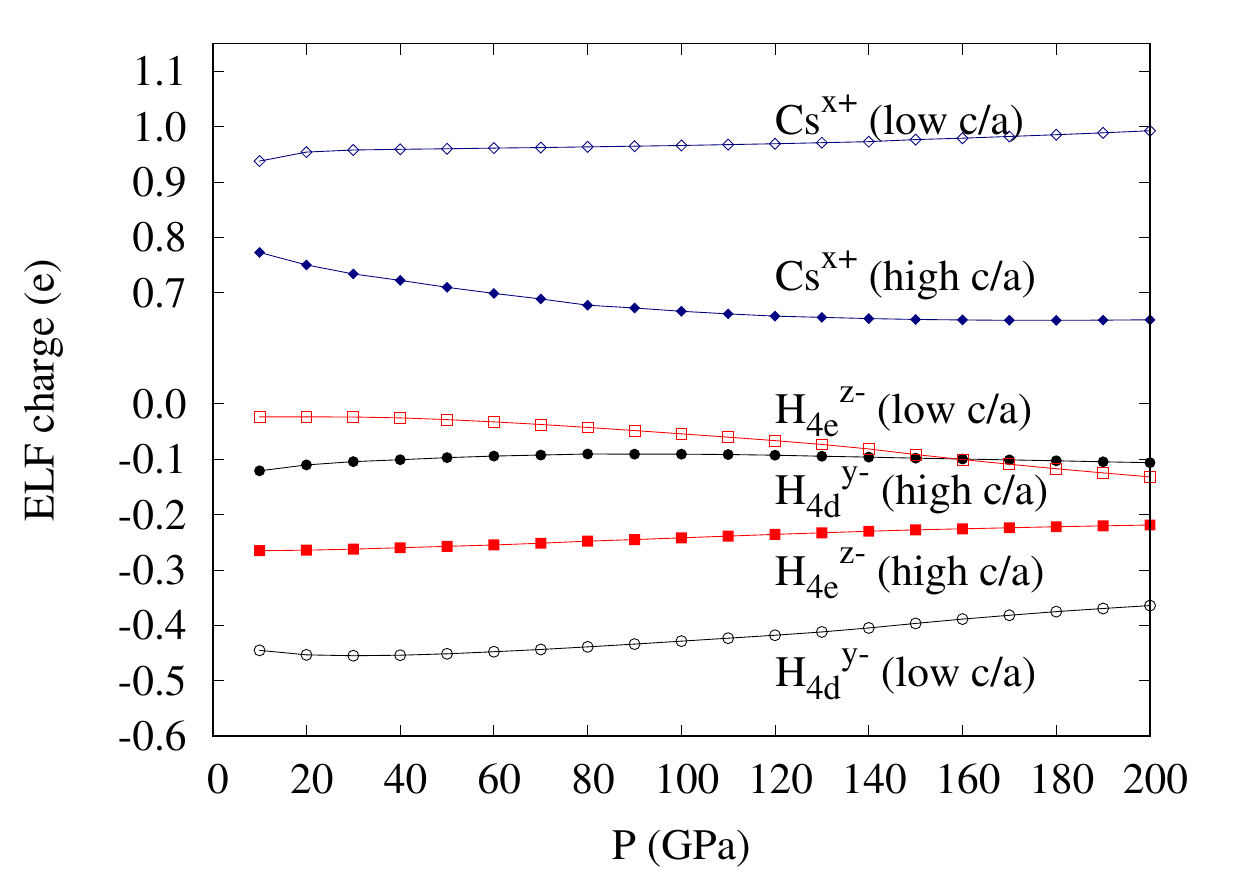}
\includegraphics[width=0.49\columnwidth]{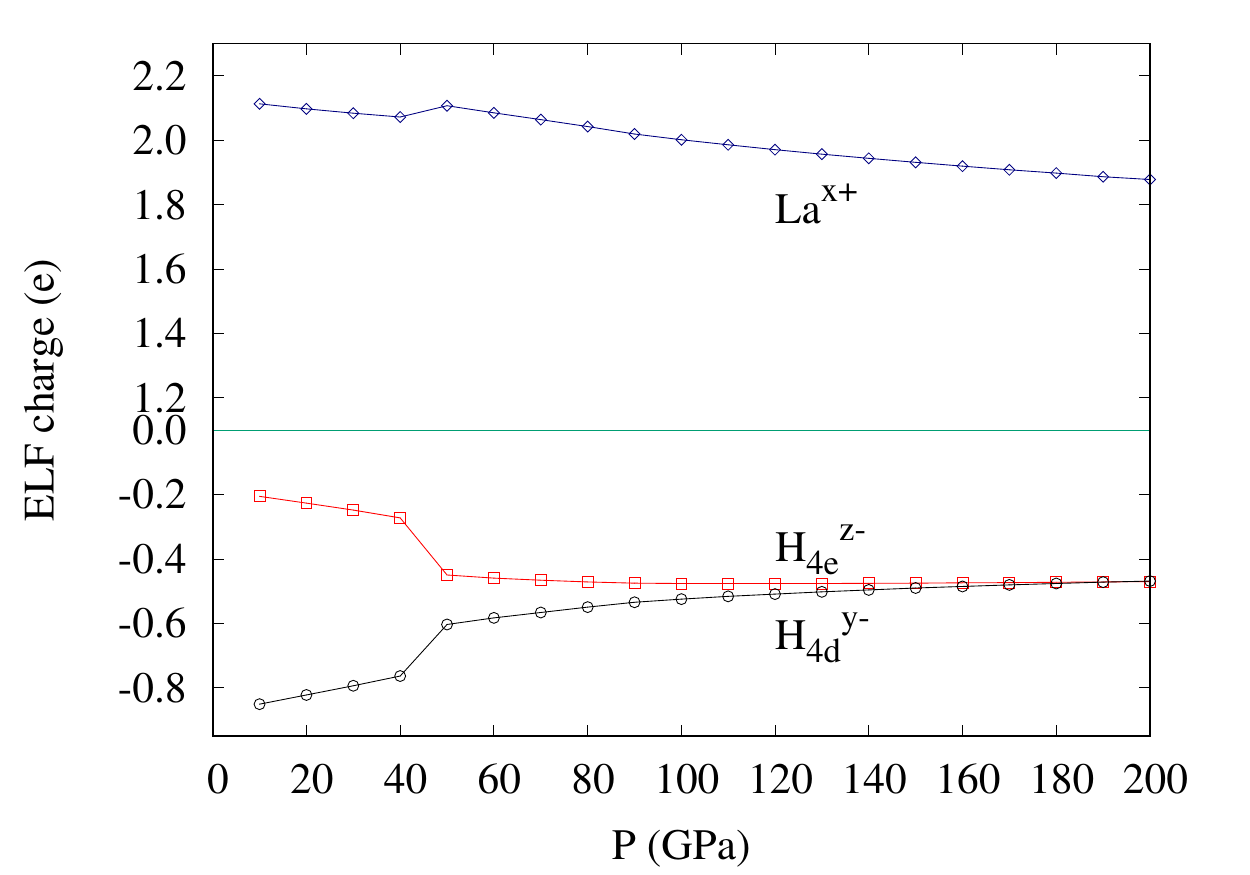}
\caption{a) Charges as a function of pressure in $I4/mmm$ compounds based on the ELF topology for the metal and hydrogen atoms located on 4$e$ and 4$d$ sites \label{fig:charges}
}
\end{figure}

\subsection{S3.1 Comparing I4/mmm materials}
We considered $I4/mmm$ structures for
SrH$_4$,  LaH$_4$, CsH$_4$ and RaH$_4$ to determine trends across the periodic table making no definitive claim about the stability of the calculated compounds.  C/a ratio, lattice parameters and ELF-basin populations are shown in Sup. Figs. \ref{fig:ca}, \ref{fig:canda}
and \ref{fig:charges}.

{\bf BaH$_4$} is discussed in detail in the main text.  It is unusual in that the low c/a structure calculated to be stable in DFT is different from the high c/a ratio structure seen in the experiment.  Nevertheless, both structures are local energy minima.  Many other materials exhibit the same theoretical coexistence of two energy minima in the same I4/mmm structure.

\begin{figure}[t!]
\includegraphics[width=1\columnwidth]{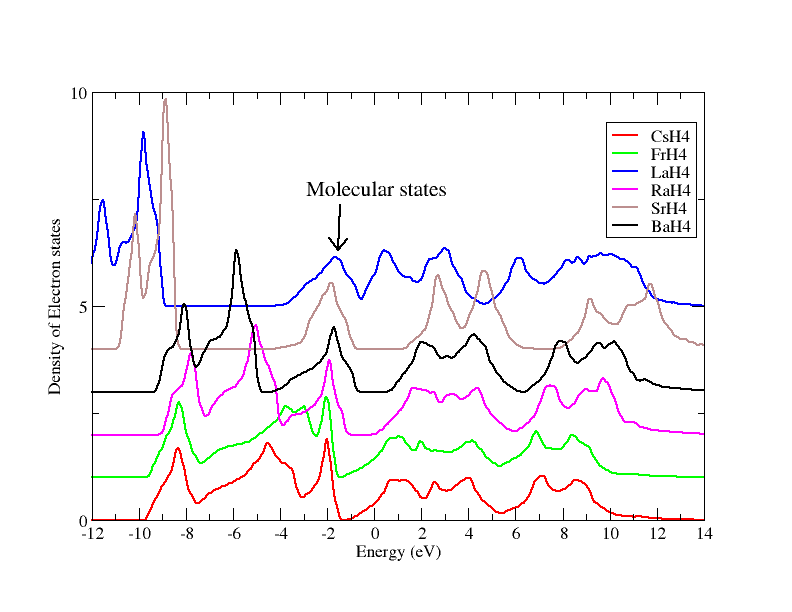}
\caption{Band structures at 50GPa in SrH$_4$, BaH$_4$, LaH$_4$, CsH$_4$ and RaH$_4$. The zero of energy has been offset to place the peak of the H$_2$ "band" together, as indicated by the arrow.
The calculated peaks contain 2 (not shown), 6 (double peak, widely separated e.g. in Cs), 2 and 4 (double peak) electrons per formula unit respectively.  In BaH$_4$ these can be interpreted chemically as Ba 5s,  Ba 5p, molecular H$_2 \sigma$, and atomic H 1s.  The Fermi energy lies in the gap at +6 eV and the conduction band has 5d and 6s character. 
\label{fig:XH4dos}}
\end{figure}

{\bf SrH$_4$}:
Strontium is above Barium on Group 2A, and calculations show the atomic and molecular forms do adopt separate enthalpy minima, with the atomic form only becoming stable above 185GPa. Experiments on SrH$_4$ have not reached these pressures.  The molecular form is non-metallic, e.g. at 50GPa there is a bandgap of about 2eV (Sup. Fig.\ref{fig:XH4dos})
 
{\bf CsH$_4$}:
Caesium lies beside Barium, and calculations show the atomic and molecular forms do adopt separate enthalpy minima, but the atomic form is unstable up to 200GPa. This is despite the fact that, being monovalent, the material is calculated to be metallic with the Fermi level in the atomic hydrogen band. Being so chemically unfavoured suggests that this structure will not be stable, probably decomposing into CsH \cite{ahuja1998theoretical} and hydrogen or a polyhydride.     

{\bf RaH$_4$}:
Radium lies below Barium in the periodic table, and is highly radioactive so the nature of its high-pressure hydrides is unlikely ever to be tested.  Calculations suggest that at all pressures up to at least  200GPa the low c/a molecular structure is stable.  The band structure is similar to BaH$_4$ with somewhat broader bands and a DFT/PBE band gap which closes at 50GPa,

{\bf LaH$_4$}:
Lanthanum is next to Ba in the periodic table, and typically forms trivalent ions.  Previous work \cite{bi2021electronic} studied the high-pressure form of LaH$_4$.   The pressure dependence of the bond in the LaH$_4$ shows a trend towards steady lengthening with pressure, with no distinct high and low variants.   At 50GPa two short H-H separations are already 1.5\AA apart in the density of states, and the characteristic ``molecular" peak has vanished (Sup. Fig.\ref{fig:XH4dos})  The   LaH$_4$ is always metallic, even in the molecular case, with the Fermi level lying in the La $sd$ conduction band. Given this, it is likely that LaH$_4$ is unstable with respect to higher polyhydrides.

The band structures reveal significant similarities between the different materials , and differences between atomic and molecular forms  (see Density of states at 50GPa).  For BaH$_4$, we calculate five distinct groups of state: low-lying Ba 5s (-22eV) and 5p orbitals (-10eV).  Then at -4eV come the H$_2$ (0.79\AA\,) molecular bonding orbitals.  Closest to the Fermi energy are the atomic hydrogen H$^-$ hybridized with Ba 5d: population analysis in atomic basis sets suggests that the Ba 6s are unoccupied.

In the Sr compound, this allocation is even clearer. In Ra, the electronic bands are broader but energy gaps in the density of state mean there is no hybridization. At 50GPa the H$_2$  bond in both Sr and Ra is 0.79(1)\AA\,

Overall, the comparison of the adjacent materials indicates that Ba is unusual in showing the atomic-molecular transformation at the lowest pressure.  The band structures are remarkably similar across all materials, with the position of the Fermi energy determined by the cation valence. We have not done extensive structure searching: our molecular dynamics simulations indicate that the molecular hydrogens are orientationally disordered at room temperature, while quantum nuclear effects are likely to be important at zero temperature.  Given the emerging picture that these materials can be thought of as dihydrides with interstitial H$_2$  it is unlikely that the metallic compounds with monovalent and trivalent cations are stable.

\begin{table}[hbt]
    \centering
    \begin{tabular}{lcccccc}
    \hline
    System & \multicolumn{3}{c}{Bader charge} & \multicolumn{3}{c}{ELF basin population} \\
          & on M (e) & on H$_o$ (e) & on H$_t$ (e) & on M (e) & on H$_o$ (e) & on H$_t$ (e) \\ 
    \hline
 $I4/mmm$-BaH$_4$ (low) & +1.174 & -0.041 & -0.546 & 8.446 & 1.047 & 1.731 \\
 $I4/mmm$-BaH$_4$ (high) & +1.036 & -0.253 & -0.265 & 8.591 & 1.367 & 1.337 \\
 $I4/mmm$-SrH$_4$ (low) & +1.247 & -0.049 & -0.574 & 8.355 & 1.050 & 1.772 \\
 $I4/mmm$-SrH$_4$ (high) & +1.143 & -0.250 & -0.321 & 8.444 & 1.362 & 1.416 \\
 $I4/mmm$-RaH$_4$ (low) & +1.278 & -0.047 & -0.591 & 8.350 & 1.052 & 1.773 \\
 $I4/mmm$-RaH$_4$ (high) & +1.136 & -0.277 & -0.291 & 8.497 & 1.390 & 1.362 \\
 $I4/mmm$-CsH$_4$ (low) & +0.691 & -0.031 & -0.315 & 8.040 & 1.029 & 1.451 \\
 $I4/mmm$-CsH$_4$ (high) & +0.538 & -0.194 & -0.075 & 8.290 & 1.258 & 1.098 \\
 $I4/mmm$-LaH$_4$  & +1.536 & -0.322 & -0.446 & 8.893 & 1.604 & 1.450 \\
 \hline
\end{tabular}
\caption{Bader charges and ELF basin populations on the metal atoms (M), hydrogens on  4$e$ sites (forming hydrogen pairs) and hydrogens on 4$d$ sites (tetrahedral sites) for the $I4/mmm$ structures at 50 GPa \label{badercharges}}
\end{table}

\subsection{S3.2 Can the atomic and molecular forms coexist in BaH$_4$?}

When two different forms of hydrogen can be found in a single structure, it opens the question of whether the real material should contain a mixture of each.  This could naively be treated as a mean field two-state system, with the probability of finding a given state being $e^{-\Delta G/kT}/[1+e^{-\Delta G/kT }]$ with $\Delta G$ being the free energy difference.   At room temperature this would be close to 50\%, and no discontinuous phase transformation could be observed.

However, the strong coupling between the atomic/molecular transition and the strain suggests that this view is overly naive.  Stabilization of the atomic form occurs only in concert with a large enlongation of the octahedral interstice, which is incompatible with a molecular form in an adjacent site.  A more appropriate model is that of Bragg and Williams (BW) which explicitly includes an interaction enthalpy between adjacent sites.  BW considered the interaction to be mainly a short-ranged bonding effect, however here it seems that strain provides the major contribution.

To test this we ran a series of molecular dynamics type calculations on a number of cells.  The purpose of these was to sample the available phase space, rather than study dynamical trajectories.  The partition function is independent of the atomic masses, we set the H$_2$ and Ba masses to be equal.   The results were monitored by
tracking all dihydrogen pairs, defined geometrically as atoms less than 1\AA\, apart.  We record each occasion when these connections change from one timestep to the next as an indication of how often the bonds break.  
We also calculated the radial distribution functions, the mean squared displacements, by visualisation of the trajectories using vmd, and by relaxation of selected snapshots.  

Long MD runs of single unit cells, just 10 atoms, followed by static relaxation was used to track dynamic instability against $\Gamma$-point phonons, it was found that starting from the low-c/a:
\begin{itemize}
\item
Molecular I4/mmm 50GPa with 300K rotated molecules and rebonded, relaxation gives structures with H$_3^-$ units. 
\item Molecular I4/mmm 50GPa with 600K transformed to a Cmcm-H3-like structure  by breaking and reforming bonds rather than rotating molecules. 
\item Molecular I4/mmm 200GPa 300K transforms to the atomic phase, with molecules diffusing in the plane by a bond-making and breaking mechanism. 
\end{itemize}
These small simulations with unphysical masses allow us to span the phase space. We also ran larger cells with 80 atoms and correct masses (137:1) to investigate the dynamics. We ran NPT simulations with a Parrinello-Rahman barostat which allows the cell to fluctuate but only permits small relative  movements of the Ba atoms.  At 50GPa MD shows that the hydrogen molecules retain their identity and rotate rather than remain orientated along the $c$ axis.  A consequence of that is further reduction in c/a so that the mean lattice parameters are a=3.6 (3.38), b=3.8 (3.38) c=5.2 (6.07), with 0K values in brackets.   At 600K a similar scenario plays out, with molecules breaking and reforming on a sub-picosecond timescale. 

Constant features across all MD on BaH$_4$  are the creation of close to one H$_2$ pair per formula unit, and the non-metallic bandstructure of the relaxed structures.

\begin{figure}[t!]
\includegraphics[width=0.5\columnwidth]{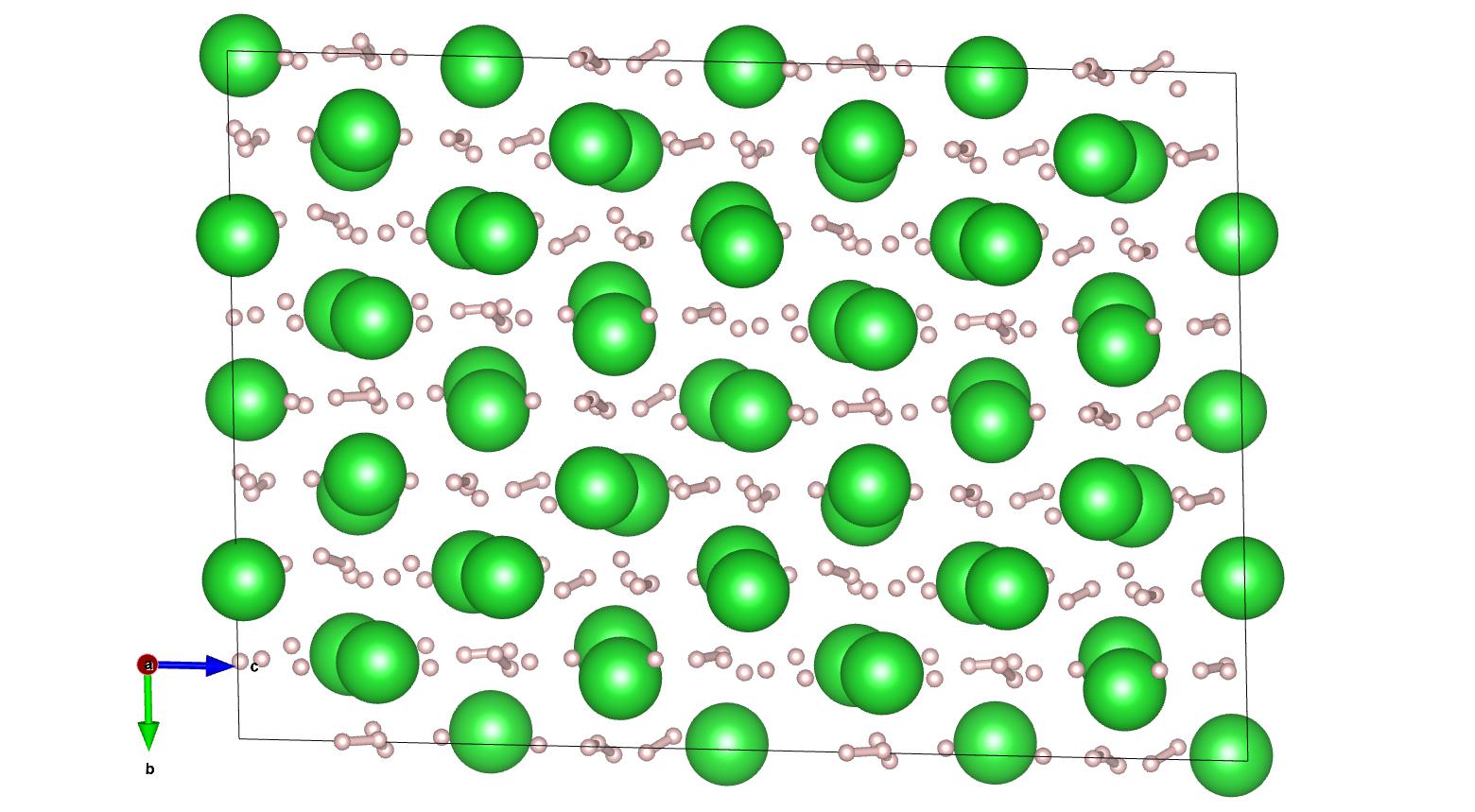}
\includegraphics[width=0.25\columnwidth]{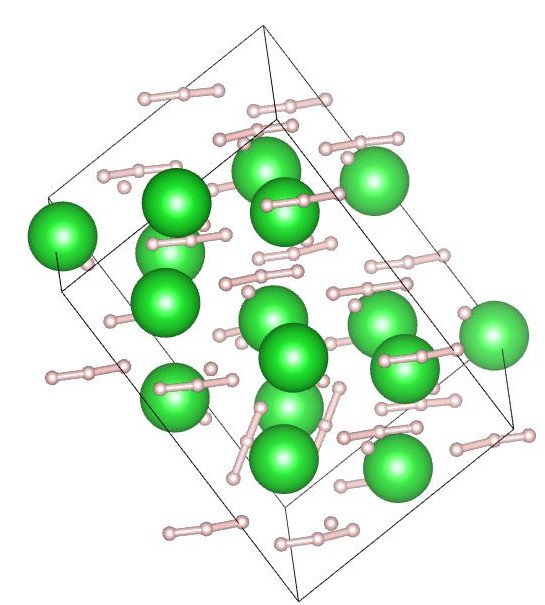}
\includegraphics[width=0.2\columnwidth]{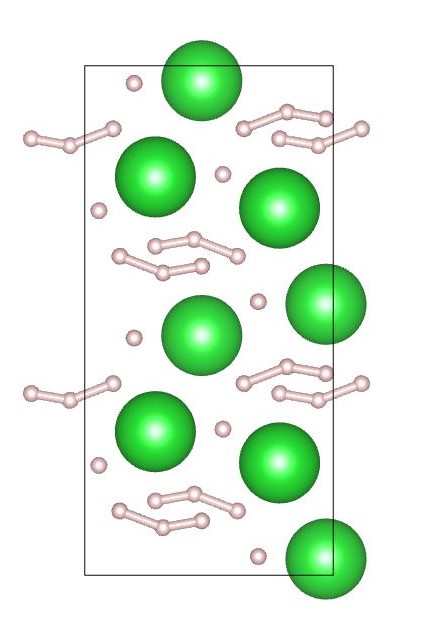}
\caption{Structures from $I4/mmm$ MD simulations. (left: MD snapshot 50GPa/300K; centre: snapshot after relaxation with H$_3^-$ units; right: 200GPa snapshot after relaxation with H$_2$ units)\label{fig:I4mmmsnap}
}
\end{figure}
\begin{figure}[t!]
\includegraphics[width=0.45\columnwidth]{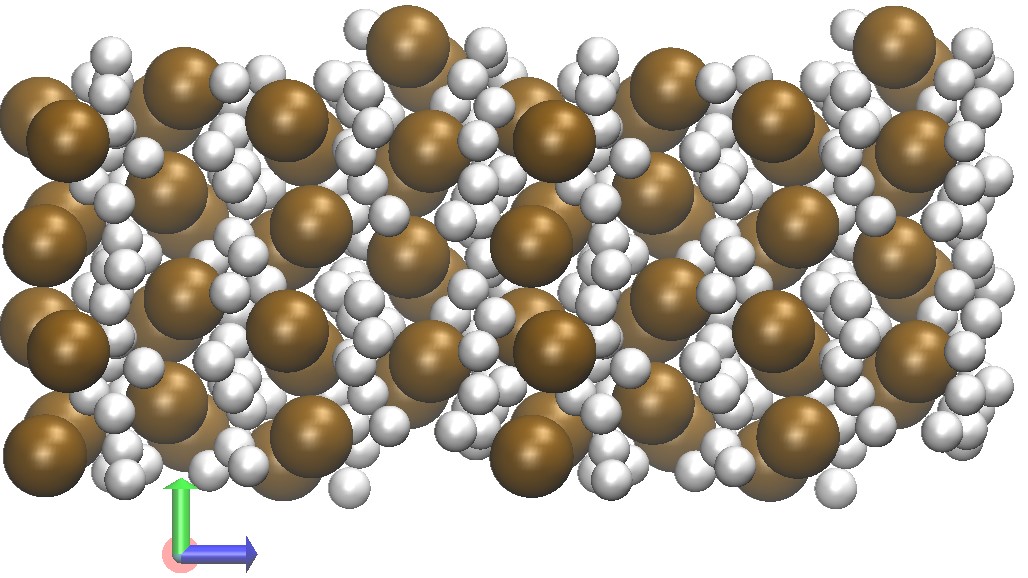}
\includegraphics[width=0.51\columnwidth]{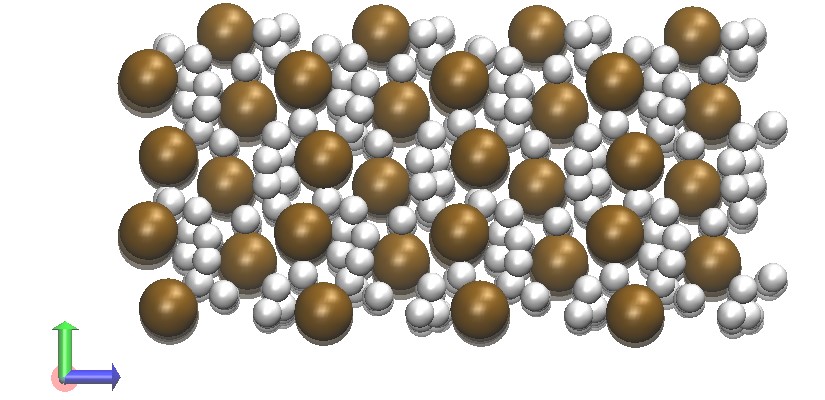}
\caption{ \label{fig:200GPa}
{\bf At room temperature both high and low c/a structures of I4/mmm BaH$_4$ contain molecules}
Snapshots from PBE MD simulation of (a) high $<a>=5.38, <b>=5.40, <c>=12.29$ and (b) low MD simulations at 200GPa and 300K  $<a>= 5.807,  <b>=5.50,  <c>=11.2$. Analysis of the number of dihydrogen pairs at less than 1\AA\, shows a distinct difference. The low c/a version averages 0.99 molecules per formula unit across the simulation.  The high c/a has "only" 0.85 pfu, however this is very high considering that the relaxed structure which has zero. 
Much of the difference arises from the more rapid breaking and remaking of bonds in the high c/a ratio case.  These bonds typically last for, of order, 0.1ps, consistent with 300cm$^{-1}$ lifetime broadening in Raman peaks.
The different nature of the two structures is evident from the pictures, the high c/a looks more like alternating Ba and H layers, which the low c/a has hydrogen primarily in interstitial locations.  }
\end{figure}

\begin{figure}[t!]
\includegraphics[width=0.48\columnwidth]{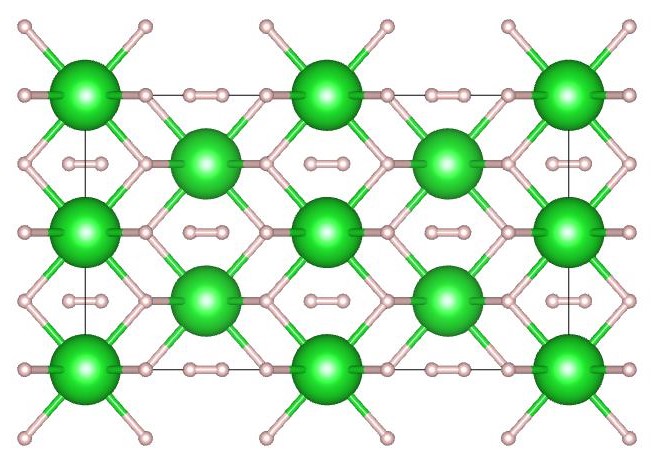}
\includegraphics[width=0.48\columnwidth]{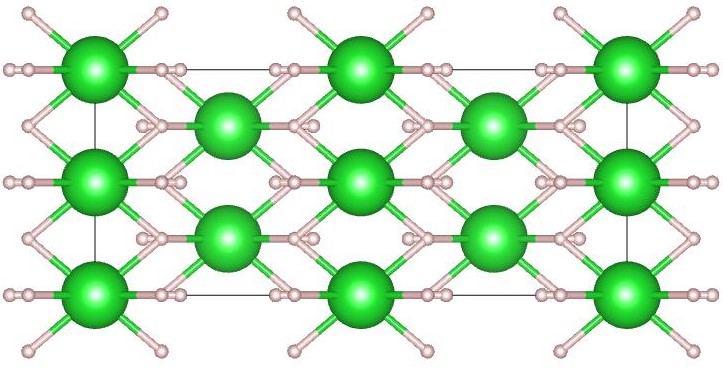}
\includegraphics[width=0.48\columnwidth]{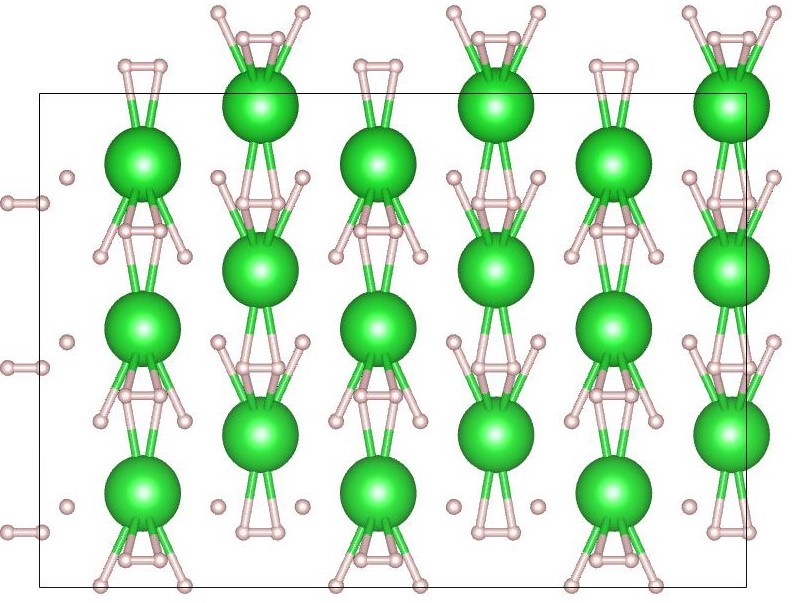}
\includegraphics[width=0.48\columnwidth]{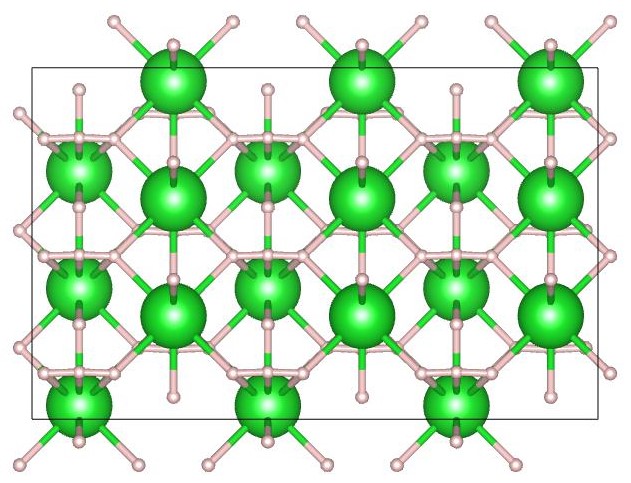}
\caption{ \label{fig:BaH4structures} Crystal Structures for BaH$_4$, I4/mmm low and high c/a (top); Cmcm-H2 and Cmcm-H3 (bottom)
Bonds are shown for Ba-H less than 2.4\AA, H-H less than 1\AA. }
\end{figure}

\subsection{S3.3 Stability of $I4/mmm$ BaH$_4$}

Molecular dynamics runs (Sup. Figs. \ref{fig:I4mmmsnap} and \ref{fig:200GPa}) show that the hydrogens in I4/mmm are dynamically disordered and only obey I4/mmm symmetry on average.  Moreover, geometry optimisation from molecular dynamics runs finds different symmetries, including the Cmcm-H3 which appears to be to low temperature ground state (Sup. Fig.\ref{fig:BaH4structures}).

\subsection{S3.4 Substoichiometric $I4/mmm$ materials}

\begin{figure}[t!]
\includegraphics[width=\columnwidth]{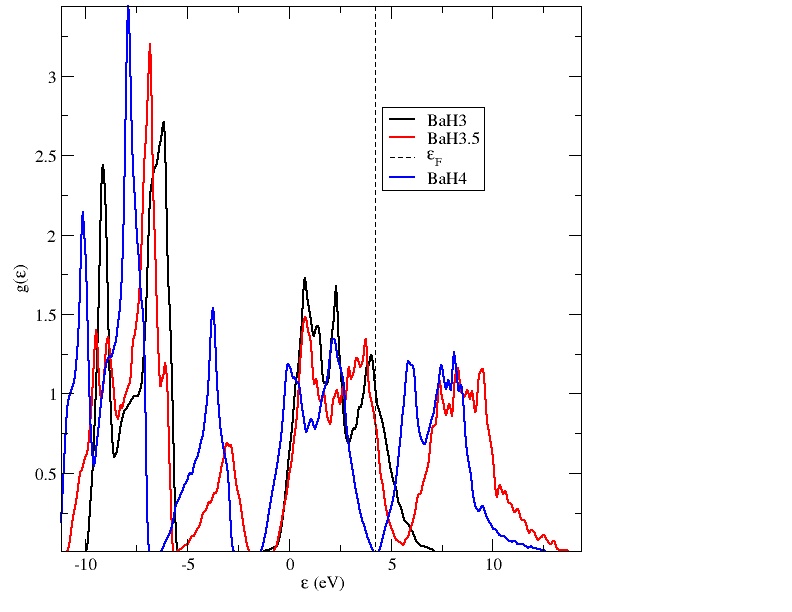}
\caption{ \label{fig:vacsdos}
Density of states of $I4/mmm$ BaH$_4$ at 50GPa with 50\% (red) and 100\% (black) of H$_2$ units converted to H.  The covalent states around -4 eV are entirely absent when all molecules are removed. The added H atoms contribute 2 states and 1 electron to the H$^-$ bands, leading to metallic structures.}
\end{figure}

In experiments, BaH$_4$ is formed from BaH$_2$ in a hydrogen-rich environment.  
We therefore investigated substoichiometric versions of the $I4/mmm$ structure, replacing the H$_2$ unit with a single H molecule. 

With all H$_2$ units changed to H, the stoichiometry is BaH$_3$ and the c/a ratio relaxed to  $\sqrt{2}$. the hydrogens remain in the high symmetry tetrahedral and octahedral sites, such that the overall symmetry of this structure is Fm-3m and the Ba are located on the sites of an fcc lattice.   With 50\% of H$_2$ changed to H, we find the c/a at 1.6.

The band structure of these off-stoichiometric structures have a clear signature of the disappearance of the H$_2$ bands and a broadening of the H$^-$ band leading to closure of the bandgap  (Sup. Fig.\ref{fig:vacsdos}).  

Nevertheless the fcc all-atomic BaH$_3$ structure  in DFT is highly unstable to molecule formation - by about 0.7eV per formula unit, whether one compares decomposition into cotunnite BaH$_2$ and BaH$_4$, or a BaH$_3$ structure with one molecule and one empty octahedral site. 

In molecular dynamics simulation with a single "missing" H we see that the hydrogen can diffuse on a sub-picosecond timescale.  The mechanism for this diffusion is the transfer of a hydrogen atom to molecule forming an 
H$_3^{-}$ complex, 
with one of the original pair moving away to become a free atom, leaving a new pairing.   We observed this migration mechanism by inspection in vmd and by monitoring the graph of connections 
$<$1\AA, between molecules in the MD.

\section{S4 Empty lattices}

\begin{figure}[!h]
\includegraphics[width=0.7\columnwidth]{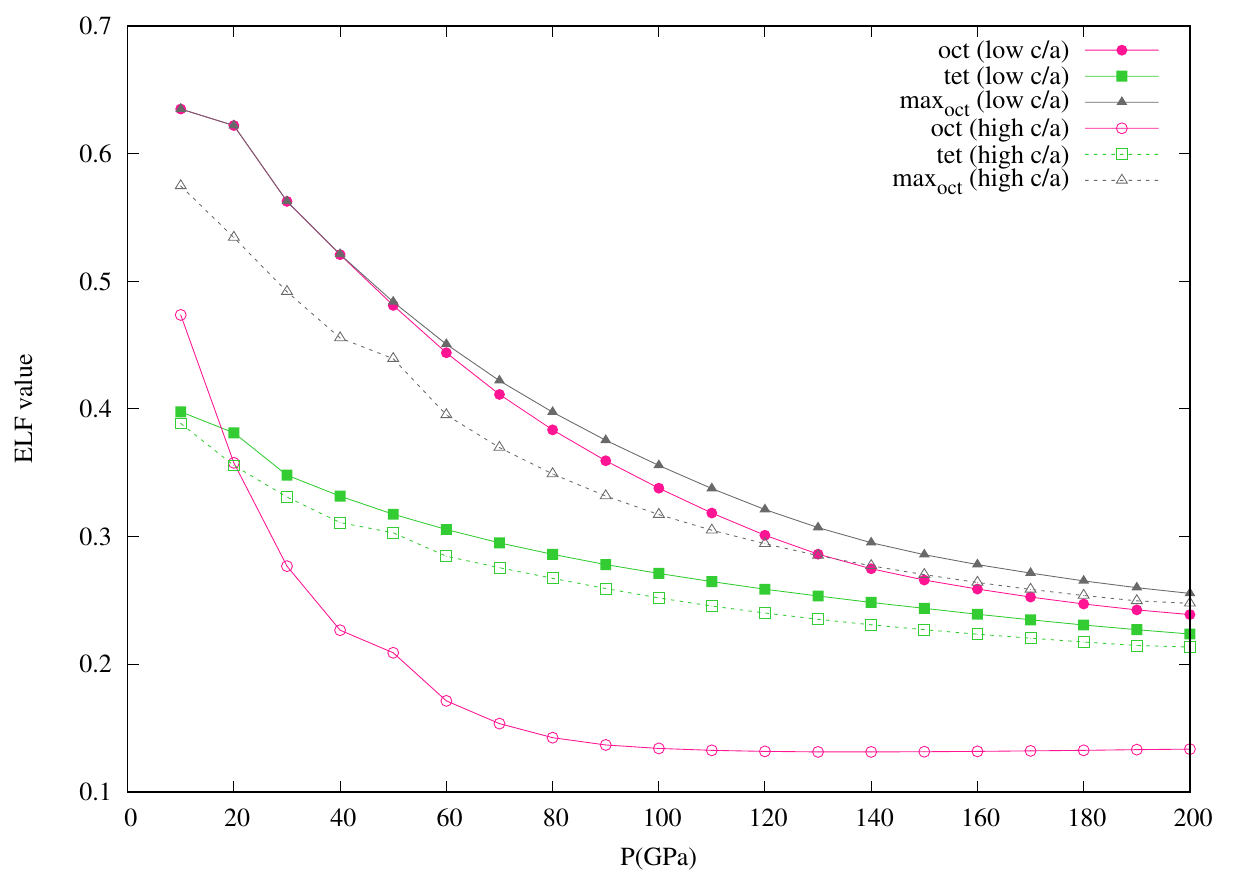}
\caption{ELF value at the 2$b$ site (oct), 4$d$ site (tet) and the ELF maxima on 4$e$ sites (max$_{\rm oct}$) with the 2$b$ site as middle point for the empty Ba lattices of low and high $I4/mmm$-BaH$_4$. }
\label{fig:ELFempty}
\end{figure}

\begin{figure}[!h]
\includegraphics[width=0.49\columnwidth]{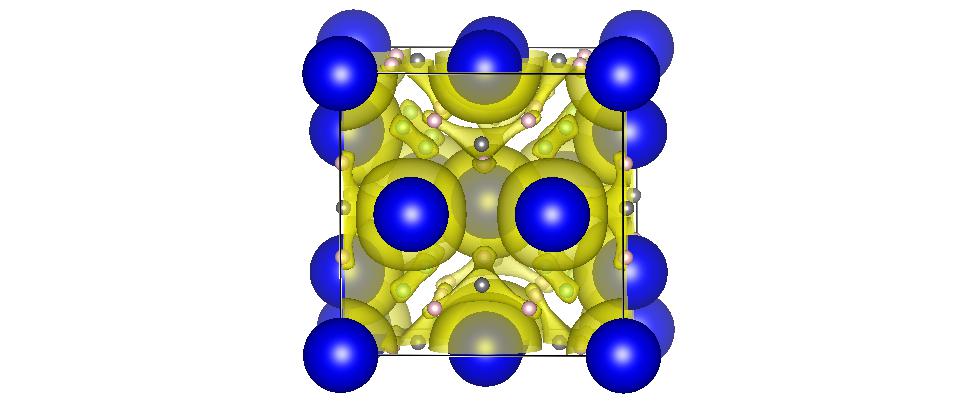}
\includegraphics[width=0.49\columnwidth]{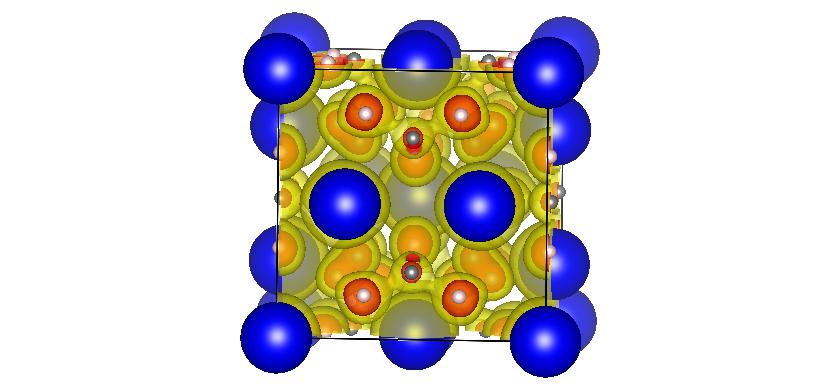}
\caption{a) ELF isosurface (ELF=0.32) for the pure $Pm\bar{3}n$ Ba sublattice of BaH$_{5.75}$ at 50 GPa (in yellow). Ba atoms represented as blue spheres and ELF maxima as green, pink and grey spheres. b) ELF isosurfaces, ELF=0.95 (in orange) and ELF=0.65 (in yellow) for $Pm\bar{3}n$ BaH$_{5.75}$ at 50 GPa. Ba atoms represented as blue spheres and the 3 non-equivalent H atoms as green, pink and grey spheres, respectively. \label{fig:pm3niso}
}
\end{figure}

We considered the ELF in the BaH$_4$ structures with the H$_2$ removed at a range of pressures (Sup. Figs.\ref{fig:ELFempty} and \ref{fig:pm3niso} ).

\section{S5 Other BaH$_x$ compounds}

\begin{figure}[t!]
\includegraphics[width=0.45\columnwidth, angle=-90]{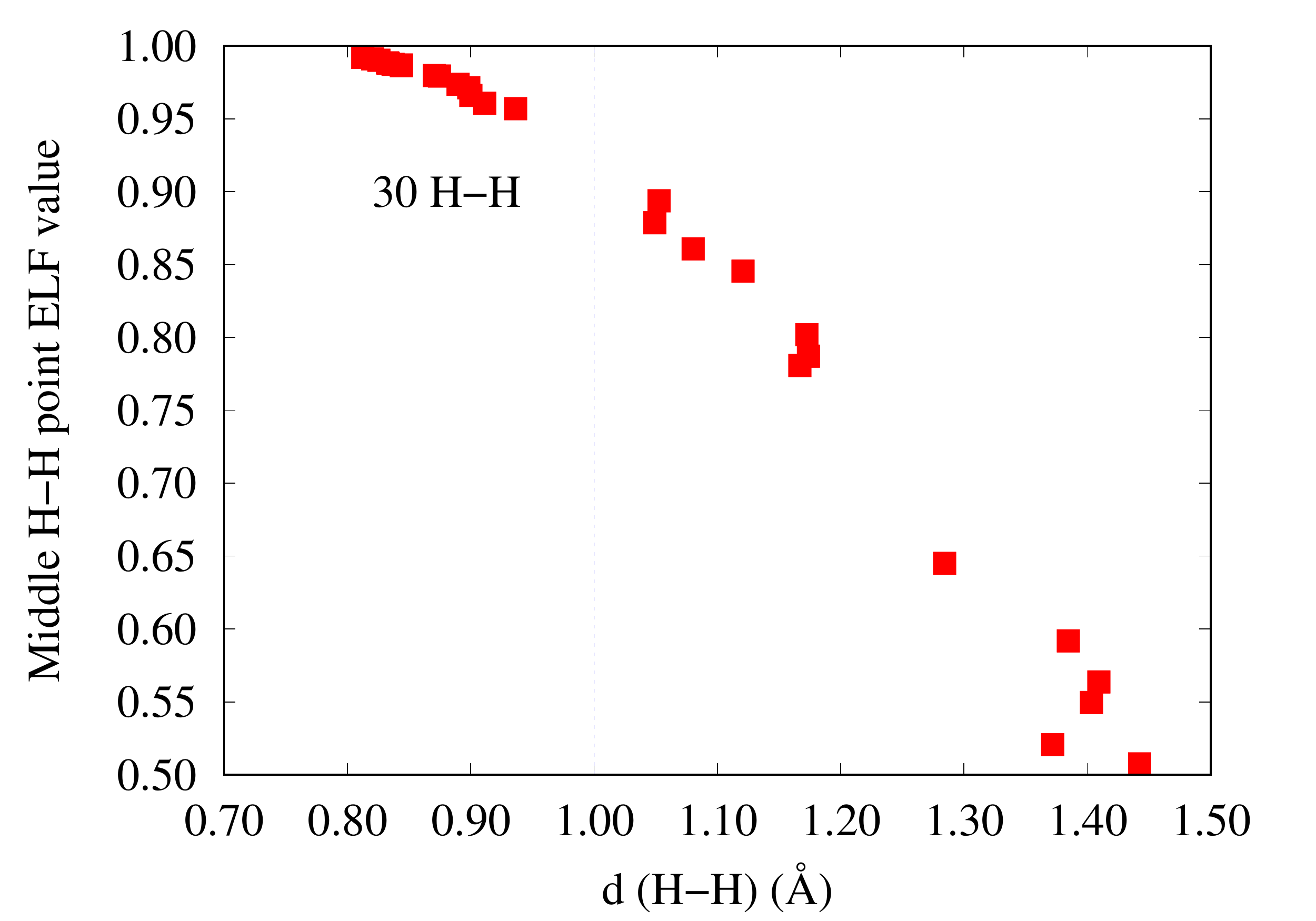}
\includegraphics[width=0.45\columnwidth, angle=-90]{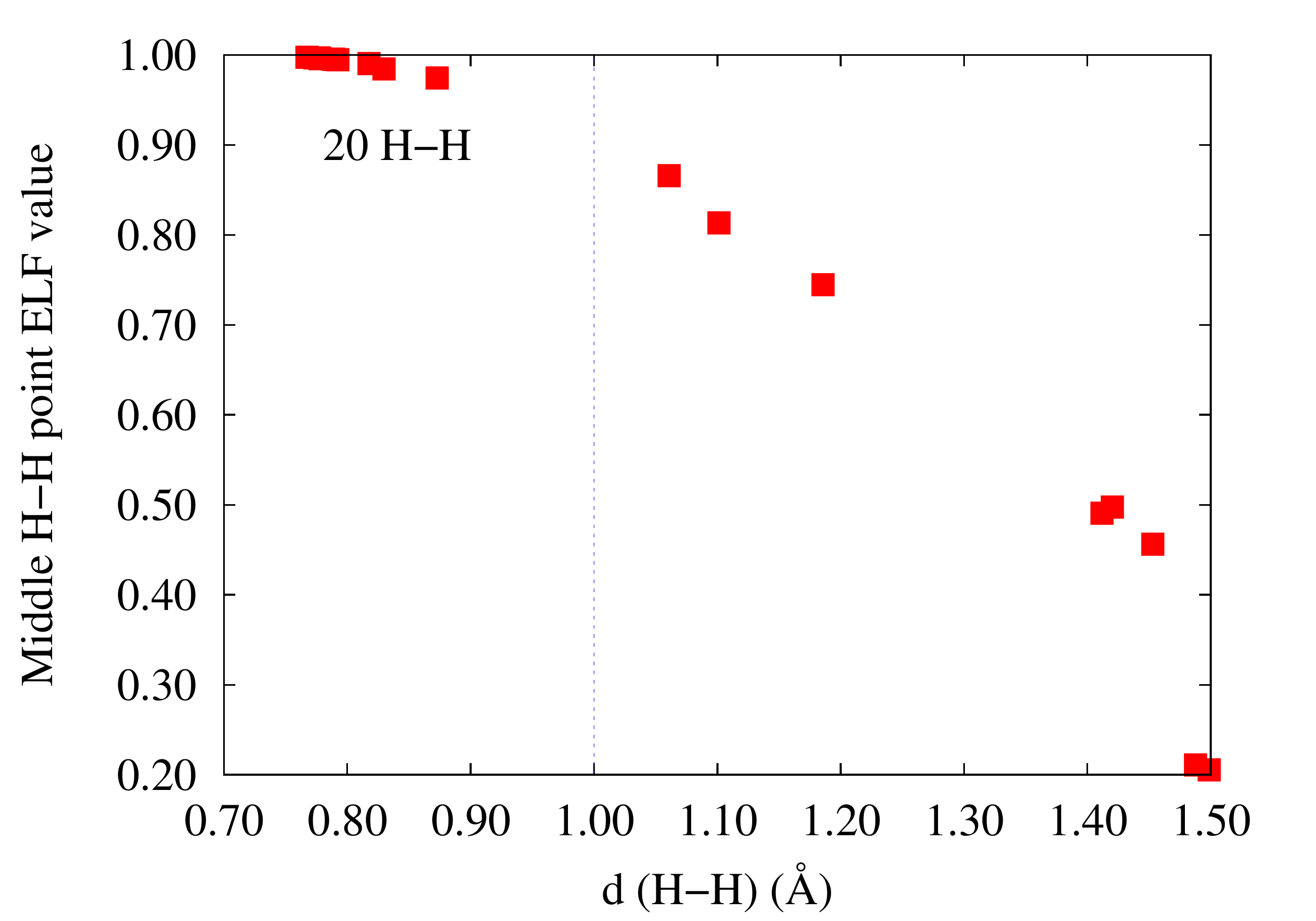}
\caption{ELF value at the middle H-H point for H-H distances up to 1.5 \AA at 50 GPa. a) BaH$_{5.75}$, b) BaH$_{12}$}
\label{fig:midELF}
\end{figure}

$Cmcm$-H3 is an interesting structure where we identify H$_3$ units (H (8$f$)-H (4$c$)$_1$-H (8$f$)). In this structure, Ba donates electrons to both, H$_3$ (4$c$)$_2$ and the H$_3$ units, that hold similarly high negative charges. This structure should be understood as BaH(H$3$). We also calculated the ELF for the high-symmetry structures
 BaH$_{5.75}$ and its empty-lattice equivalent, and for the high-pressure superconductor BaH$_{12}$ (Sup. Fig. \ref{fig:pm3niso}, \ref{fig:midELF}). 

\begin{table}[hbt]
    \centering
    \begin{tabular}{lccccc}
    \hline
    Atom & x & y & z & Bader charge & ELF charge \\ 
    \hline
 Ba (4$c$) & 0.000 & 0.117 & 0.250 & +1.164 & +1.485 \\
 H (4$c$)$_1$ & 0.000 & 0.402 & 0.250 & +0.085 & +0.320 \\
 H (8$f$) & 0.000 & 0.610 & 0.565 & -0.354 & -0.540 \\
 H (4$c$)$_2$ & 0.000 & 0.810 & 0.250 & -0.541 & -0.732 \\
 H$_3$ & & & & -0.623 & -0.754 \\
 \hline
\end{tabular}
\caption{Bader charges and ELF basin populations for the $Cmcm$-H3 structure at 50 GPa (a=3.462 \AA. b=7.740 \AA, c= 5.080 \AA). \label{baderchargescmcm}}
\end{table}

\section{S6 Details of Molecular Dynamics of specific materials}

\subsection{S6.1 MD of $I4/mmm$ BaH$_4$}

We ran a series of MD simulations on BaH$_4$ with high and low c/a ratios, at 50GPa and 200GPa, and at 300K, 600K and 1000K.  

\begin{figure}[t!]
\includegraphics[width=0.9\columnwidth]{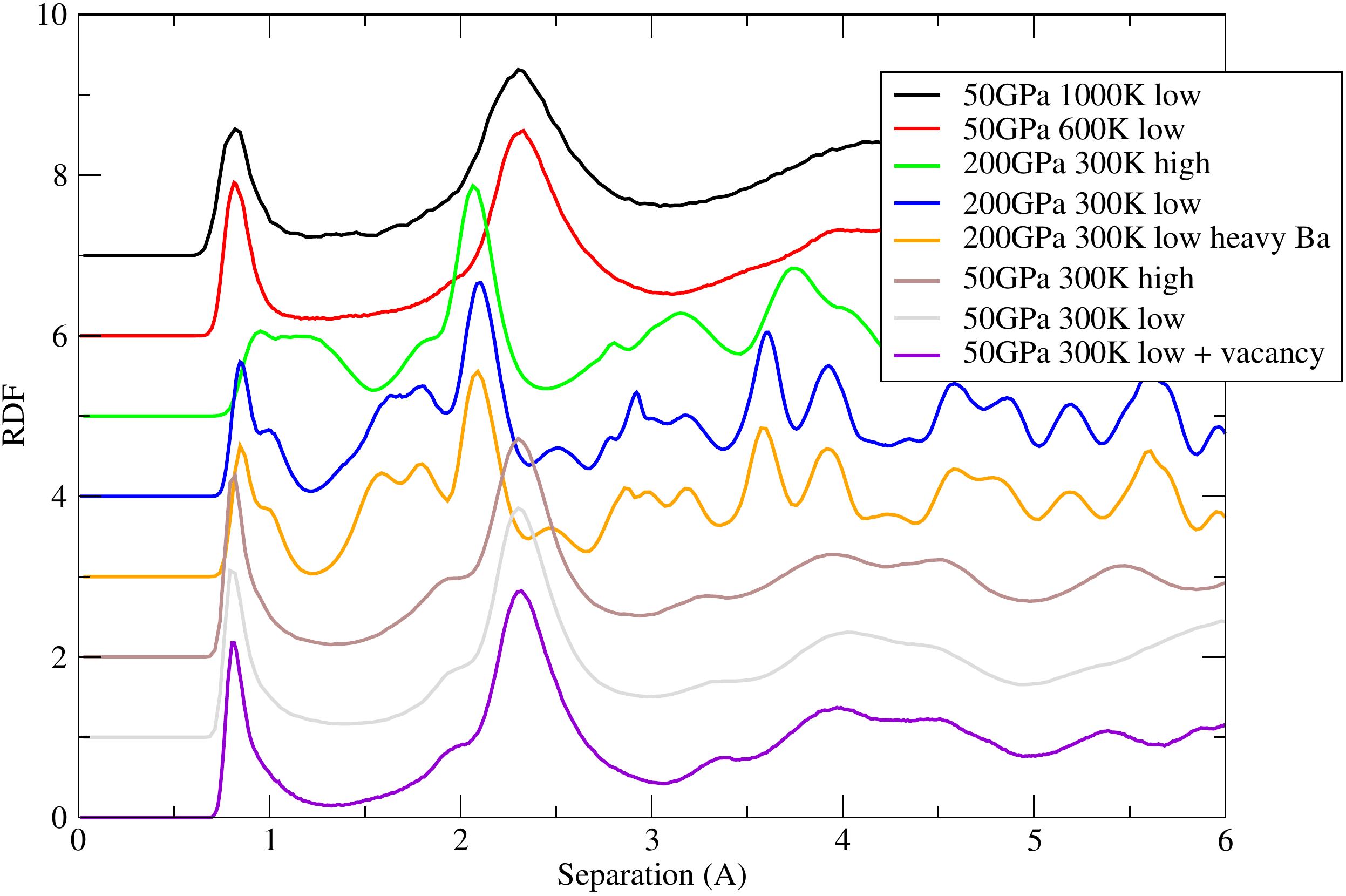}
\caption{Radial distribution functions from molecular dynamics simulations in $I4/mmm$ BaH$_4$ at various conditions. Each graph shows total g(r) normalised to total atomic density.  \label{fig:I4mmmrdf}
}
\end{figure}

At 50GPa/300K, (Sup. Fig.\ref{fig:I4mmmsnap}) the I4/mmm simulation box transformed to a distorted structure, which after relaxation has a cell 
    (   a = 7.078 , $\alpha$ =   86.9,
                    b =      7.593       ,   $\beta$  =   89.6,
           c =     10.624        , $\gamma$ =  101.7)
this contrasts with molecular dynamics with "correct" Ba mass where the Ba motion was too slow to allow for the transformation and rotation of the hydrogen molecules was observed.  A distinctive feature of this structure is the $H_3^-$ units, as shown in figure \ref{fig:I4mmmsnap}. This type of ``polyhydrogen" chain has been associated with insufficient k-point sampling\cite{magduau2017charge}, but here it is stable in grids up to 12x12x6 in an 80-atom supercell. In the RDF (Sup. Fig.\ref{fig:I4mmmrdf}) these units manifest as a shoulder on the first H$_2$ peak.
In figure \ref{fig:I4mmmsnap} we show that these H$_3^-$ units in the relaxed structure are largely parallel, but including two which have different orientations. The  H$_3^-$ units were previously noted in a metastable $C2c$ structure, so the relaxed MD structure appears to be this structure with misoriented defects.  The bond-breaking  analysis detects many bond breaking events, but on closer inspection these are almost all within the H$_3$ triplets.  This means of bond-count is remarkably stable since both H$_3^-$ units and H$_2$ molecules contain 2 atoms with one neighbour, so if a third atom approaches a molecule, or an H$_3^-$ unit breaks, the measure is unchanged.

For same 50GPa/300K conditions starting with high c/a=2.323, the c/a ratio remained roughly constant.  Despite this, the radial distribution function is remarkably similar to the low c/a version, dominated by H$_2$ molecules, with an H$_3^-$ shoulder below 1 \AA and the Ba-H peak at 
around 2.3 \AA.  

At 200GPa/300K  the molecules, initiated along the z-direction, rotated into the x-direction and rebonded, the  simulation box remained close to orthorhombic.  
The H$_2$ molecules retained their integrity, although in the relaxed final configuration each H$_2$ bond (0.87 \AA\,, Mulliken bond population $\equiv 1$) is associated with a third hydrogen (1.02  \AA\,, Mulliken bond population $\equiv 0.5$).  These objects seem to be intermediate between the symmetric H$_3$ units and independent H$_2$ and H$^-$.  At these pressures the molecular rotation is strongly suppressed, giving rise to a much more structured radial distribution function.  The high c/a version RDF is completely different the H$_2$ peak is poorly defined and now at 0.95 \AA. Bond lifetimes are half that of the low c/a version

At 50GPa/600K the molecules started to rotate and c/a ratio dropped abruptly from the static value of 1.76 to 1.32, with $b<a$ by about 10\%.  Taking the mean of 2c/(a+b) across the run gives 1.41, the ratio at which the Ba atoms lie on an fcc lattice.

\begin{figure}[t!]
\includegraphics[width=0.49\columnwidth]{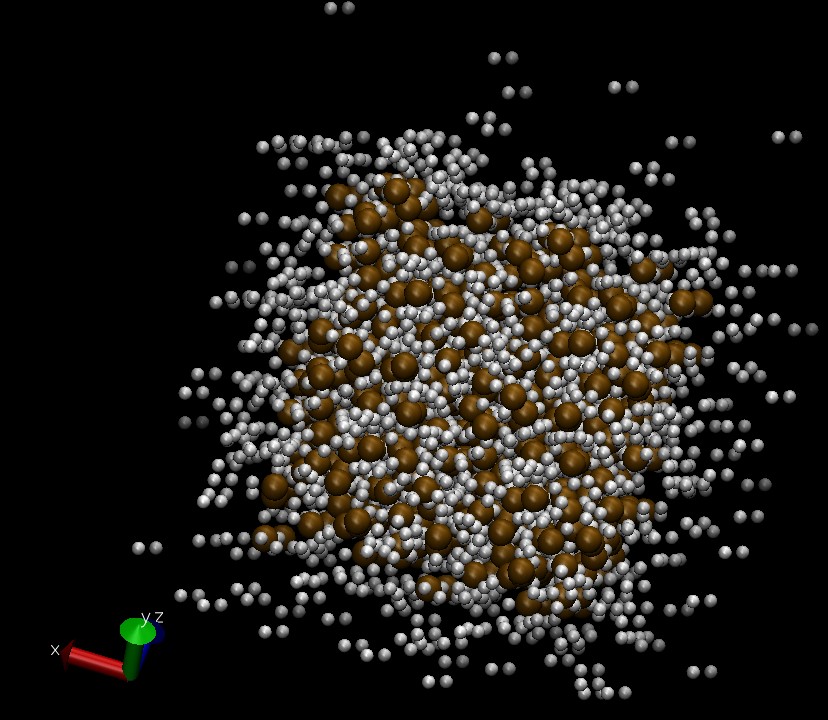}
\includegraphics[width=0.49\columnwidth]{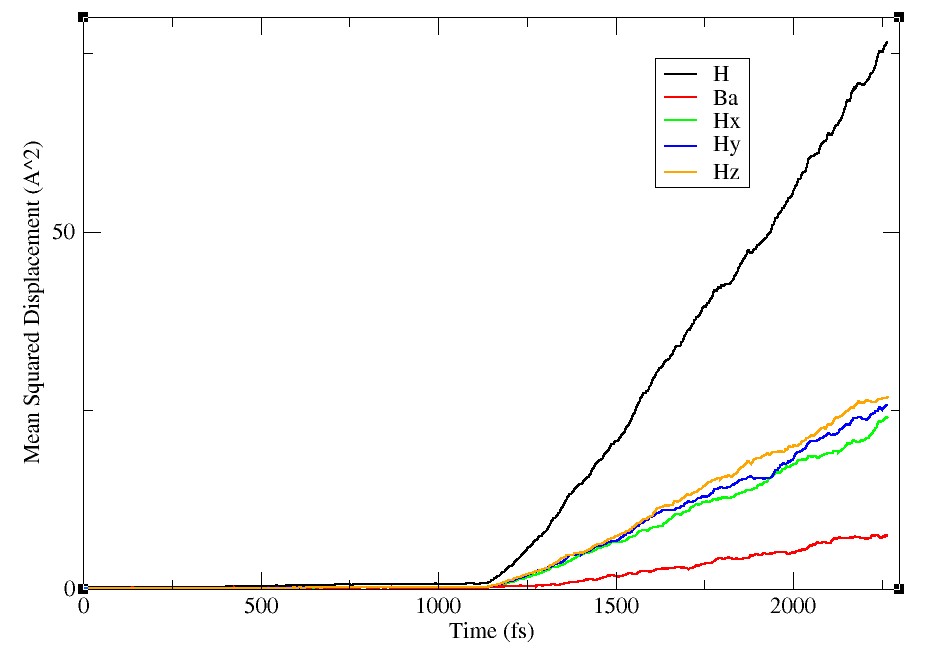}
\caption{ \label{fig:melt575}
{\bf Equal-mass MD can be used to demonstrate melting : molten Ba }Details from the "equal-mass" MD simulation in BaH$_{5.75}$ which melts at 1000K 100GPa a) Snapshot of the molten configuration (doubled cell, periodic boundary conditions suppressed).
b) Mean squared displacements of the "light" Ba (red) and Hydrogen (black, also x,y,z components showing homogeneous motion)}
\end{figure}

\begin{figure}[t!]
\includegraphics[width=0.49\columnwidth]{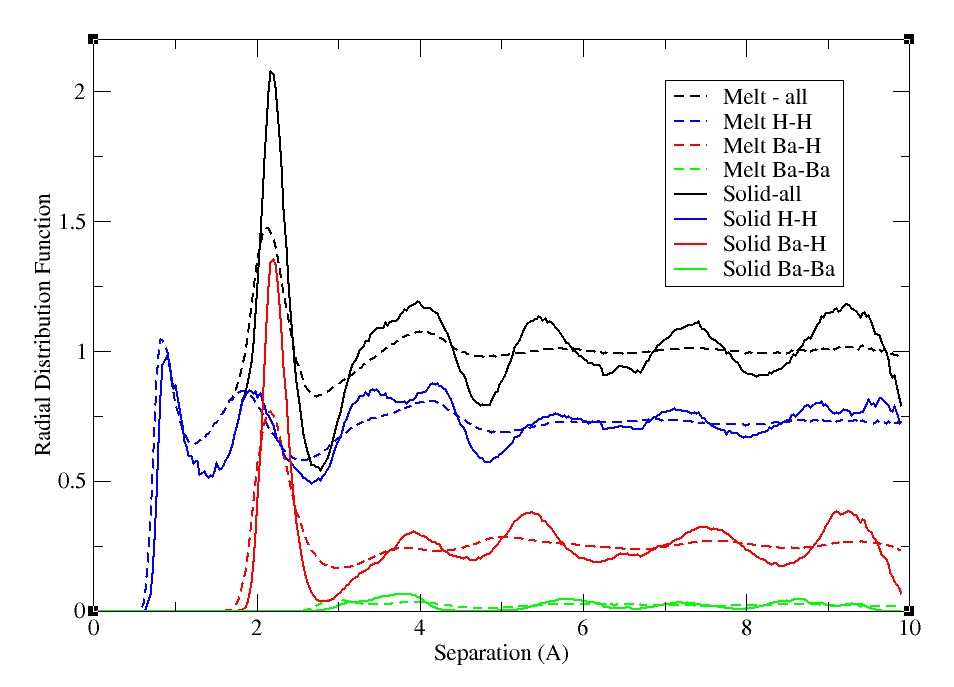}\includegraphics[width=0.49\columnwidth]{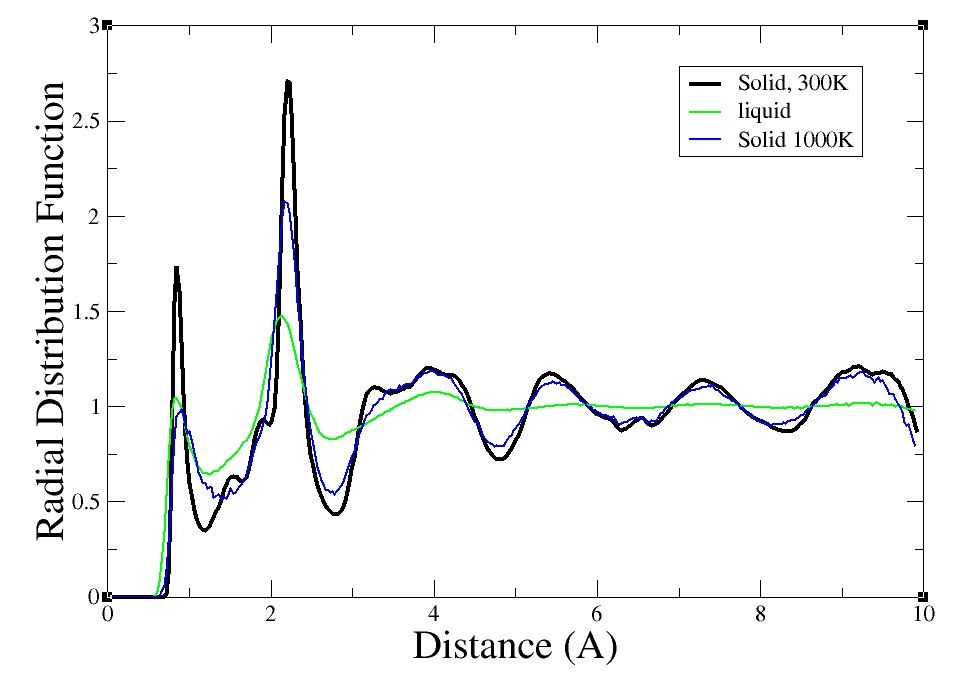}
\caption{{\bf BaH$_{5.75}$ as BaH$_2($H$_2)_x$}
(left) Radial distribution function of the 100GPa/300K simulation, compared with those at 1000K in liquid and solid form.
(right) Pre- and post-melting radial distribution functions, and the breakdown by species.
Analysis of the bonding shows that the MD in typically 94\% fully bonded, compared to the low symmetry, relaxed structure BaH$_2$(H$_2)_x$  (227 compared to the ideal 240/368 hydrogens in bonds).
The lifetime of these bonds, as measured by counting changes in the bond list, is given in Table 3. \label{fig:rdf575}
}
\end{figure}
\begin{figure}[t!]
\includegraphics[width=0.39\columnwidth]{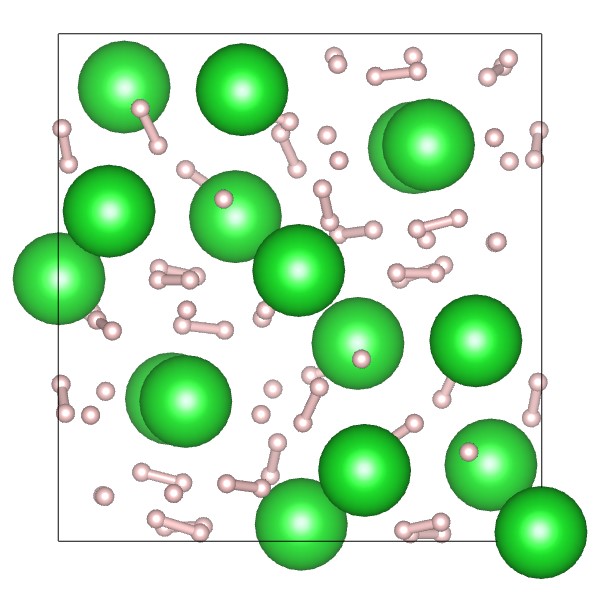}\includegraphics[width=0.59\columnwidth]{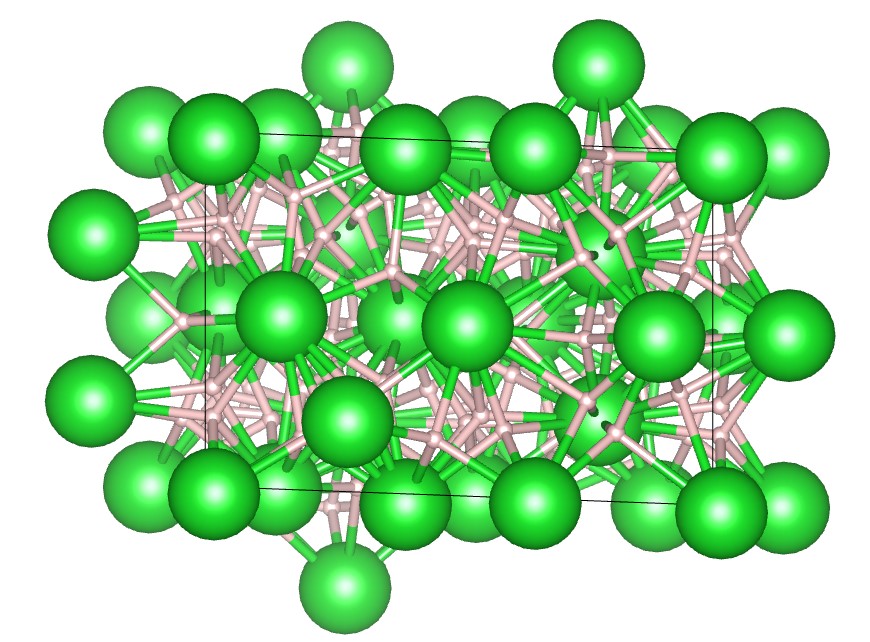}
\caption{The most stable found structure (Pc - Ba$_{16}$H$_{92}$) of BaH$_{5.75}$ relaxed at 50GPa
(left) Showing 30 H$_2$ bonds pfu in range 0.813-0.936\AA. (i.e. BaH$_2($H$_2)_x$ - next shortest is 1.049\AA. 
(right) Showing Ba-H bonds less than 3\AA, illustrating that each hydrogen has precisely 4 Ba neighbours (i.e. is confined to a tetrahedron).\label{fig:P1575}
}
\end{figure}

\begin{table}[hbt]
    \centering
    \begin{tabular}{|c|c|c|c|c|c|c|c|}
    \hline
        \textbf{System} & Time (ps) & T (K) & P (GPa) &  Mass & bonds & Events N,$ps^{-1},$ & Bond \\
        \hline
        $Cmcm$-H2 (Ba$_{16}$H$_{64}$) & 2.5  & 600 & 50 & 4 & 32,0 & 715 $\implies$ 4.5 & 0.80 \\
         $Cmcm$-H2 (Ba$_{16}$H$_{64}$) & 10  & 300 & 50 & 4 & 32,0 & 12 $\implies$ 0.01 & 0.79 \\
        $Cmcm$-H3 (Ba$_{16}$H$_{64}$) & 5  & 300 & 50 & 4 & 32,7 & 16317 $\implies$ 25.5  & 0.91 \\
     $I4/mmm$ (high, Ba$_{16}$H$_{64}$)
     & 5.6 & 300&200& 2& 27,2& 21058 $\implies$ 69.2 &0.95 \\
         $I4/mmm$ (high, Ba$_{16}$H$_{64}$)
     & 5.0 & 300& 50& 2& 32,2& 6080$\implies$19.0 & 0.81 \\
            $I4/mmm$ (low, Ba$_{16}$H$_{64}$) & 5.0 & 300&200& 2& 32,6& 12154 $\implies$ 38.0 & 0.84  \\
       $I4/mmm$ (low, Ba$_{16}$H$_{64}$) & 4.4  & 1000 & 50 & 2 & 32,1 & 8901 $\implies$15.8 & 0.80 \\
        $I4/mmm$ (low, Ba$_{16}$H$_{64}$) & 3.9  & 600 & 50 & 2 & 32,6 & 4590$\implies$ 18.4 & 0.82 \\
        $I4/mmm$ (low, Ba$_{16}$H$_{64}$) & 7.0  & 300 & 50 & 2 & 32,0 & 8019$\implies$  15.9 & 0.81 \\
        $I4/mmm$ (low, Ba$_{16}$H$_{63}$) & 4.8  & 300 & 50 & 2 & 31,1 & 5805 $\implies$19.5 & 0.81 \\
        WP(Ba$_{64}$H$_{368}$) & 6.5  & 300 & 100 & 137 & 227,10 & 75761$\implies$ 25.7 & 0.85 \\
             WP(Ba$_{64}$H$_{368}$) & 0.5  & 700 & 100 & 2 & 227,10 & 4462$\implies$ 19.7 & 0.80 \\ 
                          WP(Ba$_{64}$H$_{368}$) & 0.3  & 700 & 0 & 2 & 238, 0 &1538 $\implies$ 10.8 & 0.76 \\ 
        WP(Ba$_{64}$H$_{368}$) & 1.2  & 1000 & 100 & 2 & 175,10 & 36685$\implies$ 67.4 & 0.89\\  
        liquid (Ba$_{64}$H$_{368}$) & 4.5  & 1000 & 100 & 2 & 202,17 &   80194 $\implies$ 44.1 & 0.83 \\
     WP(Ba$_{16}$H$_{92}$) & 5.0  & 300 & 50 & 2 & 59,1 &  7514 $\implies$ 12.8 & 0.82 \\
        Ba$_8$H$_{96}$ & 2.1 & 300 & 50 & 2 & 80,1 & 1855 $\implies$ 5.5 & 0.77\\
        Ba$_8$H$_{96}$ & 2.1 & 300 & 100 & 2 & 80,4 &3526 $\implies$ 10.6 & 0.78\\
        Ba$_8$H$_{96}$ & 5.0 & 300 & 150 & 2 & 80,8 &10387 $\implies$ 13.0 & 0.78\\
        Ba$_8$H$_{96}$ & 7.0 & 300 & 250 & 2 & 77,8 & 26254 $\implies$ 24.4 & 0.80 \\
        \hline
    \end{tabular}
    \caption{Details of molecular dynamics simulations. WP is the Weaire-Phelan structure, Mass refers to the cation, bonds is the average number of hydrogen atoms within 1\AA\,\,  of precisely (one,two) other atoms. Broke is the number of times and atoms move into or out of the 1\AA\,\, range of another, and this is normalise with time and number of bonds to give a number of such "events" per bond per ps. 
    Relating this to the lifetime of the bond is complicated.  e.g. harmonic bond oscillating beyond 1\AA\, will record four "events" (two atoms, making and breaking) without being a genuine bond breaking.  The common proton exchange process also records four "events" (one atom makes a bond, one breaks a bond, the central atom both makes and breaks). However in most of the recorded H$_3$ creation events, the "reaction" is "unsuccessful" and the initially-atomic hydrogen leaves as an atomic hydrogen. Taken together, a rough estimate is that bond breaking occurs with a frequency one order of magnitude less than the "Event" rate.
     Bondlength is the position of the first peak in the radial distribution function.}   
\end{table}

\subsection{S6.1 MD of other BaH$_4$ candidates}

\begin{figure}[t!]
\includegraphics[width=0.9\columnwidth]{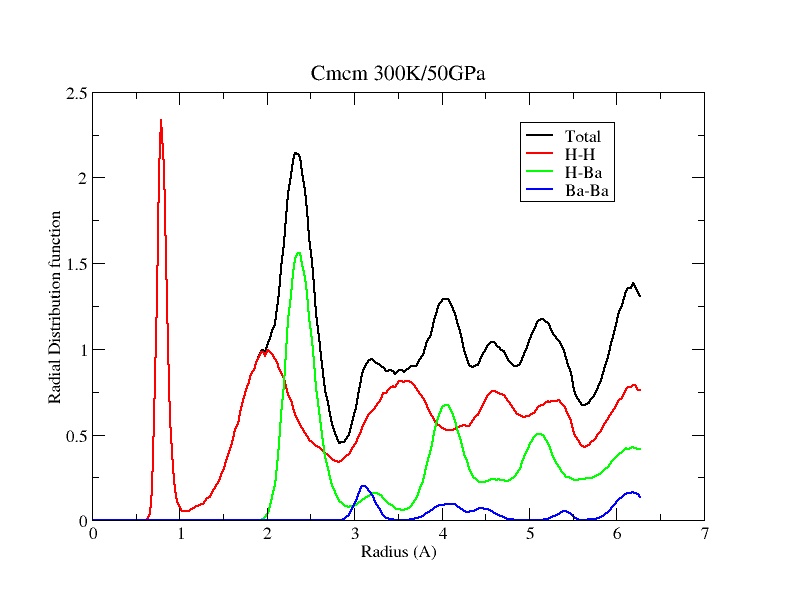}
\caption{Radial distribution functions from molecular dynamics simulations starting in the $Cmcm$H2 structure.  Each graph shows total g(r) normalised to atomic density, and contributions from each bond type. 
\label{fig:rdfCmcm}
}
\end{figure}

MD simulation in Cmcm-H2 at 50GPa shows this structure remaining stable for at least 5ps, with well-defined H$_2$ and H$^-$ (Sup. Fig. \ref{fig:rdfCmcm}) .  Nevertheless, there are hints of the asymmetric H$_3^-$, with about 10\% of H$_2$ dimers being associated with a third H  within the first minimum of the radial distribution function. 
Likewise, the Cmcm-H3, started in the stable state with H$_3^-$ units remains very close to the original structure, with 32 molecule throughout.  Our bond-breaking analysis marks this structure as having frequent bond breaking, but this is largely an artefact of fructuations in the  H$_3^-$ units, which have mean bondlengths above 0.9\AA\, and are therefore susceptible to H-H distances greater than 1\AA\, appearing in thermal fluctuations.

\subsection{S6.2 MD of BaH$_{5.75}$}

The experimentally-observed BaH$_{5.75}$ $Pm\bar{3}n$ structure is stabilized by temperature, static relaxation calculations suggest a symmetry breaking distortion is which the "atoms" pair up to become molecular.  At 50GPa, the minimum energy structure has only Pc symmetry, with a clear separation between the 15 H-H separations below 0.94\AA\, and with bond population above 0.73e$^-$, and the 16th closest dihydrogen pair (1.05\AA\,/0.5e$^-$).  These numbers demonstrate that  in conventional chemistry terms the Ba$_8$H$_{46}$ structure could be regarded as a solid solution 8BaH$_2$+15H$_2$, rather than a Barium hydride.   Further evidence for this comes from the electronic density of states which strongly resembles that in molecular I4mmm BaH$_4$, with distinctive bands associated with low-lying Ba 5s (-22eV) and 5p orbitals (-10eV), the H$_2$ molecular bonding orbitals.  Closest to the Fermi energy are the hybridised bands of the Ba 5d and atomic hydrogen: population analysis in atomic basis set suggests that the Ba 6s are unoccupied.    

The Weaire-Phelan topologically close-packed BaH$_{5.75}$ phase was run in an MD simulation with 64 formula units (Ba$_{64}$H$_{368}$ at 100GPa at 300K with full-mass Ba and 100GPa at 700K and 1000K with equal-mass MD.  The 700K run was continued at 0 GPa.
Molecular dynamics of the Ba$_{64}$H$_{368}$ supercell (100GPa/300K) reveals a different picture from the static case (Sup. Figs.\ref{fig:melt575} and \ref{fig:rdf575}).   The molecularization to approximately Ba$_{64}$H$_{128}$(H$_2)_{120}$ occurs, but the covalent bonds are continuously breaking and reforming.  
The topological close packing means that all interstitial regions are tetrahedral, and there is precisely one hydrogen atom in each tetrahedron;  However, the hydrogens are able to pair up with covalent bonds forming through  the triangular faces of the tetrahedra, and analysis shows that on average 114 bonds exist at any time. This number follows a skewed distribution with a maximum at 120: as expected for $x$ from  the formula (BaH$_2$)(H$_2)_x$.  The radial distribution function does not drop to zero, and so this is somewhat sensitive to R$_{cut}$.   With 368 hydrogen atoms in the cell, on average there are 133.5 unbonded H units compared to 128 expected in the relaxed ground state structure. The simulation ran for 6.5ps during which time all hydrogen atoms in the cell spent some time as molecules, most forming bonds at some point with more than one of the four atoms in neighbouring tetrahedra.  No hydrogen atom "escaped" from its original tetrahedron.

Heating to 700K with the equal mass sampling gives a similar result, while continuing the simulation at 0GPa/700K reduces the rebonding and increases the molecule count to an average of 119.

A longer run with 100GPa/1000K melted after a lengthy equilibration period.  The melt structure has a rather similar first peak to the solid  H-H radial distribution, showing that the majority of hydrogen is still in molecular form: the exact number depends of the radius chosen for a molecules.   A distinct Ba-H rdf peak is evident around 2.2\AA\, in both liquid and solid, but it is less distinct in the liquid.  The Ba-Ba liquid rdf has an unusual double peak at 2.8\AA\, and 3.9\AA\, This is quite unlike the solid.  We can associate the first peak with adjacent Ba$^{2+}$ ions (textbook ionic radius 1.35\AA\,), and the second peak with two Ba's separated by hydrogen. It is notable that, despite the low Ba mass, the mean-squared displacement of the Ba is still much lower than that of hydrogen, indicating the role of size in diffusion.

We attempted to determine stable ground-state structure by relaxing MD snapshots.  Relaxed structures were always significantly more stable than the $Pm\bar{3}n$ structure reported from X-ray diffraction, although the Ba atoms are always close to that structure (Fig.\ref{fig:P1575}).

Relaxing a snapshot from the large MD at 100GPa to 0GPa gave precisely 120 bonds by the sub 1\AA criterion,  with a gap in observed bondlengths between 0.92 and 1.26\AA, coinciding with a sharp drop in the Mulliken bond charge from 0.8 to 0.2.

Relaxing to 50GPa had a similar effect, with slightly longer bonds and a less sharp division between bonded and unbonded.  Again, the Mulliken charge drops from 0.8 to 0.2 between 0.9\AA\, and 1.2\AA, but now there are several bonds with distances (and charges) between.

Other snapshots showed similar results, but no unique structure emerged. It is evident from the relaxed structures that the H$_2$ molecules are oriented perpendicular to the vector the short BaH vector.  This follows from the fact that the bond links hydrogens in different tetrahedra, and therefore goes through a triangle of three bariums. 

\begin{figure}[t!]
\includegraphics[width=0.59\columnwidth]{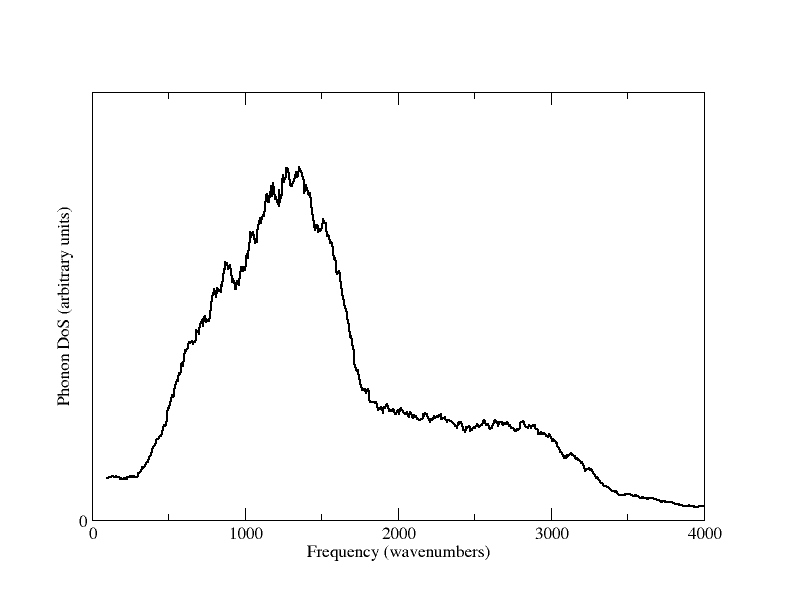}
\caption{ \label{fig:WPphonons}Fourier transform of the velocity autocorrelation function from MD of the BaH$_{5.75}$ structure (100GPa/300K).  The H$_2$ vibrations show a broad but unresolvable region between 2000 and 3000 wavenumbers. \label{fig:FT575}}
\end{figure}

These unit cells are too large for lattice dynamics calculation of Raman spectra, but we can get some indication of the vibrations in the system from the velocity autocorrelation function (Sup. Fig.\ref{fig:FT575}  This shows a large bump between 300 and 1800 cm$^{-1}$ corresponding to lattice and librational motion, and a broad shoulder extending up to 3300 cm$^{-1}$.  This breadth is due to both the range of H$_2$ environments, giving rise to multiple vibrational frequencies, and their short lifetime which broadens each peak. 

\subsection{S6.3 MD of BaH$_{12}$}

BaH$_{12}$ was reported by Chen et al. to be a molecular metallic barium hydride, identified and pseudocubic,  and that  ab initio calculations show that newly discovered semimetallic BaH$_{12}$ contains H$_2$ and H$_3$  molecular units and detached H chains which are formed as a result of a Peierls-type distortion of the cubic cage structure.
\begin{figure}[t!]
\includegraphics[width=\columnwidth]{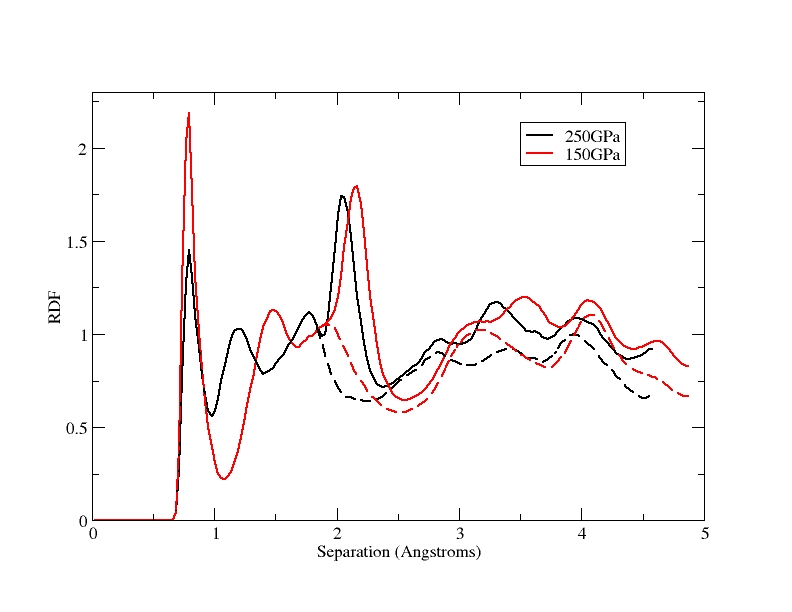}
\caption{Radial distribution functions from molecular dynamics on BaH$_{12}$. Solid lines are total RDF, dashed lines count only the H-H distances.  The sharp first peak at 0.8\AA\, indicates the preferred molecular bondlength seen in these simulations and is almost independent of pressure.  The compression with pressure is evident only in the second peak - the intermolecular distance. The second largest peak is the BaH nearest-neighbour separation, which can be seen to reduce with pressure.   The overall crystal structure can be described as pseudocubic. \label{fig:MD12}
}
\end{figure}

\begin{figure}[t!]
\includegraphics[width=\columnwidth]{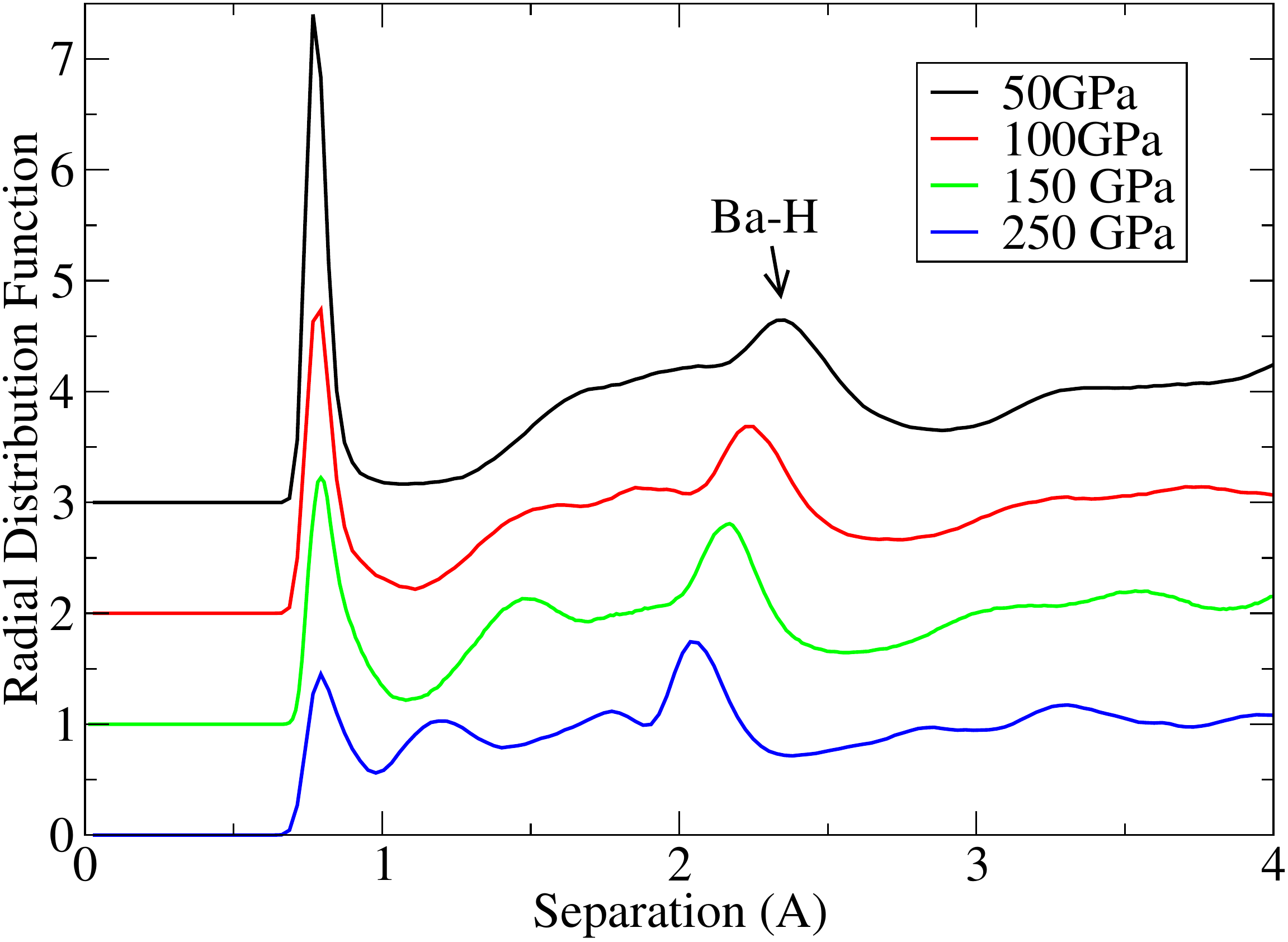}
\caption{Radial distribution functions at all pressures considered (300K) from molecular dynamics on BaH$_{12}$. \label{fig:rdf12} }
\end{figure}

\begin{figure}[t!]
\includegraphics[width=0.32\columnwidth]{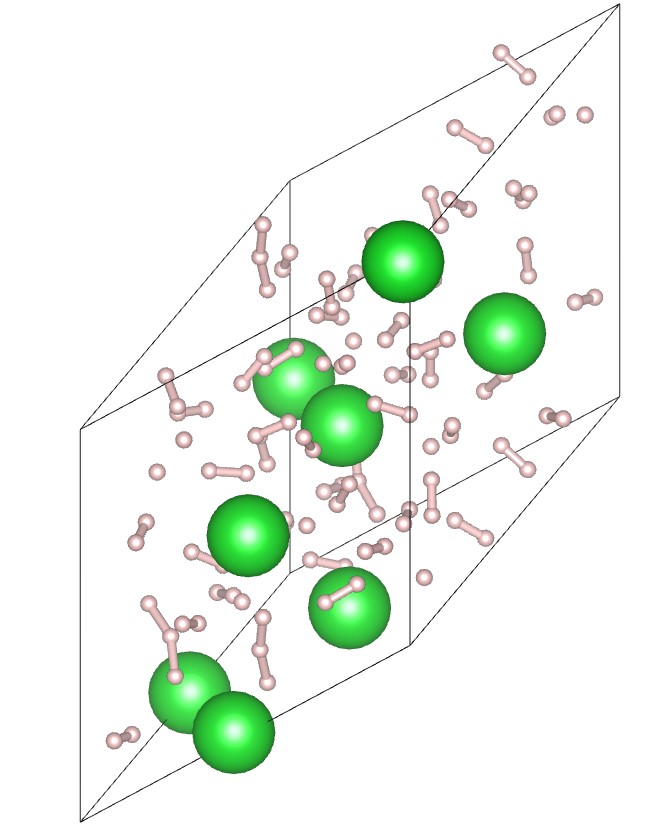}\includegraphics[width=0.32\columnwidth]{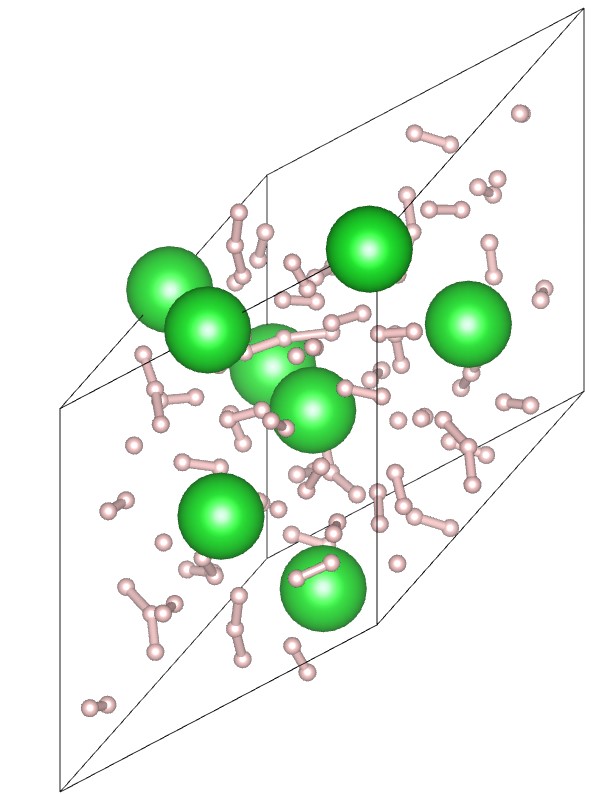}\includegraphics[width=0.32\columnwidth]{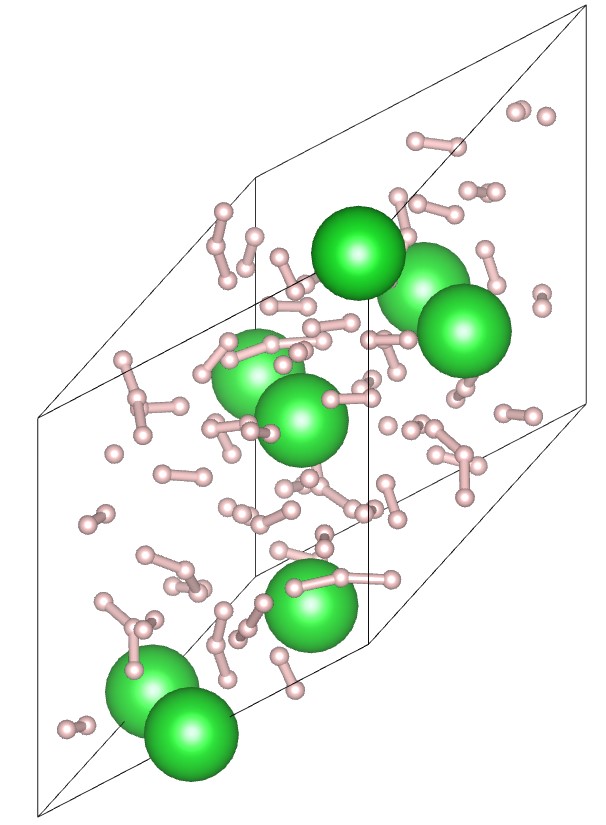}
\caption{{\bf BaH$_{12}$ as BaH$_2($H$_2)_5$}
Relaxed P1 BaH$_{12}$ structures at 50, 100 and 150GPa (left to right) Note the 100GPa case two Ba atoms have moved across the periodic boundary conditions.
Analysis of molecular dynamics simulations from this structure reveals that the number of "H$_2$ bonds" (criterion: atoms with one neighbour within 1\AA\,) is 5 per formula unit.  This conclusion is remarkably robust to reasonable choice of the bondlength.  By contrast, the more-easily calculated bond criterion "number of H-H distances less than 1\AA\, " is very sensitive to the choice of cutoff.
This is because there is a significant proton exchange via the H$_2$+H$^- \rightarrow$H$_3^-\rightarrow $H$^-+$H$_2$ mechanism.}\label{fig:P112}
\end{figure}

We ran molecular dynamics at 300K/250GPa 300K/150GPa and 300K/50GPa and found it to be stable (Sup. Figs. \ref{fig:MD12}, \ref{fig:rdf12} and \ref{fig:P112}   We used a supercell containing 8 formula units which in principle allows up to 48 molecules to form.
 For this high hydrogen concentration  bond making and breaking occurred frequently, invariably via an H$_3$ transient. Once again, the material is revealed as BaH$_2$(H$_2)_x$.  Typical calculations with  Ba$_8$H$_{96}$ generate precisely 40 molecules. 
 The sharp H-H molecular peak is present at all pressures, becoming notably less sharp, and developing a pronounced tail at higher pressures.  The second peak, corresponding to non-bonded H-H, is distinctive and moves to shorter distances with pressure, as does the BaH separation.

There is some slow but  significant diffusion of hydrogen (MSD at $\equiv 2\AA$ in 5ps).  Interestingly, the diffusion is slowest at 50GPa, and fastest at 150GPa, with 250GPa being intermediate suggesting that both the rate of bond breaking and the space available for rotation are important.  

The mean squared displacement plateaus, as expected for a solid. The Ba atoms move very little from their lattice sites (around 0.1\AA\,), despite the reduced mass. However theBarium hydrogens are fairly mobile, with a MSD around 1.5\AA\,.  This is consistent with making and breaking of bonds, plus molecular rotations.  Detailed analysis of the trajectories using vmd and inspection of lists of neighbours shows a slow diffusion of hydrogens at a timescale of tens of picoseconds associated with H+H$_2\rightarrow$H$_3\rightarrow $H$_2+$H processes followed by molecular rotation.

We used the {\it 'Atoms with precisely one neighbour within R$_{cut}$=1\AA\,'} criterion to define molecules.  If molecules were forming randomly, one would expect to see a Gaussian distribution about some average value, with strong dependence on R$_{cut}$.

In 10000 timesteps of MD with Ba$_8$H$_{96}$ at 150GPa/300K, 91\% of snapshots showed 40 molecules, 8\% showed 39 and 1\% showed 38. Only at one timestep (0.01\%) was a 41 molecule configuration reported.  The sub-40 molecule configurations typically lasted for a few fs, indicating that they are a fast physical process rather than noise.  The precise number of non-40 molecule snapshots is sensitive to the choice of R$_{cut}$, but the massive preponderance is of BaH$_2$(H$_2)_5$ configurations.  

At 250GPa, the bond-breaking is more frequent, with 40 molecules appearing only 16\% of the time, but still 99\% of snapshots show between between 36 and 40 neighbours.

In Chen's paper they present results for fcc, I4/mmm, and Cmc21 structures, although the most stable of their presented structure is pseudocubic P2$_1$.  In molecular dynamics calculation we find many snapshots with lower enthalpy, all with P1 symmetry and BaH$_2$(H$_2)_5$ character.  H$_3^+$ units are observed transiently at high T, and can be captured in relaxed structures.  We find no evidence for longer chains or hydrogen molecular units with more than three atoms.

\section{S7 Relation to superconductivity}

There has been a tremendous interest in high temperature superconductivity in polyhydrides.  This follows from Ashcroft's observation that metallic hydrogen could become a high-Tc superconductor\cite{ashcroft1968}.  The pressures required to make metallic hydrogen mean this has yet to be demonstrated.    However, many hydrogen-rich compounds have shown high Tc superconductivity at achievable pressures.  When hydrogen is mixed with a cation which donates electrons either to an atomic  hydrogen conduction band (favourable) or a molecular antibonding orbital (unfavourable).  In this way the molecular-atomic transition moves to lower pressures.  However, as in BaH$_4$,  the atomic form is much denser than the molecular form, so pressure is an essential stabilising factor via the PV term in the Gibbs free energy.

The rationale for high Tc polyhydride superconductors is that the temperature required to break the electron-phonon coupling is related to the phonon energy, and light hydrogen atoms leads to high phonon frequency and energy.    

This leads to a dichotomy at low pressures.  The hydrogens need to be close enough to form an extended phonon, but not so close that they form H$_2$ molecules.    Another is that  the atomic hydrogens in the conduction band are spaced by  the hydrogens in molecules.  These molecules nevertheless play a significant role in the Cooper-pair forming lattice phonons.   BaH$_4$ illustrates this mechanism: here the molecules are H$_2$ but other spacer molecules are possible provided their vibration couples to the lattice phonon. 

A final consideration is that the molecule formation is an alternative electron-pairing mechanism to Cooper-pair formation.  In both cases the associated band gap opening stabilises the structure.  In our Born-Oppenheimer MD there is no quantum coupling between electronic and nuclear degrees of freedom, so if the superconducting state was lower in energy than the molecularised one, we would not know. 

\providecommand{\latin}[1]{#1}
\makeatletter
\providecommand{\doi}
  {\begingroup\let\do\@makeother\dospecials
  \catcode`\{=1 \catcode`\}=2 \doi@aux}
\providecommand{\doi@aux}[1]{\endgroup\texttt{#1}}
\makeatother
\providecommand*\mcitethebibliography{\thebibliography}
\csname @ifundefined\endcsname{endmcitethebibliography}
  {\let\endmcitethebibliography\endthebibliography}{}

\end{document}